\documentclass[11pt]{article}
\usepackage{jheppub}

\usepackage{amsmath}
\usepackage{amssymb}
\usepackage{braket}
\usepackage{bbold}
\usepackage{pifont}
\usepackage{color}
\usepackage{physics}
\usepackage{wasysym}

\usepackage{romannum}

\usepackage{graphicx}
\usepackage{ytableau}
\usepackage{multirow}

\def\beq{\begin{equation}}
\def\eeq{\end{equation}}

\def\Tr{\mathop{\rm Tr}}

\fontfamily{cmss}

\def\crbig{\\\noalign{\vspace {3mm}}}

\newcommand{\md}{\mathrm{d}}
\def\ov{\overline} 

\usepackage{array} 
\usepackage[skip=0.333\baselineskip]{caption}

\title{Extracting Bigravity from String Theory}

\author[\eighthnote,\twonotes]{Dieter L\"ust}
\author[\twonotes]{, Chrysoula Markou}
\author[\twonotes]{, Pouria Mazloumi}
\author[\twonotes]{, Stephan Stieberger}

\affiliation[\eighthnote]{Arnold Sommerfeld Center for Theoretical Physics, \\ Ludwig--Maximilians--Universit\"at M\"unchen \\
Theresienstra{\ss}e 37, 80333 Munich, Germany} 

\affiliation[\twonotes]{Max--Planck--Institut f\"ur Physik (Werner--Heisenberg--Institut)\\
F\"ohringer Ring 6, 80805 Munich, Germany}

\emailAdd{dieter.luest@lmu.de, cmarkou@mpp.mpg.de}
 \emailAdd{\hskip1.4cm pmazlomi@mpp.mpg.de, stephan.stieberger@mpp.mpg.de}
 
\abstract{The origin of the graviton from string theory is well understood: it
corresponds to a massless state in closed string spectra, whose low--energy
effective action, as extracted from string scattering amplitudes, is that of
Einstein--Hilbert. In this work, we explore the possibility of such a
string--theoretic emergence of ghost--free bimetric theory, a recently proposed
theory that involves two dynamical metrics, that around particular backgrounds
propagates the graviton and a massive spin--2 field, which has been argued to be a
viable dark matter candidate. By choosing to identify the latter with a massive
spin--2 state of open string spectra, we compute tree--level three--point string
scattering amplitudes that describe interactions of the massive spin--2 with itself
and with the graviton. With the mass of the external legs depending on the
string scale, we discover that extracting the corresponding low--energy effective
actions in four spacetime dimensions is a subtle but consistent process and proceed
to appropriately compare them with bimetric theory. Our findings consist in
establishing that string and bimetric theory provide to lowest order the same set of two--derivative terms
describing the interactions of the massive spin--2  with itself and with the graviton,
albeit up to numerical coefficient discrepancies, a fact that
we analyze and interpret. We conclude with a mention of future investigations.}

\begin{document}

\begin{flushright}
\hfill{MPP--2021--79}

\hfill{LMU-ASC 15/21} 

\vspace{15mm}
\end{flushright}
\maketitle
\flushbottom

\ytableausetup
{boxsize=.65em}
\ytableausetup
{aligntableaux=center}

\section{Introduction}

In the year 1974 Scherk and Schwarz \cite{Scherk:1974ca} suggested that the massless spin--2 closed string excitation has a natural interpretation in terms
of the graviton, i.e. the field quantum of gravity. This means that the fundamental mass scale of string theory, the string tension  $\alpha'$, is related
to the characteristic mass scale of gravity, namely the Planck mass $M_P$. Subsequently it was also realized that the massless spin--2 graviton of string theory interacts according to the
covariance laws of general relativity, which was explicitly verified by computing the closed string interactions of the massless graviton and  taking the limit of $\alpha'\rightarrow 0$ 
\cite{Green:1982sw,Gross:1986mw} (see also \cite{Blumenhagen:2013fgp}).
Therefore, the low--energy effective action of the massless spin--2 excitation perfectly agrees with the Einstein action of general relativity.

In addition to its massless spectrum, string theory contains an infinite number of massive, in general higher spin, excitations.
Their presence implies
that the effective gravity action, as an expansion in powers of $\alpha'$, contains an infinite number of higher curvature corrections (for some earlier work
on higher curvature terms up to four derivatives see e.g. \cite{Stelle,Boulware,David,Horowitz,Gross:1986iv,Deser,tHooft,Maldacena,LPS}).  
In fact, the $\alpha'$  corrections to the low--energy effective of the massless fields occur when integrating out all massive string excitations, which is
a legitimate procedure 
 at low momentum transfer. However, at 
higher momenta the effective action must also contain the massive string excitations with masses
of the order of the string scale $M_s\simeq \sqrt{1/\alpha'}$ as dynamical, propagating  fields, and also their self--interactions as well as the interactions among light  and massive fields.

Among many 
modifications of general
relativity, massive gravity models (for reviews see  \cite{Hinterbichler:2011tt,deRham:2014zqa}) as well as the  DGP model \cite{Dvali:2000hr} 
are attracting a significant amount of attention. A closely related relative  of massive gravity is bimetric gravity.
The recent interest in this kind of
extensions of general relativity has been, to a considerable extent, due to the discovery that the Boulware--Deser ghost instability (BD), associated with nonlinear extensions of the Fierz--Pauli spin--2 mass term (FP), can be avoided in the context of what has come to be known as dRGT theory \cite{deRham:2010kj}. This development enabled the construction \cite{Hassan:2011tf, bimetric_1, Hassan:2013pca, Deser:2013bs, DelMonte:2016czb} (see \cite{Schmidt-May:2015vnx} for a review) of the first, and to date unique, ghost--free bimetric theory, which involves two dynamical  Lorentzian metrics $g_{\mu\nu}$ and $f_{\mu\nu}$, the kinematics of which are governed by two distinct Einstein--Hilbert actions that are associated with two different Planck scales and two independent copies of diffeomorphism invariance. The two  metrics interact nonlinearly via a potential which breaks explicitly the two copies of diffeomorphisms to their diagonal subgroup and whose very specific form guarantees the absence of the BD ghost. This second feature of the potential is essentially the main differentiating property of ghost--free bimetric theory with respect to earlier bimetric considerations by Rosen \cite{Rosen:1940zz, Rosen:1975kk}, Aragone and Deser \cite{Aragone:1971kh} and Isham, Salam and Strathdee \cite{Isham:1971gm}.

Around Minkowski backgrounds, bimetric theory propagates the massless graviton as well as a massive spin--2 field. In cosmology, the massive spin--two field  has been discussed as a possibility to be a viable candidate for dark matter in the universe, cf.  \cite{Bernard:2015gwa,  Heisenberg_1, Aoki:2016zgp, Babichev:2016bxi, Babichev:2016hir}. While the baryons (e.g. representing the full standard model of particle physics) are coupled in the standard way to the metric $g_{\mu\nu}$, species of dark matter particles couple differently to the two metrics $g_{\mu\nu}$ and $f_{\mu\nu}$ \cite{Bernard:2015gwa}. These couplings  and the coupling of massive  spin--two fields to matter can be probed by scattering amplitudes and   consistently be derived from string theory. Since  bimetric theory represents essentially a unique consistent deformation of general relativity and  gravitational polarization and modified Newton dynamics appear as natural consequences in this theory it is very interesting to study this setup within 
string phenomenology to obtain a more fundamental theory of dark matter.

It is the aim of this work to investigate how bimetric gravity originates as an effective field theory from (compactified) string theory.
Indeed,  the presence of massive spin--two fields among the 
infinite Regge excitations of a given spin  is generic and well known in string theory. Actually, a massive spin--2 excitation can occur both in the massive closed string sector and also as a massive open string state describing D--brane excitations  in type II theories.
In the first case of a massive closed string spin--2 excitation, one could extract bimetric gravity as a double copy from two open string sectors \cite{LMMS_next}.
This option would be in close analogy to the well-known double copy construction of Einstein (super) gravity as product of two open string Yang-Mills gauge theories, cf. \cite{Bern:2019prr} for a through account on double copy constructions; the possibility of a double copy construction for massive gravity has also very
recently been investigated from a field--theory perspective in \cite{Momeni:2020vvr, Johnson:2020pny}. Proceeding in a similar way, the massive closed string 
spin--2 state then emerges as the product of two massive spin-one vector bosons. The double copy construction of massive higher spin string states was considered in \cite{Ferrara:2018iko}, to which we will also refer in section 3.

However, in this work we will follow an alternative path, namely we will regard the massive spin--2 state of bimetric gravity as a particular open string excitation. In fact, it was already discussed in \cite{Ferrara:2018wlb} that bimetric gravity finds  a very natural realization in D-brane models, where the standard massless  graviton lives as closed string in the
 higher dimensional bulk, but the massive spin--2 state arises as open string on the world-volume of a D3-brane.
Part of the motivation for the open string realization of bimetric gravity comes from the observation \cite{Gording:2018not}  that the lowest order interactions of the massive spin--2 field can be rewritten via a field redefinition
as a Einstein plus (Weyl)$^2$ action. Actually the number of propagating degrees of freedom of the Einstein plus (Weyl)$^2$ action  agrees with the number of degrees of freedom
in bimetric gravity, although the massive spin--2 mode of the former is a ghost. Furthermore, in massive Weyl supergravity the  maximal number of supersymmetries, namely ${\cal N}=4$ in four dimensions, precisely agrees with the maximal amount of supersymmetry
of the massive open string sector on the the world-volume of D3-branes.

There further exists a holographic way to realize massive gravity from string theory. In particular, suitable couplings in the form of double trace derformations of (boundary) conformal field theories are long known to result in a linear combination of the original energy--momentum tensors ceasing to obey a conservation law, with the respective combination of the (bulk) gravitons of the dual AdS vacua thereby aquiring a non--zero mass \cite{Porrati:2003sa, Kiritsis:2006hy, Aharony:2006hz, Kiritsis:2008at}. More recently, such setups where shown to give again rise to massive AdS gravity when the couplings are instead accounted for by messenger fields, as well as the case in which there is no coupling, but the CFTs involved are thought of as being boundaries of a higher--dimensional CFT, with the resulting graviton mass understood as the holographic ``leakage'' of the energy--momentum tensor in the extra dimension \cite{Bachas:2017rch, Bachas:2018zmb, Bachas:2019rfq}. Our own setup is ab initio conceptually different, in that the (masless or massive) gravitons we consider are already present in the spectrum of type II string theory and do not have to be associated with two initially decoupled ``universes''/CFTs. Moreover, we are interested in the open question of the interactions of these two modes, with the hope of extracting information on the potential of ghost--free bimetric theory; we address this problem by computing, for the first time for such modes, physical observables, namely amplitudes describing their scattering. As we will see, the ones we compute are model--independent and our expressions are valid in any vacuum (before we restrict ourselves to Minkowski) and free from non--localities or quantized graviton mass constraints, which may arise in holographic realizations.

In order to derive the effective action for the massive open string spin--2 excitations themselves and also for the massless closed string excitation (the graviton), we will employ the techniques
of computing string amplitudes with massive vertex operators.  In particular, 
we will systematically set up the formalism to compute 
massive string amplitudes among massive open and massless closed string excitations. 
In this way, we will close the gap for computing tree--level scattering amplitudes of massless and massive spin--2 excitations at the first mass level in  string theory. 
As it is standard for string interactions, the amplitudes are evaluated as world--sheet integrals of correlation functions of vertex operator insertions on an appropriately chosen worldsheet, with the vertex operators corresponding to the external string states of the scattering. While it is clear that the massless spin--2,  the graviton, is described by a known closed string vertex operator,  the massive spin--2 is 
represented by an open  string vertex operator with several
interesting properties. Consequently, while the graviton is treated as a bulk field, the massive spin--2 is to be thought of as being localized on the world-volume of a D--brane. 
The relevant worldsheet topology is then that of a disk, whose boundary is attached to the brane worldvolume, with the vertex operators inserted in the bulk or on the boundary for closed or open strings respectively.

In this way, we compute, using generic boundary conditions, the superstring three--point amplitude describing 
the scattering of one massless closed string graviton and two massive open string spin--2 states. Furthermore, we compute the three--point amplitude involving three massive open string spin--2 states. 
Given that the expressions of the vertex operators are valid both in the critical and less dimensions, we are then able to restrict the resulting expressions to a four--dimensional Minkowski background and further 
take their suitable and perhaps surprisingly subtle low--energy limit, thereby deriving the corresponding low--energy effective  action, which is of course unique only up to field redefinitions. 
We then compare this effective string action with the ghost--free bimetric action, expanded around backgrounds proportional to Minkowski 
for the two metrics, to cubic order in the fields. In light of this comparison, we discover an agreement between the string and the field predictions in regard to the structure of the interactions of the massive spin--2 with the graviton and with itself, but a discrepancy in the respective numerical coefficients, which we examine and interpret, along with discussing the possibility of shedding light on the origin of the bimetric parameters from $\alpha'$ and the string  coupling.

Crucially, the construction of the vertex operators under the condition of BRST invariance imposes that they are on--shell and so all string amplitudes 
we can compute are inevitably on--shell: to compare with the bimetric action, we are thus forced to consider only the on--shell terms in the expansion of the latter and it is clear that on--shell string amplitudes can only take us so far. Let us also mention an important subtlety at the end of the introduction:
in principle, all string excitations essentially become relevant at the string scale and consequently, describing the interactions of a massless graviton with a single massive spin--2 field in this context seems at first glance 
problematic. However, as we will discuss in the course of the paper, we will be able to extract from the massive string amplitudes the terms that are relevant for the interactions of the bimetric 
massive spin--2 field by carefully defining the limit that has to be taken on the external momenta of the massless and massive fields. We will return to the issue of the (absence of an appropriate) mass gap between the various states of string towers, and its impact on the extracted low--energy effective actions, in \cite{LMMS_next}, where we assume the massive spin--2 state of bimetric theory to be part of the closed string spectrum.

The paper is organized as follows. In section 2, we review the rudiments of bimetric theory and focus on its expansion around Minkowski backgrounds; imposing appropriate on--shell conditions on its cubic vertices, we bring them to a form that can be comparable with string effective actions. In section 3, we scan the string spectrum for suitable string states and argue in favour of the identification of the massive spin--2 field with an open string state and also give the known vertex operators that we will make use of. In section 4, after introducing the tools and techniques necessary for the calculation of tree--level disk amplitudes, we compute the one corresponding to the scattering of one graviton and two massive spin--2 states, as well as the one corresponding to the scattering of three massive spin--2 states. In section 5, we construct a consistent method for extracting effective actions from string amplitudes with both massive and massless states in the external legs, derive the effective Lagrangian with one graviton and two massive spin-2 fields as well as the one involving three massive spin--2 fields and compare them with the bimetric predictions. Section 6 contains our conclusions, followed by appendices A and B, where we present several intermediate formulae and steps taken in our calculations of the two amplitudes.

\textbf{Conventions}. We use the mostly plus metric signature. The notation $[\mathbb{A}]\equiv \Tr(\mathbb{A})$ is used for a matrix $\mathbb{A}$, so that, in the Minkowski background that we will soon restrict ourselves to, we have for example that 
\beq
\begin{array}{ccl}
[G]=\eta_{\mu \nu} G^{\mu \nu} \quad , \quad [GM^2]= G_{\mu \nu} M^{\nu \lambda} M^\mu_\lambda\,.
\end{array}
\eeq
When it comes to phrasing, with ``eom'', ``dof'', ``diffeos'' and ``reps'' we refer to equations of motion, degrees of freedom, diffeomorphisms and representations respectively. Let us also note that we will systematically ignore overall unphysical numerical prefactors in the amplitudes and Lagrangians.

\section{Spin--2 field theory}

\subsection{Ghost--free bimetric theory}

In this subsection we summarize the rudiments of the Hassan--Rosen ghost--free bimetric theory. The action in four spacetime dimensions and without matter sources reads \cite{Hassan:2011tf}
\beq \label{HR}
S= m_g^2 \int \md^4x \sqrt{g} R(g) +  \alpha^2 m_g^2 \int \md^4x \sqrt{f} R(f)  - 2\, \alpha^2 m_g^4  \int \md^4x \sqrt{g}\, V(S;\beta_n)\,.
\eeq
In the above, $g_{\mu \nu}$ and $f_{\mu \nu}$ are two different metrics, whose kinematics are governed by two distinct Einstein--Hilbert terms, with $m_g$ and $\alpha m_g$ being the respective Planck masses and $\alpha$ a dimensionless parameter quantifying their ratio. They interact via the potential
\beq \label{pot}
V(S;\beta_n) =  \sum\limits_{n=0}^4 \beta_n e_n(S) \,,
\eeq
where $\beta_n$ are five free dimensionless parameters and $e_n$ are the elementary symmetric polynomials that can be defined via
\beq 
e_n(S)= S^{\mu_1}_{\hphantom{\mu_1} [\mu_1} \dots S^{\mu_n}_{\hphantom{\mu_n} \mu_n]} \,.
\eeq
Their argument is the matrix $S^\mu_{\hphantom{\mu}\nu}$, that is defined via
\beq \label{square_root}
S^\mu_{\hphantom{\mu}\nu} S^\nu_{\hphantom{\mu}\rho} = g^{\mu \sigma} f_{\sigma \rho} \quad \textrm{namely}  \quad S^\mu_{\hphantom{\mu}\nu} = (\sqrt{g^{-1}f})^\mu_{\hphantom{\mu}\nu}   \,. 
\eeq
For clarity, we note that we use the parametrization of \cite{Babichev:2016bxi, Babichev:2016hir} for the mass and interaction scales that appear in (\ref{HR}).

$V$ breaks explicitly the two distinct diffeomorphism invariances, associated with $g_{\mu \nu}$ and $f_{\mu \nu}$ respectively, to their diagonal subgroup; (each term of) the action (\ref{HR}) is namely invariant under diffeomorphisms with parameter $\xi^\mu$ under which both metrics transform in the \textit{same} manner. Infinitesimally, these take the form
\beq \label{infdd}
\delta_\xi g_{\mu \nu}= -2g_{\rho(\mu} \nabla_{\nu)} \xi^{\rho}  \quad , \quad \delta_\xi f_{\mu \nu}= -2f_{\rho(\mu} \widetilde{\nabla}_{\nu)} \xi^{\rho}\,,
\eeq
where $\nabla_\mu$ and $\widetilde{\nabla}_\mu$ are the covariant derivatives with respect to $g_{\mu \nu}$ and $f_{\mu \nu}$ respectively. By means of a Hamiltonian analysis in ADM variables it can be shown that the structure of $V$ guarantees the absence of the Boulware--Deser ghost at the full nonlinear level and that, of the initial $20$ dof of $g_{\mu \nu}$ and $f_{\mu \nu}$, only $7$ are propagating in total \cite{Hassan:2011tf, Hassan:2011ea}. As the notions of mass and spin may be bereft of meaning in non--maximally symmetric spacetimes, identification of the ``massless'' or ``massive'' dof among the propagating seven may not be generically possible, but around maximally symmetric background solutions it can be shown that the mass eigenstate basis consists of a massless and a massive spin--2 field.

Several comments are in order:
\begin{itemize}
 
 \item While $m_g$ and $\alpha m_g$ are the obvious scales that appear in (\ref{HR}), there exists a third one, namely the interaction scale that is determined by $m_g$, $\alpha$ and the arbitrarily small or large $\beta_n$. In the effective field theory description, the cutoff depends a priori on all $m_g$, $\alpha$ and $\beta_n$ and may be determined for a specific background upon inspection of the expansion of the action around it.
 
 \item The appearance of the square root matrix $S^\mu_{\hphantom{\mu}\nu}$ can be traced back \cite{deRham:2010kj, Hassan:2011vm} to the Goldstone boson analysis of massive gravity of Arkani--Hamed, Georgi and Schwartz \cite{ ArkaniHamed:2002sp}. While its presence complicates the Hamiltonian analysis at the full nonlinear level, it is actually crucial  the absence of the BD ghost.
 
 \item Given the symmetry properties of $e_n$,  it is straightforward to show that the structure of (\ref{HR}) is symmetric under the interchange $g_{\mu \nu} \leftrightarrow f_{\mu \nu}$  \cite{Hassan:2011tf}. This ceases to be the case when it comes to matter couplings: as coupling the same energy--momentum tensor to both metrics may excite ghost dof, a choice of either $g_{\mu \nu}$ or $f_{\mu \nu}$ without loss of generality has to be made \cite{Yamashita:2014fga, deRham:2014naa}.

 \item Neither $g_{\mu \nu}$ nor $f_{\mu \nu}$ is massive or massless; identification of the massive and massless states may be meaningful only around specific backgrounds.

\item For an arbitrary reference, namely non--dynamical, metric $f_{\mu \nu}$, (\ref{pot}) with (\ref{square_root}) is the unique potential that gives mass to a dynamical $g_{\mu \nu}$, governed by an Einstein--Hilbert kinetic term, in a ghost--free manner \cite{deRham:2010kj, Hassan:2011vm}. Given such an action of ghost--free massive gravity, the only, currently known to be possible, kinetic term that may be written for $f_{\mu \nu}$, upon the promotion of the latter to a dynamical field, is a kinetic term of the Einstein--Hilbert type \cite{deRham:2013tfa, deRham:2015rxa, Matas:2015qxa}. It is in this sense that (\ref{HR}) is \textit{unique}.

\end{itemize}

Choosing $g_{\mu \nu}$ as the metric to which matter, for example Standard Model fields $\Phi$ with energy--momentum tensor $T_{\mu \nu}$, couples, a term of the form
\beq
\int \md^4x \sqrt{g}\, \mathcal{L}_{\textrm{m}}(g,\Phi)
\eeq
is added to (\ref{HR}). The equations of motion for $g_{\mu \nu}$ and $f_{\mu \nu}$ take then respectively the form
\beq \label{eom}
\begin{array}{ccl}
\mathcal{G}_{\mu\nu}(g)+\alpha^2 m_g^2 \, V^g_{\mu\nu}(g,f;\beta_n) &=&\frac{1}{m^2_g} T_{\mu\nu} 
\crbig
\mathcal{G}_{\mu\nu}(f)+ m_g^2 \, V^f_{\mu\nu}(g,f;\beta_n) &=& 0 \,,
\end{array}
\eeq
where $\mathcal{G}_{\mu\nu}(g)$ and $\mathcal{G}_{\mu\nu}(f)$ are the Einstein tensors of $g_{\mu \nu}$ and $f_{\mu \nu}$ respectively and
\beq
    V^g_{\mu \nu} \equiv -\frac{2}{\sqrt{g}}\frac{\delta (\sqrt{g} V)}{\delta g^{\mu \nu}} \quad , \quad     V^f_{\mu \nu} \equiv -\frac{2}{\sqrt{f}}\frac{\delta (\sqrt{g} V)}{\delta f^{\mu \nu}}    \quad ,  \quad T_{\mu \nu} \equiv -\frac{1}{\sqrt{g}}\frac{\delta (\sqrt{g} \mathcal{L}_m )}{\delta g^{\mu \nu}}\,.
\eeq

At this point it should be clear that our goal is to extract information on the potential (\ref{pot}) from string theory. It should, moreover, be evident that, to compare with an effective action derived from scatttering amplitudes, (\ref{HR}) must be written in terms of well--defined mass eigenstates around a particular background and not in the original metrics $g_{\mu \nu}$, $f_{\mu \nu}$; we have to treat (\ref{HR}) \textit{perturbatively}. To this end, we turn to the next subsection.
 
\subsection{Expanding around equal backgrounds} \label{eae}

An important class of solutions of bimetric theory in the vacuum is that of proportional backgrounds \cite{Hassan:2012wr, Hassan:2011tf}. In the following, we restrict ourselves to the subclass of equal Minkowski backgrounds; we namely expand both metrics according to
\beq
    g_{\mu\nu}=\eta_{\mu\nu}+ \delta g_{\mu\nu} \quad , \quad  f_{\mu\nu}= \eta_{\mu\nu}+ \delta f_{\mu\nu}\,.
\eeq
within the action (\ref{HR}) and adapt all relevant formulae in our references for the particular case of Minkowski backgrounds. To diagonalize the mass matrix, namely the potential (\ref{pot}), one may define \cite{Hassan:2012wr}
\beq \label{eigen}
 G_{\mu \nu}\equiv m_g \, (\delta g_{\mu \nu} + \alpha^2 \delta f_{\mu \nu}) \quad ,  \quad  M_{\mu \nu} \equiv \alpha  m_g \, (\delta f_{\mu\nu} - \delta g_{\mu \nu})\,,
\eeq
it can be shown that the \textit{only} terms quadratic in the perturbations in the expansion of the action (\ref{HR}) are  \cite{Hassan:2012wr, Hassan:2011tf}
\beq \label{lFP}
\begin{array}{ccl}
 \mathcal{L}^{(2)}(G)&=& \frac{1}{2} G^{\mu \nu}\, \hat{\mathcal{E}}_{\mu \nu}^{ \rho \sigma} G_{\rho \sigma} 
\crbig
 \mathcal{L}^{(2)}(M)&=& \frac{1}{2} M^{\mu \nu}\, \hat{\mathcal{E}}_{\mu \nu}^{ \rho \sigma} M_{\rho \sigma} -\frac{m_{\textrm{FP}}^2}{4} \, \big( [M^2]-[M]^2\big)\,,
\end{array}
\eeq
where $\hat{\mathcal{E}}^{ \rho \sigma}_{\mu \nu}$ is the Lichnerowicz operator in (our case of the) Minkowski spacetime, given by
\beq \label{lichne}
\hat{\mathcal{E}}^{ \rho \sigma}_{\mu \nu} \equiv \frac{1}{2} \Big( \eta_\mu^\rho \eta_\nu^\sigma \partial^2 - \eta_\nu^\sigma \partial_\mu \partial^\rho -\eta_\mu^\sigma \partial_\nu \partial^\rho + \eta_{\mu \nu} \partial^\sigma \partial^\rho + \eta^{\rho \sigma} \partial_\mu \partial_\nu - \eta_{\mu \nu} \eta^{\rho \sigma} \partial^2 \Big)
\eeq
and
\beq \label{mFP}
m_{\textrm{FP}}^2 \equiv m_{\textrm{Pl}}^2\,(\beta_1 + 2 \beta_2+\beta_3)\quad , \quad m_{\textrm{Pl}}^2 \equiv m_g^2 (1+\alpha^2)\,.
\eeq
It is evident, as well as anticipated, that the kinetic terms of $G_{\mu \nu}$ and $M_{\mu \nu}$ are identical to the quadratic terms obtained from the linearization of the Einstein--Hilbert Lagrangian around a Minkowski background, while the self--interaction term of $M_{\mu \nu}$ is the standard Fierz--Pauli mass term, namely the unique ghost--free mass term for a spin--2 field in Minkowski spacetime \cite{Fierz:1939ix}. $G_{\mu \nu}$ and $M_{\mu \nu}$ are namely respectively the massless and massive eigenstates of the theory at the quadratic level, with the kinematics of both of which being GR--like. We will of course assume $m_{\textrm{FP}}^2 \neq 0 $ or, equivalently, $(\beta_1 + 2 \beta_2+\beta_3) \neq 0$.

 It should be noted that generalizations of the quadratic Lagrangian (\ref{lFP}) for the massive spin-2 field around curved backgrounds have been investigated in \cite{Buchbinder:1999ar}, where it was also shown that the respective quadratic equations of motion may be matched to the effective equations of motion extracted from bosonic open string theory to lowest order in $\alpha'$. Interestingly, these generalizations imply that a coupling of the type
\beq
R^{\alpha \mu \beta \nu} M_{\alpha \beta} M_{\mu \nu}\,,
\eeq
where $R_{\alpha \mu \beta \nu}$ is the background Riemann tensor, is allowed in a Ricci--flat background, thus modifying the Fierz--Pauli mass term. Notice that this term is absent from the bimetric expansion around Minkowski (\ref{lFP}), but, as highlighted in \cite{Nikiforova:2020sac}, may appear in ghost--free and tachyon--free torsion bigravity \cite{Sezgin:1979zf, Sezgin:1981xs, Hayashi:1979wj, Hayashi:1980av, Hayashi:1980ir, Hayashi:1980qp}. The latter is a type of generalized Einstein--Cartan theory with propagating torsion and importantly has a subclass with the same spectrum as ghost--free bimetric theory around Minkowski spacetime  \cite{Nair:2008yh, Nikiforova:2009qr, Deffayet:2011uk, Damour:2019oru}.

It is further clear that the quadratic level is not sufficient for our purposes: we are interested in interactions of the mass eigenstates and so turn to the terms that are cubic in the perturbations, the totality of which being \cite{Babichev:2016bxi, Babichev:2016hir}
\beq \label{G3_off}
\begin{array}{ccl}
\mathcal{L}_{\textrm{G}^3}&=&\frac{1}{4m_{\textrm{Pl}}}\Big[G^{\mu \nu}\big(\partial_\mu  G_{\rho \sigma} \partial_\nu G^{\rho \sigma}- \partial_\mu [G] \partial_\nu  [G]+2\partial_\nu  [G] \partial^\rho G_{\mu \rho}+2 \partial_\nu G_{\mu \rho} \partial^\rho [G]
\crbig
&& -2\partial_\rho  [G] \partial^\rho G_{\mu \nu}  +2\partial_\rho  G_{\mu\nu} \partial_\sigma G^{\rho \sigma}-4\partial_\nu  G_{\rho \sigma} \partial^\sigma  G_\mu^\rho-2 \partial^\rho  G_{\nu \sigma} \partial^\sigma G_{\mu \rho}
\crbig
&& +2\partial_\sigma  G_{\nu \rho} \partial^\sigma  G_\mu^\rho \big)  + \frac{1}{2} [G]\, \big( \partial_\rho [G] \partial^\rho [G] -\partial_\rho G_{\mu \nu} \partial^\rho G^{\mu \nu} -2 \partial_\rho [G] \partial_\mu G^{\mu \rho} 
\crbig
&& +2 \partial_\rho G_{\mu \nu} \partial^\nu G^{\mu \rho}\big) \Big]
\end{array}
\eeq
\beq \label{GM2_off}
\begin{array}{ccl}
 \mathcal{L}_{\textrm{GM}^2} &=& \frac{m_{\textrm{Pl}}}{8} (\beta_1+2\beta_2+\beta_3) \,\big[ [G][M]^2 -4[M][GM] -[G][M^2]+4[GM^2] \big]
 \crbig
&& +\frac{1}{4m_{\textrm{Pl}}}\Big[ G^{\mu\nu}\big(\partial_\mu M_{\rho \sigma}  \partial_\nu M^{\rho \sigma}-\partial_\mu [M]  \partial_\nu [M]+2\partial_\nu [M]  \partial_\rho M_{\mu}^{\rho}
\crbig
&& +2\partial_\nu M_{\mu}^{\rho}  \partial_\rho [M] -2 \partial_\rho  [M] \partial^\rho M_{\mu\nu}+2 \partial_\rho M_{\mu \nu}  \partial_\sigma M^{\rho \sigma} -4 \partial_\nu M_{\rho \sigma}  \partial^\sigma M_{\mu}^{\rho}
\crbig
&&  -2 \partial_\rho  M_{\nu \sigma}  \partial^\sigma M_{\mu}^{\rho}  +2 \partial_\sigma M_{\nu \rho}  \partial^\sigma M_{\mu}^{\rho} \big) +\frac{1}{2} [G] \big( \partial_\rho [M]  \partial^\rho [M]
\crbig
&& - \partial_\rho M_{\mu\nu}  \partial^\rho M^{\mu\nu} -2 \partial_\rho [M]  \partial_\mu M^{\mu\rho}+2\partial_\rho M_{\mu \nu}  \partial^\nu M^{\mu\rho} \big) \Big]
\crbig
&&  +\frac{1}{2m_{\textrm{Pl}}} \Big[  M^{\mu\nu} \big( \partial_\mu G_{\rho \sigma}  \partial_\nu M^{\rho \sigma}- \partial_\mu [G] \partial_\nu  [M] + \partial^\rho G_{\rho \mu}  \partial_\nu [M]
\crbig
&& +\partial_\nu G_{\mu \rho}  \partial^\rho [M]- \partial_\rho G_{\mu\nu}  \partial^\rho [M] + \partial_\rho G^{\rho \sigma}  \partial_\sigma M_{\mu \nu} -2 \partial_\mu G^{\rho \sigma}  \partial_\sigma M_{\nu \rho}
\crbig
&& + \partial_\mu [G]  \partial^\rho  M_{\rho \nu}+ \partial^\rho G_{\mu \nu}  \partial^\sigma M_{\rho \sigma}-2\partial_\rho G_{\mu \sigma}  \partial_\nu M^{\rho \sigma}- 2 \partial^\rho G_{\mu \sigma}  \partial^\sigma M_{\nu \rho}
\crbig
&&+ 2 \partial^\rho G_{\mu \sigma}  \partial_\rho M_{\nu}^{\sigma} + \partial^\rho [G]  \partial_\nu M_{\mu \rho} - \partial^\rho [G ] \partial_\rho M_{\mu \nu} \big) +\frac{1}{2} [M] \big( \partial_\rho[ G ] \partial^\rho [M]
\crbig
&& - \partial_\rho G_{\mu\nu}  \partial^\rho M^{\mu \nu}- \partial_\rho [G]  \partial_\sigma  M^{\rho \sigma}- \partial_\rho  G^{\rho \sigma}  \partial_\sigma [M] +2 \partial_\rho G_{\mu\nu}  \partial^\nu M^{\mu \rho} \big) \Big]
\end{array}
\eeq
\beq \label{M3_off}
\begin{array}{ccl}
\mathcal{L}_{\textrm{M}^3}&=& \frac{m_{\textrm{Pl}}}{24 \alpha} \big[(1-2\alpha^2)\beta_1 +3(1-\alpha^2)\beta_2 +(2-\alpha^2) \beta_3  \big] \, [M]^3
 \crbig
 && - \frac{m_{\textrm{Pl}}}{24 \alpha} \big[3(2-3\alpha^2)\beta_1 +15(1-\alpha^2)\beta_2 +3(3-2\alpha^2) \beta_3  \big] \, [M][M^2]
 \crbig
 && + \frac{m_{\textrm{Pl}}}{24 \alpha} \big[(5-7\alpha^2)\beta_1 +12(1-\alpha^2)\beta_2 +(7-5\alpha^2) \beta_3  \big] \, [M^3]
 \crbig
&& \frac{1-\alpha^2}{4 \, \alpha m_{\textrm{Pl}}}\Big[M^{\mu \nu}\big(\partial_\mu  M_{\rho \sigma} \partial_\nu M^{\rho \sigma}- \partial_\mu [M] \partial_\nu  [M]+2\partial_\nu  [M] \partial^\rho M_{\mu \rho}
\crbig
&&+2 \partial_\nu M_{\mu \rho} \partial^\rho [M]-2\partial_\rho  [M] \partial^\rho M_{\mu \nu}  +2\partial_\rho  M_{\mu\nu} \partial_\sigma M^{\rho \sigma} -4\partial_\nu  M_{\rho \sigma} \partial^\sigma  M_\mu^\rho
\crbig
&&-2 \partial^\rho  M_{\nu \sigma} \partial^\sigma M_{\mu \rho}+2\partial_\sigma  M_{\nu \rho} \partial^\sigma  M_\mu^\rho \big)  + \frac{1}{2}[ M]\, \big( \partial_\rho [M] \partial^\rho [M] 
\crbig
&& -\partial_\rho M_{\mu \nu} \partial^\rho M^{\mu \nu} -2 \partial_\rho [M] \partial_\mu M^{\mu \rho}  +2 \partial_\rho M_{\mu \nu} \partial^\nu M^{\mu \rho}\big) \Big]    \,.
\end{array}
\eeq
In the above, we have repackaged all cubic vertices of \cite{Babichev:2016bxi, Babichev:2016hir}, namely \textit{all} cubic vertices  in the expansion of (\ref{HR}) around Minkowski backgrounds in three groups: $\mathcal{L}_{\textrm{G}^3}$ and $\mathcal{L}_{\textrm{M}^3}$ contain the self--interactions of $G_{\mu \nu}$ and $M_{\mu \nu}$ respectively, while $ \mathcal{L}_{\textrm{GM}^2}$ contains the interactions of one $G_{\mu \nu}$ and two $M_{\mu \nu}$ fields. Let us also note that while the metrics $g_{\mu \nu}$ and $f_{\mu \nu}$ are dimensionless, the eigenstate perturbations (\ref{eigen}) have mass dimension $1$ and all Lagrangians have mass dimension $4$ as they should. 

Let us now summarize those observations and results of \cite{Babichev:2016bxi, Babichev:2016hir} that are relevant for our purposes:
\begin{itemize}

\item Up to cubic order, the derivative self--interactions of both $G_{\mu \nu}$ and $M_{\mu \nu}$ are strictly of the standard GR form, albeit with disparate couplings, respectively $\frac{1}{ m_{\textrm{Pl}}}$ and $\frac{1-\alpha^2}{ \alpha m_{\textrm{Pl}}}\,$. It is only $M_{\mu \nu}$  that further enjoys non--derivative self--interactions, which of course originate in the bimetric potential.

\item Up to cubic order, the only kind of interactions involving both massive and massless fields are vertices with one and only one $G_{\mu \nu}$, both derivative and non--derivative; $M_{\mu \nu}$ does not decay into $G_{\mu\nu}$ (in the absence of sources). Furthermore, due to (\ref{GM2_off}), $M_{\mu \nu}$ couples to $G_{\mu \nu}$ even for small $\alpha$.

\item Similar statements hold also at the quartic level and may be argued to stay true even at the full nonlinear level: in particular, the self--interactions of $G_{\mu \nu}$ are strictly of the standard GR form,  $G_{\mu \nu}$ namely respects \cite{Boulanger:2000rq}. Moreover, upon introducing a coupling of $g_{\mu \nu}$ to matter, it is straightforward to show  \cite{Hassan:2012wr, Hassan:2011tf} that both $G_{\mu \nu}$ and $M_{\mu \nu}$ couple to $T^{\mu \nu}$, with strengths $1/m_{\textrm{Pl}}$ and $\alpha/m_{\textrm{Pl}}$ respectively: both the massless $G_{\mu \nu}$ and the massive $M_{\mu \nu}$ mediate gravitational interactions, with those of the latter being $\alpha$ suppressed for small $\alpha$. $G_{\mu \nu}$ can, therefore, be identified with the standard GR graviton, which also justifies the definition of the Planck mass in (\ref{mFP}), while $M_{\mu \nu}$ interacts only gravitationally for small $\alpha$, thereby enjoying a fundamental observational property of DM. The idea of it being a DM candidate was systematically analyzed to constrain its mass to
\beq
 1 \textrm{ TeV} \lesssim m_{\textrm{FP}}\lesssim  66 \textrm{ TeV} \,.
\eeq

\item The various couplings in (\ref{M3_off}) and (\ref{GM2_off}) depend nontrivially on the bimetric parameters $(m_{\textrm{Pl}}, \alpha, \beta_n)$. 
\end{itemize}

It is thus clear that the first non--trivial $G_{\mu \nu}$, to which we will refer to as the graviton in the following, and $M_{\mu \nu}$ interactions appear at the cubic level  and so we aim at computing the two 3--point string amplitudes the describe the tree--level scattering of
\begin{enumerate}
\item one massless and two massive spin--2 states, the corresponding effective action of which we will extract and appropriately compare with (\ref{GM2_off}), and
\item three massive spin--2 states, the corresponding effective action of which we will extract and appropriately compare with (\ref{M3_off}),
\end{enumerate}
our ultimate goal being relating the string couplings and scale to the bimetric parameters. Before that, however, further manipulation of (\ref{G3_off}), (\ref{GM2_off}) and (\ref{M3_off}) is necessary, as we will see in the next subsection.

\subsection{On--shell conditions} \label{on_shell_calc}

As already mentioned in the introduction, the vertex operators and therefore the string amplitudes we will compute are on--shell: the polarization tensors of external string states satisfy specific conditions due to BRST and Lorentz invariance. As will see explicitly in subsection \ref{vert_build}, for the graviton and the massive spin--2 state this means that they 
\begin{itemize}
\item have on--shell mass, which reflects in momentum space the (massless or massive) Klein--Gordon eom satisfied by the respective fields, and
\item are transverse and traceless (``TT'').
\end{itemize}
with the appopriate number of dof propagating in total. Consequently, we have to impose these conditions on $G_{\mu \nu}$ and $M_{\mu \nu}$ that appear at the cubic level in (\ref{G3_off}), (\ref{GM2_off}) and (\ref{M3_off}) in order to be able to compare field with string effective actions. To see how, let us consider the quadratic eom derived from (\ref{lFP})
\beq \label{eom1_G}
\begin{array}{ccl}
 \hat{\mathcal{E}}_{\mu \nu}^{ \rho \sigma} G_{\rho \sigma} &=&0 
\end{array}
\eeq
\beq\label{eom1_M}
\begin{array}{ccl}
  \hat{\mathcal{E}}_{\mu \nu}^{ \rho \sigma} M_{\rho \sigma} - \frac{1}{2}m_{\textrm{FP}}^2 \, (M_{\mu \nu} -\eta_{\mu \nu}[M]) &=&0 \,.
\end{array}
\eeq
Clearly, the eom (\ref{eom1_G}) and (\ref{eom1_M}) are invariant under
\beq \label{inv1}
\delta G_{\mu \nu} = 2 \partial_{(\mu} \widetilde{\xi}_{\nu)} \quad , \quad \delta M_{\mu \nu} =0 \,,
\eeq
namely one copy of linearised diffeomorphisms with parameter $\widetilde{\xi}^\mu$ as anticipated and as can also be seen from the relations (\ref{infdd}) and (\ref{eigen}).

There are obviously various ways of fixing the gauge invariance (\ref{inv1}), with one such example being the transverse--traceless \textit{gauge}
\beq \label{TT1}
\partial^\mu G_{\mu \nu} =0 \quad , \quad [G] =0 \,,
\eeq
in which the eom (\ref{eom1_G}) of $G_{\mu \nu}$ takes the form of the massless Klein--Gordon equation
\beq \label{eom2}
\Box G_{\mu \nu} =0\,.
\eeq
There exist, however, residual gauge transformations, namely those that leave (\ref{TT1}) invariant, which can be invoked to finally leave $2$ propagating dof in $G_{\mu \nu}$, namely the two helicity states of the graviton. On the other hand, it can be shown (in the absence of matter sources and for $m_{\textrm{FP}}^2 \neq 0$, as we assume)  that the eom (\ref{eom1_M}) of $M_{\mu \nu}$ can be brought to the form of the massive Klein--Gordon equation
\beq \label{eom3}
(\Box - m_{\textrm{FP}}^2) M_{\mu \nu} =0\,.
\eeq
subject to the \textit{constraints} of transversality and tracelessness
\beq \label{constr1}
\partial^\mu M_{\mu \nu} =0 \quad , \quad [M] =0 \,,
\eeq
leaving thus $5$ propagating dof in $M_{\mu \nu}$, which cannot be reduced further in number. The helicity decompositions of $G_{\mu \nu}$ and $M_{\mu \nu}$ are respectively 
\begin{equation}
\pm2 \textrm{ and }\pm 2\oplus\pm1\oplus 0\, .
\end{equation}

The equations (\ref{TT1}), (\ref{eom2}) for $G_{\mu \nu }$ and (\ref{eom3}), (\ref{constr1}) for $M_{\mu \nu }$, as written in momentum space, are precisely the on--shell conditions that the respective vertex operators satistfy in the string side, as will see in subsection \ref{vert_build}. We, therefore, impose them on the cubic Lagrangians (\ref{G3_off}), (\ref{GM2_off}) and (\ref{M3_off}) and by means of partial integration obtain up to total derivatives
\beq \label{G3}
\begin{array}{ccl}
 \mathcal{L}_{\textrm{G}^3}&=& \frac{1}{m_g \sqrt{1+\alpha^2}} \, G^{\mu \nu}\big(\partial_\mu  G_{\rho \sigma} \partial_\nu G^{\rho \sigma}  -2\partial_\nu  G_{\rho \sigma} \partial^\sigma  G_\mu^\rho\big)      
\end{array}
\eeq
\beq \label{GM2}
\begin{array}{ccl}
 \mathcal{L}_{\textrm{GM}^2}&=& \frac{1 }{m_g\sqrt{1+\alpha^2}}\,  \Big[ G^{\mu\nu}\big(\partial_\mu M_{\rho \sigma}  \partial_\nu M^{\rho \sigma}  -4 \partial_\nu M_{\rho \sigma}  \partial^\sigma M_{\mu}^{\rho}   \big)
\crbig
&& \qquad \qquad + 2 M^{\mu\nu} \big( \partial_\mu G_{\rho \sigma}  \partial_\nu M^{\rho \sigma}-\partial_\rho G_{\mu \sigma}  \partial_\nu M^{\rho \sigma} \big) \Big]
\end{array}
\eeq
\beq \label{M3}
\begin{array}{ccl}
 \mathcal{L}_{\textrm{M}^3}&=&\frac{ (-\beta_1+\beta_3) \,(1+\alpha^2)^{3/2} m_g}{6\, \alpha}\,   [M^3]
\crbig
&& + \frac{(1-\alpha^2) }{m_g\alpha \sqrt{1+\alpha^2}}  \, M^{\mu \nu}\big(\partial_\mu  M_{\rho \sigma} \partial_\nu M^{\rho \sigma}  -2\partial_\nu  M_{\rho \sigma} \partial^\sigma  M_\mu^\rho\big)    \,,
\end{array}
\eeq
We have also used that for general equal backgrounds the bimetric eom (\ref{eom}) (without sources) reduce to two copies of Einstein's equations, each with cosmological constant \cite{Hassan:2012wr}
\beq \label{Lamb1}
\begin{array}{ccl}
\Lambda_g =\alpha^2 m_g^2 \, (\beta_0+3\beta_1+3\beta_2+\beta_3)  \quad , \quad \Lambda_f = m_g^2 \, (\beta_1+3 \beta_2+3 \beta_3+\beta_4)\,,
\end{array}
\eeq
where consistency forces
\beq
\Lambda_g = \Lambda_f \equiv \Lambda\,,
\eeq
while for the Minkowski backgrounds we have chosen
\beq \label{Lamb2}
\Lambda= 0\,,
\eeq
which means that two parameters among $\alpha$ and $\beta_n$ can be eliminated: we have chosen to fix $\beta_0$ and $\beta_4$ without loss of generality. It then turns out that the couplings of the vertices (\ref{G3}), (\ref{GM2}) and (\ref{M3}) finally depend on only four bimetric parameters, namely $m_g$ (or equvalently $m_{\textrm{Pl}}$), $\alpha$ and $\beta_1$, $\beta_3$.

We now make the following observations:
\begin{itemize}
\item The only case in which non--derivative vertices survive the on--shell conditions is that of the self--interactions of $M_{\mu \nu}$. 

\item The Lagrangians (\ref{G3}), (\ref{GM2}) and (\ref{M3}) contain \textit{all} possible Lorentz invariant three--point vertices involving massless and massive TT spin--2 fields that may be written, up to two derivatives, up to partial integrations, \textit{excluding} non--GR--like self--interactions of the graviton and, more importantly, $M_{\mu \nu}$ decay into gravitons, the absence of the latter being a discriminatory property of ghost--free bimetric theory. This means that these terms are not entirely unique to ghost--free bimetric theory, but what \textit{is} is the particular dependence of the various couplings on the bimetric parameters $m_g$, $\alpha$ and $\beta_1$, $\beta_3\,$. In principle we expect the string to yield all these cubic vertices, with the respective couplings depending on the string scale and couplings.

\item As we are restricted to on--shell amplitudes, the string computations will never yield information on off--shell versions of the extracted effective actions; these can be derived by means of field redefinitions and are not unique. We can namely attempt to extract (\ref{G3}), (\ref{GM2}) and (\ref{M3}), but not (\ref{G3_off}), (\ref{GM2_off}) and (\ref{M3_off}).

\end{itemize}


Let us now move on to string theory.

\section{From fields to strings}

\subsection{Identification of the relevant string states}\label{spectrum}

Our first task is to identify the massless helicity--$\pm2$ $G_{\mu \nu}$ and the massive  spin--2 $M_{\mu \nu}$ with apposite string states. Because of the tachyon's presence in the spectrum of the bosonic string, we will focus on the superstring; nevertheless the former case will prove to be of use and  so we consider both. With the critical dimension of the bosonic and of the super--string being $26$ and $10$ respectively, we are seeking those states that propagate the number of dof that correspond to symmetric and traceless massless helicity--$\pm2$  and massive spin--2 fields in the given dimension. The relevant counting is naturally given by the formula
\beq
\begin{array}{ccl}
\textrm{number of on--shell dof} =  \begin{cases} \frac{1}{2}(D-2)(D-1) -1\,, & \textrm{for}\quad  G_{\mu \nu}\\
\frac{1}{2}D(D-1) -1\,, & \textrm{for} \quad M_{\mu \nu}\,, \end{cases}
\end{array}
\eeq
with the results summarized in table \ref{dof_dim}.
\begin{table} 
\centering 
\renewcommand{\arraystretch}{1.5}
  \begin{tabular}{ c || c | c | c  | c }
   Field & Little group &  $D=4$ & $D=10$ & $D=26$\\ \hline
   $G_{\mu \nu}$ & $SO(D-2)$ & $2$  & $35$ & $299$ \\ \hline
   $M_{\mu \nu}$ & $SO(D-1)$ &  $5$ & $44$ & $324$
  \end{tabular}
\renewcommand{\arraystretch}{1}
\caption{Propagating dof in various dimensions}\label{dof_dim}
\end{table}
Notice that, as expected, the difference in the number of propagating dof between massive and massless states reflects the longitudinal dof of the massive state, namely those of a (massless) vector and (massless) scalar in the specified dimension, that add up to $(D-2)+1$.

Knowing the number of propagating dof we are looking for, and bearing in mind that they should correspond to rank--$2$ tensors of the respective little group, we can now scan the first few mass levels of the bosonic and super--string spectra in search of the desirable states. The various mass levels of the bosonic string spectrum are built using those transverse ($i=1,2,\dots,24$), right and left, bosonic oscillators $\alpha^i_{-n}$ and $\ov{\alpha}^i_{-n}$ that act as creation operators ($n \in \mathbb{N}$) in the light--cone gauge, while in the case of the superstring, using those transverse (now $i=1,2,\dots,8$), right and left, bosonic $\alpha^i_{-n}$ and $\ov{\alpha}^i_{-n}$ and fermionic $b^i_{-m}$ and $\ov{b}^i_{-m}$ oscillators, that act as creation operators in the light--cone gauge and yield bosonic states ($n \in \mathbb{N}$, $m \in \mathbb{N}_0+\frac{1}{2}$). Following \cite{Blumenhagen:2013fgp}, the lowest levels of the spectra in question are (in the superstring after the GSO projection that removes the tachyon and renders the spectrum manifestly supersymmetric in spacetime):

\begin{itemize}

\item open bosonic string: 

\begin{enumerate}
\item The ground state $| 0 \rangle  $ is a tachyon with mass $m^2=-\frac{1}{\alpha'}$ and a singlet of $SO(25)$.
\item The first excited state $\alpha^i_{-1} | 0 \rangle $ is massless and a vector of $SO(24)$.
\item The second excited level contains the states $a^i_{-2} | 0 \rangle  \,$, $ a^i_{-1} a^j_{-1} | 0 \rangle  $. These are tensors of $SO(24)$ and combine uniquely to form a symmetric representation of $SO(25)$ that propagates  $324$ dof and has mass $m^2=\frac{1}{\alpha'}$.
\end{enumerate}

\item open superstring, NS sector: 

\begin{enumerate}
\item There exists one massless state, $b_{-1/2}^i | 0 \rangle$, which is a vector of $SO(8)$ and propagates $8$ dof
\item The first massive states that appear are $b^i_{-3/2} | 0 \rangle\,$,  $\alpha^i_{-1} b^j _{-1/2} | 0 \rangle$ and $b^i_{-1/2} b^j_{-1/2} b^k_{-1/2} | 0 \rangle \,$, which are tensors of $SO(8)$ and uniquely combine to form a symmetric rank--2 tensor and an antisymmetric rank--3 tensor of $SO(9)$, both with mass $m^2=\frac{1}{\alpha'}$, that propagate $44$ and $84$ dof respectively.
\end{enumerate}

\end{itemize}

Several comments and observations are in order:
\begin{itemize}
\item For both the bosonic and the superstring case, we have assumed the presence of spacetime--filling D--branes in the case of the open string spectra, namely D$25$-- and D$9$--branes respectively; $D$--dimensional Poincar\'e invariance remains \textit{unscathed}.

\item The graviton, namely $G_{\mu \nu}$ in our case, is universally identified with the massless symmetric rank--2 tensor of the \textit{closed} string spectrum. In particular, the massless level $\ov{b}_{-1/2}^i | 0 \rangle_L  \times b^j_{-1/2} | 0 \rangle _R $ ($\alpha_{-1}^i \ov{\alpha}_{-1}^j | 0 \rangle $) of the NS--NS sector of the closed superstring (bosonic) spectrum contains a symmetric, an antisymmetric and a singlet representation of $SO(8)$ ($SO(24)$), which propagate $35$ ($299$), $28$ ($276$) and $1$ dof and are thus identified with the graviton, Kalb--Ramond field and the dilaton respectively. As highlighted in the introduction, that the graviton state enjoys  self--interactions of the Einstein--Hilbert type at tree--level was first confirmed to quartic order in the heterotic string by extracting the low--energy effective action from three-- and four--point amplitudes with this state in the external legs \cite{Green:1982sw,Gross:1986mw}.

\item In the \textit{open} string spectrum of both the bosonic and of the superstring, we observe a rank--2 tensor with mass
\beq \label{massR}
m^2=\frac{1}{\alpha'}\, ,
\eeq
propagating the correct number of dof in the critical dimension in accordance wth table \ref{dof_dim}. Motivated by a recent interpretation of the ghost spin--2 mode within Eistein$+$Weyl$^2$ supergravity as precisely this massive spin--2 brane state \cite{Ferrara:2018wlb}, and given that the sum of Weyl$^2$ gravity and infinite higher--derivative terms has been shown to be classically equivalent to ghost--free bimetric theory (for a large class of solutions of the latter) \cite{Gording:2018not}, we choose to identify $M_{\mu \nu }$ with the former case in the present paper and are currently investigating further options in \cite{LMMS_next}. Comparing the mass--squared (\ref{massR}) of the string state with Fierz--Pauli mass (\ref{mFP}) as predicted by ghost--free bimetric theory, we also identify
\beq \label{idfirst}
m_g^2 (1+\alpha^2) \, ( \beta_1 + 2 \beta_2+\beta_3) \overset{!}{=} \frac{1}{\alpha'}
\eeq
We observe that a large Planck mass $m_g^2(1+\alpha^2)$ in the LHS can be made compatible with a small string scale in the RHS for not too small values of the $\beta_n$.

\item It is instructive to appreciate how the various string states contribute to yield the correct dof of $M_{\mu \nu}$\,. In particular, the vector component is due to $ b^i_{-3/2}|0\rangle$ ($\alpha^i_{-2}  | 0 \rangle $) in the superstring (bosonic string) case, while the rest of the components lie in $b^i_{-1/2}\alpha^j_{-1}|0\rangle $ ($\alpha^i_{-1} \alpha^j_{-1} | 0 \rangle $). Notice that the tensor product of two massless vectors of the open string spectrum is not sufficient to account for the full set of propagating dof of the massive spin--2. In four dimensions, this product would amount to the following helicity decomposition
\begin{equation}\label{one}
\pm 1\otimes\pm 1=\pm 2\oplus2(0)\,,
\end{equation}
from which the two $\pm1$ helicity components of the massive spin--2 are missing, while one of the two scalar components of which is redundant.

\end{itemize}

To summarize, we choose to identify $G_{\mu \nu}$ and $M_{\mu \nu }$ with massless and massive states of the closed and open string spectrum respectively, namely treat the former as a bulk and the latter as a brane string excitation, with the little group Young tableaux given in red and blue of table \ref{ytb} correspondingly; all states of the relevant mass levels can be found in the same table, as can the Young tableaux of their $SO(D-2)$ representations, with $D$ being the appropriate critical dimension.

\begin{table}
\centering 
\renewcommand{\arraystretch}{1.5}
  \begin{tabular}{ c || c | c  | c | c | c }
   Type &  $\alpha' m^2 $ & Case & String states & $SO(D-2)$ reps & little group reps \\ \hline \hline
   \multirow{2}{*}{closed} & \multirow{2}{*}{0}  & bosonic & $ \alpha_{-1}^{i} \ov{\alpha}_{-1}^{j} | 0 \rangle $ & \multirow{2}{*}{$\ydiagram{1} \times \ydiagram{1}$} &  \multirow{2}{*}{$\textcolor{red}{\ydiagram{2}} + \ydiagram{1,1} + \bullet$}  \\ \cline{3-4}
   & & superstring &  $\ov{b}_{-1/2}^i | 0 \rangle_L  \times b^j_{-1/2} | 0 \rangle _R $ &  &   \\ \cline{1-6}
   \multirow{5}{*}{open} & \multirow{5}{*}{$+1$} & \multirow{2}{*}{ bosonic}   & $ \alpha^i_{-2} | 0 \rangle   $ & $\ydiagram{1}$ & \multirow{2}{*}{\textcolor{blue}{\ydiagram{2}} } \\ \cline{4-5}
   & &   & $ \alpha^i_{-1} \alpha^j_{-1}| 0 \rangle $ & $\ydiagram{2} + \bullet$  &  \\ \cline{3-6}
 & &\multirow{3}{*}{superstring}  &  $b^i_{-3/2} | 0 \rangle $  &  \ydiagram{1} &   \multirow{3}{*}{$\textcolor{blue}{\ydiagram{2}} +\ydiagram{1,1,1}$} \\ \cline{4-5}
 & & &  $\alpha^i_{-1} b^j _{-1/2} | 0 \rangle $  & $\ydiagram{2} + \ydiagram{1,1} + \bullet$ &   \\ \cline{4-5}
 & &   & $b^i_{-1/2} b^j_{-1/2} b^k_{-1/2} | 0 \rangle $  &  $ \ydiagram{1,1,1}$ &  
  \end{tabular}
\caption{The relevant bosonic and superstring mass levels ($i=1,2,\dots, D-2$).} \label{ytb}
\renewcommand{\arraystretch}{1}
\end{table}

\subsection{Disk amplitudes involving massless and massive open and closed strings} \label{vert_build}

As is well known, in the string $S$--matrix approach around a fixed background (with a conformal field theory description) the scattering of string states is described by a perturbative expansion in the (dimensionless) string coupling $g_s$ and the (dimensionful) parameter $\alpha'$; in this work we will compute tree--level amplitudes, which will be expansions in $\alpha'$. As is clear from the previous section, we are interested in particular in the interactions of closed and open string states at tree--level.
Tree--level amplitudes ${\cal A}(N_0,N_c)$ involving $N_o$ open and $N_c$ closed strings are described by a disk world--sheet, which is an oriented manifold with one boundary.
The latter can be conformally mapped to the unit disk, or, equivalently, the upper complex half--plane $\mathcal{H}_+= \{z \in \mathbb{C} | \Im(z)>0\}$, where $z$ is here the worldsheet complex coordinate; its boundary is the real line parametrized by $x \in \mathbf{R}$. As is custom in the literature, $\mathcal{H}_+$ will be referred to as the disk in the following.

By means of the operator--state correspondence, asymptotic closed and open string states are represented by vertex operators which are thought of as being inserted in the interior and the boundary of the disk respectively. A generic physical state with momentum $p^\mu$ is created at the worldsheet location $(z,\ov{z})$ by a vertex operator $V(z,\ov{z})$ described by a conformal field in the two--dimensional worldsheet conformal field theory. In this way an open string state $j$  and a closed string state $i$ are represented by an operator $V_o(x_j)$ inserted at a point $x_j$ and an operator $V_c(z_i, \ov{z}_i)$ inserted at a point $z_i$ in $\mathcal{H}_+$, respectively. The tree--level disk amplitude involving $N_o$ open and $N_c$ closed strings takes the form 
\beq \label{genampl}
\mathcal{A}(N_o,N_c) = \sum_{\sigma } \Bigg( \int_{\mathcal{I}_\sigma} \displaystyle\prod_{j=1}^{N_o} \md x_j \, \displaystyle\prod_{i=1}^{N_c} \int_{\mathcal{H}_+} \md^2 z_i  \Bigg) \, V_{\textrm{CKG}}^{-1} \ \, \langle \displaystyle\prod_{j=1}^{N_o}:V_o(x_j): \, \displaystyle\prod_{i=1}^{N_c}:V_c(z_i, \ov{z}_i): \rangle_{\mathbb{D}_2}\,,
\eeq
see for  \cite{Stieberger:2009hq} for the generic setup and its relation to pure $N_o+2N_c$--point open string amplitudes. The correlator is evaluated by Wick contractions and using the conformal field theory 
on the disk $\mathbb{D}_2$ to be specified in the sequel. General and explicit expressions 
for (\ref{genampl}) have been worked out in \cite{Stieberger:2015vya}. Due to conformal invariance all the vertex operators  positions are integrated over the whole worldsheet. For the open string vertex operators, the integration regions $\mathcal{I}_\sigma$ are due to all possible cyclic inequivalent orderings $\sigma \in S_{N_o-1}$ of the $V_o(x_j)$ along the boundary, which are summed over. The (inverse of the) volume $V_{\textrm{CKG}}$ of the worldsheet conformal Killing group, in the case of the disk the group being $PSL(2,\mathbf{R})$, also appears as a prefactor in (\ref{genampl}); otherwise the amplitude diverges.

Amplitudes involving open and closed strings describe excited D--branes emitting and absorbing bulk closed string states. The computation of scattering amplitudes off D--branes  has been initiated for massless external states in 
\cite{Klebanov:1995ni,Gubser:1996wt,Garousi:1996ad,Hashimoto:1996kf}, for intersecting D--branes (D--branes with background flux) in \cite{Lust:2004cx}, while the scattering of one massive state and other massless states  in four--dimensional D--brane compactifications has been accomplished  in \cite{Feng:2010yx}.
A recent investigation of open string  amplitudes involving only one massive state and any number of massless ones through field--theory techniques has been performed in 
\cite{Guillen:2021mwp}. 
In this work we set out to compute $\mathcal{A}(2,1)$ and $\mathcal{A}(3,0)$ involving at least two massive string states and  describing inelastic scattering of closed strings off D--branes. Therefore, there are two significant differences between all past literature and in the present work, where
\begin{itemize}
\item \textit{all} external states are either helicity--2 or spin--2, and
\item at least \textit{two} external states are massive.
\end{itemize}
As we will see, the former point is associated with considerably more tedious, compared to the scattering of lower--spin states, calculations required to arrive at the full amplitudes, while the latter is the origin of a serious subtlety in deriving the corresponding effective actions. To the best of our knowledge, the subtlety in question has never been encountered in the literature before; we will explain its nature and argue as to how a consistent effective action can be extracted in such cases.

To proceed, we need the vertex operators $V_G(z,\ov{z})$ and $V_B(x)$ that create the states in table (\ref{ytb}) that describe $G_{\mu \nu}$ and $M_{\mu \nu }$ respectively. In general, a vertex operator is built using the momentum eigenstate $e^{iqX(z)}$ and other operators and commutes with the BRST charge (up to total derivatives) \cite{Friedan:1985ge}. In the case of the superstring that we will focus on, the string coordinates $X^\mu$, $\psi^\mu$, that are vectors of $SO(9,1)$ and also worldsheet superpartners, suffice to build $V_G(z,\ov{z})$ and $V_B(x)$. Let us note that a vertex operator describing a superstring state is not unique; rather, there exist infinite vertex operators for each state, each of which is written in a specific ``ghost picture'', namely carries a different ghost charge $\chi$ which appears in the form  $e^{\chi\phi}$; this degeneracy is due to the superconformal ghost system, whose bosonization involves a scalar field $\phi$. 

In the $(0,0)$--ghost picture, the graviton vertex operator is given by (cf. e.g. \cite{Mayr:1993vu})
\beq \label{gravert}
V_G^{(0,0)}(z,\ov{z},\varepsilon,q) =-\frac{2 g_c}{\alpha'}\ \varepsilon_{\mu \nu} \, \Big[ i\ov\partial X^\mu +\frac{\alpha'}{2} (q\widetilde{\psi})\widetilde{\psi}^\mu(\ov{z}) \Big] \Big[ i \partial X^\nu +\frac{\alpha'}{2} (q \psi) \psi^\nu(z) \Big]  \, e^{iqX(z, \ov{z})}\,,
\eeq
subject to the  conditions
\beq \label{on1}
\begin{array}{ccl}
\varepsilon_{\mu \nu} q^\mu = \varepsilon_{\mu \nu} q^\nu = 0 \quad ,  \quad q^2 =0 
\crbig
\varepsilon_{\mu \nu} = \varepsilon_{\nu \mu} \quad , \quad \varepsilon_{\mu \nu} \eta^{ \mu \nu}=0\,,
\end{array}
\eeq
where $\varepsilon_{\mu \nu}$ and $q$ the polarization tensor and the momentum of the graviton state, respectively.
Above we have the closed string coupling  $g_c$. Usually, the latter is related to the gravitational coupling constant as $g_c=\frac{\kappa}{2\pi}$.

In the $(-1)$--ghost picture, the vertex operator that creates the massive spin--2 in the superstring case of table (\ref{ytb}) is given by \cite{Koh:1987hm,Feng:2010yx,Feng:2012bb}
\beq \label{vertop}
V_M^{(-1)}(x,\alpha,k)= \frac{g_{o}}{(2\alpha')^{1/2}}\ T^a\ e^{-\phi(x)}\, \alpha_{\mu \nu} \,i \partial X^\mu (x) \psi^\nu(x) \ e^{ikX(x)}\,,
\eeq
with  $T^a$ being the Chan--Paton factor referring to generators of the respective group on the brane and subject to the on-shell conditions
\beq\label{on2}
\begin{array}{ccl}
\alpha_{\mu \nu} k^\mu = 0 \quad ,  \quad k^2=-\frac{1}{\alpha'}
\crbig
\alpha_{\mu \nu} = \alpha_{\nu \mu} \quad , \quad \alpha_{\mu \nu} \eta^{ \mu \nu}=0\,,
\end{array}
\eeq
where $\alpha_{\mu \nu}$ and $k$ are respectively the polarization tensor and the momentum of the massive spin--2 state. 
Above we have the open string coupling  $g_o$, which usually is related to the 
gauge coupling constant as $g_o=(2\alpha')^{1/2} g_{\textrm{YM}}$. 
In the zero--ghost picture the vertex operator  for the massive spin--2 state  is given by \cite{Bianchi:2015yta}
\beq \label{vertop2}
\begin{array}{ccl}
V_M^{(0)}(x,\alpha,k) &=&\displaystyle{ \frac{g_{o}}{(2\alpha')}\  T^{a} \  \alpha_{\mu \nu} \big[ i \partial X^\mu (x) \partial X^\nu (x) -2i\alpha' \partial\psi^\mu (x) \psi^\nu(x)}
\crbig
&&\displaystyle{ \qquad  \qquad  \qquad  \qquad  +2\alpha' \, (k\psi)(x)\, \psi^\nu(x) \partial X^\mu(x) \big]\  e^{ik X (x)} \,,}
\end{array}
\eeq
where we have kept $\alpha'$ arbitrary and reinstated the correct normalization.

We shall also need the vertex operator that creates the massive spin--2 in the bosonic case of table (\ref{ytb}), which is given by \cite{Bianchi:2015yta}
\beq \label{bosv2}
\widetilde{V}_M(x,\alpha,k)= \frac{g_{o}}{(2\alpha')}\ T^a \ \alpha_{\mu \nu}\, i\partial X^\mu (x) i \partial X^\nu (x)\ e^{ikX(x)} \,,
\eeq
subject to the same conditions (\ref{on2}). It is obvious that the concept of the ghost picture carries no meaning in the bosonic case, as there is no superghost system.

Several comments are in order:
\begin{itemize}

\item It is now clear that the conditions (\ref{on1}) are precisely the equation of motion (\ref{eom2}) and the TT gauge (\ref{TT1}) of $G_{\mu \nu}$ from the field--theory perspective, while the conditions (\ref{on2}) are the equation of motion (\ref{eom3}) and the constraints (\ref{constr1}) of $M_{\mu \nu}$. From the string theory perspective, the condition of transversality of the first line of (\ref{on1}) and of (\ref{on2}) is a consequence of BRST invariance, while those of the second line are due to the properties of the respective Lorentz representation. We would, therefore, like to highlight that transversality and tracelessness are seen either as a gauge choice or a constraint related to diffeos in field theory, while they are respectively imposed by BRST and Lorentz invariance in string theory. In the following we will refer to the full sets (\ref{on1}) and (\ref{on2}) as the on--shell conditions.

\item A vertex operator for an arbitrary mass level is constructed by writing a linear combination of all possible operators with conformal dimension $(1,1)$ or $1$ (for the closed or the open string respectively) and imposing BRST invariance. For the case of the superstring, let us recall that the conformal dimension of $e^{\chi\phi}$ is $-\frac{1}{2}\chi^2-\chi$. Up to a suitable overall normalization, the full vertex operators for $m^2=\frac{1}{\alpha'}$ are \cite{Koh:1987hm}
\beq \label{vertop_all}
\begin{array}{ccl}
V^{(-1)}(x,\alpha,k) &=&  \displaystyle{\frac{g_{o}}{(2\alpha')^{1/2}}\  \,e^{-\phi(x)}\, \big[ E_{\mu \nu \rho} \psi^\mu \psi^\nu \psi^\rho + ( \alpha_{\mu \nu} +\sigma_{\mu \nu}) \,i \partial X^\mu \psi^\nu }
\crbig
&& + H_\mu \partial \psi^\mu + \xi_\mu \partial \phi \,\psi^\mu \big] \, e^{ikX}
\end{array}
\eeq
for the superstring in the $-1$ ghost picture and
\beq \label{bosv2_all}
V(x,\alpha,k)=\frac{g_{o}}{(2\alpha')}\ \big( B_\mu \, i \partial^2X^\mu + \alpha_{\mu \nu}\, i\partial X^\mu i \partial X^\nu \big) e^{ikX} 
\eeq
for the bosonic string, where $ E_{\mu \nu \rho}$ and $\sigma_{\mu \nu}$ are antisymmetric. Ignoring the contribution of the term $\xi^\mu$ as it can be accounted for by two of the other terms and a total derivative, let us note that each of the terms in (\ref{vertop_all}) and in (\ref{bosv2_all}) corresponds to each of the $SO(D-2)$ states at the given mass level as presented in table (\ref{ytb}). Due to BRST invariance, it can be shown that (\ref{vertop_all}) reduces to (\ref{vertop}) subject to (\ref{on2}) and the vertex operator of a massive transverse (completely antisymmetric) three-form, while (\ref{bosv2_all}) reduces to (\ref{bosv2}) subject to (\ref{on2}). It is, therefore, clear, that the resulting vertex operators create precisely the representations of the little group at this mass level as given in table (\ref{ytb}).

\item All these considerations are obviously true in the critical dimension. However, it can be argued that, \textit{independently} of the details of the compactification to the four--dimensional spacetime that is of interest to us from the point of view of bimetric theory, the vertex operators (\ref{gravert}), (\ref{vertop}) and (\ref{bosv2}), where now $\mu,\nu, \dots=0,1,2,3$  describe $G_{\mu \nu}$ and $M_{\mu \nu}$ equally well in $D\!=\!4$.
In other words, since the vertex operators under consideration do not involve fields referring to the internal compactification our results hold for any string compactification 
with D--brane and orientifold background, which allows for a conformal field theory description  \cite{Blumenhagen:2006ci} (see also \cite{Ferrara:2018wqd} for the case of the massive spin--2 vertex operator).
\end{itemize}

We are now ready to start computing the string amplitudes $\mathcal{A}(2,1) $ and $\mathcal{A}(3,0)$ describing the scattering involving $G_{\mu \nu}$ and $M_{\mu \nu}$ in four--dimensional spacetime. The ghost picture in which the vertex operators are written within an amplitude should be such that the total background ghost charge of $+2$ on the disk is cancelled. With the ghost charge being additive and its distribution among the operators within a single amplitude being arbitrary, while all possible inequivalent operator orderings at the boundary should be taken into account, we may write without loss of generality
\beq \label{ampl}
\begin{array}{ccl}
\mathcal{A}(2,1) &=&  \displaystyle  \int_{\mathbf{R}}\int_{\mathcal{H}_+} \frac{  \md x_1 \md x_2 \,  \md^2z }{V_{\textrm{CKG}} }
\crbig
&&  \quad \langle :V_M^{(-1)}(x_1,\alpha_1,k_1): \, :  V_M^{(-1)}(x_2,\alpha_2,k_2): \, :V_G^{(0,0)}(z, \ov{z},\varepsilon,q):\rangle_{\mathbb{D}_2}
\end{array}
\eeq
and
\beq \label{ampl2_1}
\begin{array}{ccl}
\mathcal{A}(3,0) &=&\displaystyle  \int_{\mathbf{R}}   \frac{   \prod\limits_{i=1}^{3} \md x_i}{V_{\textrm{CKG}}}   \,  \langle :V_M^{(-1)}(x_1, \alpha_1,k_1): \, :  V_M^{(-1)}(x_2, \alpha_2,k_2):\, :V_M^{(0)}(x_3, \alpha_3,k_3):\rangle_{\mathbb{D}_2}
\crbig
&& \qquad + (2\leftrightarrow 3)\,,
\end{array}
\eeq
where the permutation of labels $2$ and $3$ takes into account the fact, that on a disk
the conformal Killing group does not change the cyclic ordering of the vertex operators positions $x_i$. For (\ref{ampl2_1}) we also need the vertex operator (\ref{vertop2}) for the massive spin--2 state in the zero--ghost picture. Finally, for the bosonic string we simply have:
\beq \label{genampl2}
\widetilde{\mathcal{A}}(3,0) = \displaystyle  \int_{\mathbf{R}}   \frac{   \prod\limits_{i=1}^{3} \md x_i}{V_{\textrm{CKG}}}\ \langle \displaystyle\prod_{i=1}^{3}:\widetilde{V}_B(x_i,\alpha_i,k_i):  \rangle_{\mathbb{D}_2} + (2\leftrightarrow 3)\,,
\eeq

\subsection{Disk correlators involving open and closed string fields} \label{Cor}

Due to the boundary at the real axis there are non--trivial correlators between
left-- and right--movers. In order to compute the amplitudes, it is convenient to use
the ``doubling trick,'' \cite{Gubser:1996wt,Garousi:1996ad} to convert disk correlators to the standard holomorphic ones.
This method  accommodates the boundary conditions by extending the definition of holomorphic fields to the entire complex plane
such that their operator product expansions (OPEs) on the complex plane reproduce all the OPEs among holomorphic and anti--holomorphic fields on ${\cal H}_+$.
By the boundary conditions on the D--brane world--volume the open string momenta $k_i$ are restricted to lie within the world--volume directions. On the other hand, the closed string momenta $q_r$  have generic
directions. 
Since D--branes are infinitely heavy objects they can absorb momentum in the transverse direction, which in turn implies that only along the world--volume directions momentum conservation is furnished. This can be taken into account by introducing the matrix
 $D$ 	accounting for the specific boundary conditions
in $d$ space--time dimensions.
Then, the longitudinal closed string momentum is given by
\begin{equation}\label{para}
q^\parallel=\frac{1}{2}\ (q+Dq)\ ,
\end{equation}
while normal to the brane we have the remaining momentum
\begin{equation}\label{perp}
q^\perp=\frac{1}{2}\ (q-Dq)\ ,
\end{equation}
and total momentum conservation along the D--brane world volume reads:
\begin{equation}\label{momcons}
\sum_{i=1}^{N_o}k_i+\sum_{r=1}^{N_c}q_r^\parallel=0\ .
\end{equation}
Typically, in flat space--time the matrix $D^{\mu\nu}$ is a diagonal matrix, equal to Minkowski metric $\eta^{\mu\nu}$ in directions along the D--brane
(Neumann boundary conditions) and to $-\eta^{\mu\nu}$ in directions orthogonal to the brane (Dirichlet boundary conditions). 

With these preliminaries, the relevant correlators $\langle\ldots\rangle_{\mathbb{D}_2}$ on the disk ${\cal H}_+$ read, see for example \cite{Garousi:1996ad,Hashimoto:1996kf}
\beq \label{corrs}
\begin{array}{ccl}
\langle X^\mu(z_1) X^\nu(z_2) \rangle = -\frac{\alpha'}{2} g^{\mu \nu} \, \ln(z_1-z_2) \quad ,  \quad \langle X^\mu(z_1) \widetilde{X}^\nu(\ov{z}_2) \rangle = -  \frac{\alpha'}{2} D^{\mu \nu} \, \ln(z_1-\ov{z}_2) 
\crbig
\langle X^\mu(x) X^\nu(z) \rangle  =  -  \alpha' g^{\mu \nu} \, \ln(x-z) 
\crbig
\langle X^\mu(x_1) X^\nu(x_2) \rangle = - 2 \alpha' g^{\mu \nu} \, \ln(x_1-x_2) 
\crbig
\langle \psi^\mu(z_1) \psi^\nu(z_2) \rangle  = \frac{g^{\mu \nu}}{z_1-z_2} \quad , \quad \langle \psi^\mu(z_1) \widetilde{\psi}^\nu(\ov{z}_2) \rangle  = \frac{D^{\mu \nu}}{z_1-\ov{z}_2} \quad , \quad \langle \psi^\mu(x_1) \psi^\nu( x_2) \rangle = \frac{g^{\mu \nu}}{x_1 -x_2 } 
\crbig
\langle e^{\phi(z_1)} e^{\phi(z_2)} \rangle = \frac{1}{z_1-z_2}\quad , \quad \langle e^{-\phi(x_1)} e^{-\phi(x_2)} \rangle = \frac{1}{x_1-x_2}\,,
\end{array}
\eeq
with  $x_i\in \mathbf{R}$ referring to open string positions and $z_r\in{\cal H}_+$ denoting closed string positions of the corresponding fields.
Note that the form of the boson correlators is the same in both the bosonic and the superstring. In all amplitudes of the type (\ref{genampl}), there always appears the so--called Koba--Nielsen factor upon performing the contractions \cite{Stieberger:2009hq,Stieberger:2015vya}
\beq \label{KBgen}
\begin{array}{ccl}
\mathcal{E} &\equiv& \langle \displaystyle  \prod_{j=1}^{N_o} :e^{ik_j X(x_j)}:  \, \prod_{i=1}^{N_c}  :e^{iq_iX(z_i,\ov{z}_i)}: \rangle
\crbig
&=& \displaystyle  \prod_{j_1<j_2}^{N_o} |x_{j_1}-x_{j_2}|^{2\alpha' k_{j_1}k_{j_2}} \displaystyle  \prod_{i=1}^{N_c} |z_i-\ov{z}_i|^{\alpha' q_{i\parallel}^2} 
\crbig
&& \qquad \times  \displaystyle  \prod_{j=1}^{N_o}  \prod_{i=1}^{N_c} |x_{j}-z_i|^{2\alpha' k_j q_i}  \displaystyle  \prod_{i_1<i_2}^{N_c} |z_{i_1}-z_{i_2}|^{\alpha' q_{i_1}q_{i_2}}|z_{i_1}-\ov{z}_{i_2}|^{\alpha' q_{i_1}Dq_{i_2}}
\end{array}
\eeq
subject to the momentum conservation condition (\ref{momcons}). 
Finally, in order to correctly normalize the amplitudes (\ref{genampl}), some additional factors have to be taken into account. They stem from determinants and Jacobians of certain path integrals. On the disk, the net result of those contributions is an additional factor of
\cite{Polchinski:1998rq}:
\begin{equation}
C_{D_2}=\frac{1}{\alpha'g_o^2}\ .
\end{equation}
\section{String amplitudes involving massless and massive spin--two states}

\subsection{Kinematics}

All open string states have momentum $k_i$ parallel to the brane, while the momentum $q_r$ of the closed string state splits into a parallel (\ref{para}) and orthogonal part (\ref{perp}).

For the amplitude $\mathcal{A}(2,1)$ involving two massive states subject to (\ref{on2}) and one graviton subject to (\ref{on1}) the condition of momentum conservation (\ref{momcons})  takes then the form
\beq \label{momc}
(k_1+k_2 + q_\parallel)^\mu = 0
\eeq
and one may further define the Mandelstam variables as
\beq \label{mandel}
\begin{array}{ccl}
s \equiv \alpha'(k_1+k_2)^2 = -2 +2 \alpha' k_1k_2
\crbig
t \equiv \alpha'(k_1+k_3)^2 =-1 + \alpha' k_1 q
\crbig
u \equiv \alpha'(k_1+k_4)^2=-1+\alpha'k_1 Dq\,,
\end{array}
\eeq
where we have defined
\beq \label{def_momenta}
k_3^\mu \equiv \frac{q^\mu}{2} \quad , \quad k_4^\mu \equiv \frac{(Dq)^\mu}{2}\,.
\eeq
It is also easy to see that $ k_{1,2} \cdot q = k_{1,2} \cdot Dq$, so that there is exactly one kinematic invariant $t$, just like in the scattering of three massless states, with
\beq \label{morem}
u=t \quad , \quad s = -2 -2t 
\eeq 
and (\ref{momc}) takes the form
\beq \label{consGM2}
s+t+u=-2\,.
\eeq

Furthermore,   for the amplitude $\mathcal{A}(3,0)$ involving three massive states subject to (\ref{on2}) the condition of momentum conservation (\ref{momcons}) assumes the form
\beq \label{consM3}
(k_1+k_2+k_3)^\mu =0 \, .
\eeq
By taking the square of the latter and using (\ref{on2}) we obtain
\beq \label{produs}
k_i \cdot k_j = (2\alpha')^{-1} \quad, \quad i \neq j \,.
\eeq

Notice that since in our case we assume spacetime filling D--branes, both in the original setup in the critical dimension and after the compactification, the condition of momentum conservation along the brane becomes in fact momentum conservation in the full four--dimensional spacetime. 

\subsection{Three graviton amplitude on the sphere}

For completeness and comparison let us briefly display the tree--level amplitude of three massless gravitons.
The universal three--point (massless) graviton amplitude $\mathcal{M}(3)$ on the worldsheet sphere is given by \cite{Gross:1986mw,Lust:1987xm}
\beq \label{ampl_G3}
\begin{array}{ccl}
\mathcal{M}(3)&=& g_c \,    \Big[ (q_1\cdot \varepsilon^2 \cdot q_1) (\varepsilon^3 \cdot \varepsilon^1)+(q_2 \cdot \varepsilon^3 \cdot q_2) (\varepsilon^2 \cdot \varepsilon^1)+(q_3 \cdot \varepsilon^1 \cdot q_3) (\varepsilon^2 \cdot \varepsilon^3)
\crbig
&& +2\, q_1 \cdot \varepsilon^2\cdot  \varepsilon^1\cdot  \varepsilon^3\cdot  q_2 +2 \, q_2 \cdot \varepsilon^3 \cdot  \varepsilon^2 \cdot  \varepsilon^1\cdot  q_3+  2 \, q_3 \cdot \varepsilon^1 \cdot  \varepsilon^3 \cdot  \varepsilon^2 \cdot  q_1 \Big]+\mathcal{O}(q^4)\,,
\end{array}
\eeq
with the closed string coupling $g_c$, the graviton momenta $q_i$ and polarizations $\varepsilon^i$. This is the three graviton amplitude in pure Einstein gravity valid for both heterotic and type II string theory.
On the other hand, in the presence of D--branes or for Einstein--Yang--Mills theory one needs to compute the three graviton amplitude (\ref{ampl2_1}) on the worldsheet disk, cf. also \cite{Stieberger:2014cea_1, Stieberger:2014cea_2, Stieberger:2014cea_3, Stieberger:2014cea_4}. In this case the amplitude is one power of $g_c$ higher than the sphere. This originates from the different normalization factors $\frac{C_{D_2}}{C_{S_2}}=\frac{g_c^2}{8\pi g_o^2}\sim g_c$ or Euler numbers 
$\frac{g_c^{-\chi(D_2)}}{g_c^{-\chi(S_2)}}=\frac{g_c^{-1}}{g_c^{-2}}\sim g_c$ with the open and closed string coupling constants $g_o$ and $g_c$, respectively.

\subsection{Amplitude of one graviton and two massive spin--two states}

In this subsection we will compute $\mathcal{A}(2,1)$ involving one graviton and two massive spin--two states. Substituting the vertex operators (\ref{gravert}) and (\ref{vertop}) in (\ref{ampl}), we obtain
\beq \label{ampl_2}
\begin{array}{ccl}
\mathcal{A}(2,1) &=& \frac{g_c g_o^2}{\alpha'^2}\ C_{D_2} \, \Tr \big( T^a T^b\big) \, \alpha_{\kappa \lambda}^1 \alpha_{\rho \sigma}^2 \varepsilon_{\mu \nu} \,    \displaystyle  \int_{\mathbf{R}}\int_{\mathcal{H}_+} \frac{  \md x_1 \md x_2 \,  \md^2z }{V_{\textrm{CKG}} } \langle :e^{-\phi(x_1)}: \, : e^{-\phi(x_2)}:  \rangle 
\crbig
&&  \Big\{  \langle : \partial X^\kappa (x_1) e^{ik_1X(x_1)}: \,:  \partial X^\rho (x_2) e^{ik_2X(x_2)} : \, : \ov{\partial} X^\mu (\ov{z}) e^{iq \widetilde{X}(\ov{z})} :\,: \partial X^\nu(z)e^{iq X(z)}: \rangle  
\crbig
&& \times \, \langle    \psi^\lambda(x_1) \psi^\sigma(x_2)  \rangle - \frac{i\, \alpha'}{2} \langle    \psi^\lambda(x_1) \psi^\sigma(x_2) \,:\big(q \psi(z)\big) \psi^\nu(z): \rangle 
\crbig
&& \times \, \langle : \partial X^\kappa (x_1) e^{ik_1X(x_1)}: \,:  \partial X^\rho (x_2) e^{ik_2X(x_2)} : \, : \ov{\partial} X^\mu (\ov{z}) e^{iq \widetilde{X}(\ov{z})}: \, :e^{iq X(z)}: \rangle 
\crbig
&& -\frac{i \alpha'}{2} \langle : \partial X^\kappa (x_1) e^{ik_1X(x_1)}: \,:  \partial X^\rho (x_2) e^{ik_2X(x_2)} : \, :e^{iq \widetilde{X}(\ov{z})} : \, : \partial X^\nu (z) e^{iqX(z)}: \rangle 
\crbig
&& \times \, \langle    \psi^\lambda(x_1) \psi^\sigma(x_2) \,:\big(q\widetilde{\psi}(\ov{z})\big) \widetilde{\psi}^\mu(\ov{z}): \rangle
\crbig
&&  -\frac{\alpha'^2}{4} \langle : \partial X^\kappa (x_1) e^{ik_1X(x_1)}: \,:  \partial X^\rho (x_2) e^{ik_2X(x_2)} : \, : e^{iq \widetilde{X}(\ov{z})}: \, :e^{iq X(z)}:\rangle
\crbig
&& \times \, \langle    \psi^\lambda(x_1) \psi^\sigma(x_2) \,: \big(q \widetilde{\psi}(\ov{z})\big) \widetilde{\psi}^\mu(\ov{z}) : \, : \big(q \psi (z)\big) \psi^\nu(z): \rangle \Big\}\,.
\end{array}
\eeq
Notice that we treat the closed string vertex operator as the product of two open string operators, that are \textit{normal--ordered} within the full correlation function. Using subsection (\ref{Cor}), we now compute \textit{all} correlators that appear in (\ref{ampl_2}). The process is highly laborious but straightforward with the resulting expressions being very long; we thus do not document but a sample of these in appendix \ref{sample_contrGM2}. The Koba--Nielsen factor for $\mathcal{A}(2,1)$ is given by
\beq \label{KB}
\begin{array}{ccl}
\mathcal{E}_{(2,1)} & \equiv & \langle : e^{ik_1X(x_1)} : \, :  e^{ik_2X(x_2)} : \, :e^{iq\widetilde{X}(\ov{z})}: \, : e^{iqX(z)}:  \rangle 
\crbig
&=&|x_1-x_2|^{2 \alpha' k_1k_2}
|z-\ov{z}|^{\alpha' q_\parallel^2}  |x_1-z|^{2\alpha' k_1 q}  |x_2-z|^{2\alpha' k_2 q} 
\crbig
&=&  |x_1-x_2|^{s+2}
|z-\ov{z}|^{s} \, |x_1-z|^{-s}  |x_2-z|^{-s}  \,.
\end{array}
\eeq
where we have used (\ref{KBgen}) and the details of the kinematics (\ref{mandel}), (\ref{morem}), (\ref{consGM2}).

To treat the volume factor $V_{\textrm{CKG}}^{-1}\,$, the prescription is to
\begin{enumerate}
\item use the $PSL(2,\mathbb{R})$ invariance of the disk to fix the position of any three vertex operators of the amplitude; recall that, from the point of view of the open string, $\mathcal{A}(2,1)$ involves four vertex operators. We choose
\beq \label{vert_pos}
(x_1,x_2,z,\ov{z}) = (x,-x,i,-i) \,,
\eeq
for which have that
\beq \label{KB_after}
\begin{array}{ccl}
\mathcal{E}_{(2,1)} &=&   4^{s+1}\,  |x|^{s+2}  (x^2+1)^{-s} \,.
\end{array}
\eeq
\item insert a c--ghost at each fixed position. For our choice (\ref{vert_pos}), the c--ghost contribution is
\beq \label{ghost_GM2}
\langle c(x_1)c(z) \widetilde{c}(\ov{z}) \rangle = (x_1-z)(x_1-\ov{z})(z-\ov{z})=2i(x-i)(x+i)\,.
\eeq
\end{enumerate}

Using (\ref{vert_pos}), (\ref{KB_after}), (\ref{ghost_GM2}) and all contractions, we obtain the amplitude as an integral over the real line
\beq \label{amplf}
\mathcal{A}(2,1) = \frac{g_c}{\alpha'^2} \, \Tr \big( T^a T^b\big)\, \displaystyle \sum_{i=1}^4 \mathbf{A}_i\,,
\eeq
where
\beq \label{A1_1}
\begin{array}{ccl}
\mathbf{A}_1 &=&4^s  \,  \alpha_{\kappa \lambda}^1 \alpha_{\rho \sigma}^2 \varepsilon_{\mu \nu}  \, \int\limits_{-\infty}^{\infty} dx \,  |x|^{s+2}  (x^2+1)^{-s} \frac{(x-i)(x+i)}{2x} \,\Big\{  \frac{\Theta^{\mu \nu \kappa \lambda
\rho \sigma}}{(x-i)^4(x+i)} + \frac{\Lambda^{\mu \nu \kappa \lambda \rho
\sigma}}{(x-i)(x+i)^4} 
\crbig
&& +\frac{\Xi^{\mu \nu \kappa \lambda \rho
\sigma}}{(x-i)^3(x+i)^2}+\frac{\Sigma^{\mu \nu \kappa \lambda \rho
\sigma}}{(x-i)^2(x+i)^3} -\frac{i}{2} \frac{\Gamma^{\mu \nu \kappa \lambda \rho \sigma}
}{(x-i)^4} -\frac{i}{2} \frac{\Delta^{\mu \nu \kappa \lambda \rho
\sigma} }{(x+i)^4} -\frac{i}{2} \frac{\Phi^{\mu \nu \kappa \lambda \rho
\sigma} }{(x-i)^3(x+i)}
\crbig
&& -\frac{i}{2} \frac{\Psi^{\mu \nu \kappa \lambda
\rho \sigma} }{(x-i)(x+i)^3} -\frac{i}{2} \frac{\Omega^{\mu \nu \kappa
\lambda \rho \sigma} }{(x-i)^2(x+i)^2} \Big\}
\crbig
\end{array}
\eeq
\beq \label{A2_1}
\begin{array}{ccl}
\mathbf{A}_2 &=&   4^s  \,  \alpha_{\kappa \lambda}^1 \alpha_{\rho \sigma}^2 \varepsilon_{\mu \nu}  \, \int\limits_{-\infty}^{\infty} dx \,  |x|^{s+2}  (x^2+1)^{-s} \frac{(x-i)(x+i)}{(2x)^2} \, \Big\{ \frac{ P^{\mu \nu \kappa \rho} g^{\lambda \sigma}    }{(x-i)^4} + \frac{ \widetilde{P}^{\mu \nu \kappa \rho}     g^{\lambda \sigma}  }{(x+i)^4}
\crbig
&& + \frac{  Q^{\mu \nu \kappa \rho \lambda \sigma}  }{(x-i)^3(x+i)}+ \frac{\widetilde{Q}^{\mu \nu \kappa \rho \lambda \sigma} }{(x-i)(x+i)^3} + \frac{  R^{\mu \nu \kappa \rho \lambda \sigma} }{(x-i)^2(x+i)^2}+\frac{i}{2}\, \frac{  S^{\mu \nu \kappa \rho \lambda \sigma}   }{(x-i)^3} - \frac{i}{2}\, \frac{  \widetilde{S}^{\mu \nu \kappa \rho \lambda \sigma}   }{(x+i)^3} 
\crbig
&&+ \frac{i}{2}\, \frac{ T^{\mu \nu \kappa \rho \lambda \sigma} }{(x-i)^2(x+i)}- \frac{i}{2}\, \frac{  \widetilde{T}^{\mu \nu \kappa \rho \lambda \sigma} }{(x-i)(x+i)^2}  -\frac{1}{4} \,\frac{U^{\mu \nu \kappa \rho} g^{ \lambda \sigma} }{(x-i)^2}
 -\frac{1}{4} \,\frac{  \widetilde{U}^{\mu \nu \kappa \rho} g^{ \lambda \sigma}   }{(x+i)^2} - \frac{1}{4} \, \frac{ W^{\mu \nu \kappa \rho} g^{ \lambda \sigma}  }{(x-i)(x+i)} \Big\}
\end{array}
\eeq

\beq \label{A3_1}
\begin{array}{ccl}
\mathbf{A}_3&=&   4^s  \,  \alpha_{\kappa \lambda}^1 \alpha_{\rho \sigma}^2 \varepsilon_{\mu \nu}  \, \int\limits_{-\infty}^{\infty} dx \,  |x|^{s+2}  (x^2+1)^{-s} \frac{(x-i)(x+i)}{(2x)^3}\, \Big\{ \frac{G^{\mu \nu \kappa \rho} g^{ \lambda \sigma}}{(x-i)^3} + \frac{H^{\mu \nu \kappa \rho} g^{ \lambda \sigma}}{(x+i)^3}
\crbig
&&+\frac{I^{\mu \nu \kappa \rho \lambda \sigma}}{(x-i)^2(x+i)} + \frac{J^{\mu \nu \kappa \rho \lambda \sigma}}{(x-i)(x+i)^2}  -i \frac{K^{\mu \nu \kappa \rho \lambda \sigma}}{(x-i)^2} -i \frac{L^{\mu \nu \kappa \rho \lambda \sigma}}{(x+i)^2} -i \frac{M^{\mu \nu \kappa \rho \lambda \sigma}}{(x-i)(x+i)}
\crbig
&&  + \frac{N^{\mu \nu \kappa \rho} g^{ \lambda \sigma}}{x+i}+ \frac{O^{\mu \nu \kappa \rho} g^{ \lambda \sigma}}{x-i} \Big\}
\end{array}
\eeq
\beq \label{A4_1}
\begin{array}{ccl}
\mathbf{A}_4 &=& 4^s  \,  \alpha_{\kappa \lambda}^1 \alpha_{\rho \sigma}^2 \varepsilon_{\mu \nu} g^{\lambda \sigma}  \, \int\limits_{-\infty}^{\infty} dx \,  |x|^{s+2}  (x^2+1)^{-s} \frac{(x-i)(x+i)}{(2x)^4} \,\Big\{ A^{\mu \nu  \kappa \rho }  + \frac{B^{\mu \nu  \kappa \rho }}{(x-i)(x+i)}
\crbig
&&  + \frac{C^{\mu \nu  \kappa \rho }}{(x+i)^2}+ \frac{\widetilde{\Delta}^{\mu \nu  \kappa \rho }}{(x-i)^2} + i  \frac{E^{\mu \nu  \kappa \rho }}{x+i} + i \frac{F^{\mu \nu  \kappa \rho }}{x-i} \Big\}

\end{array}
\eeq
To facilitate the computation, we have organized the various terms within $\mathcal{A}(2,1)$ in groups $\mathbf{A}_i$, with $i$ standing for the power in which $(x_1-x_2)$ appears originally in the denominator: for example, $\mathbf{A}_1$ contains all integrands that, before the fixing of the vertex operator positions, were proportional to $(x_1-x_2)^{-1}$. Moreover, the objects $ A^{\mu \nu  \kappa \rho }, B^{\mu \nu  \kappa \rho }$, etc, to which we refer from now as the ``kinematic packages'', are esentially due to the numerators of the field contractions and depend solely on $\alpha'$ and the momenta and polarizations of the external states. We compute \textit{all} these kinematic packages and document them in appendix \ref{KinPack}, where we have organized the various terms within any kinematic package in powers of $\alpha'$; it turns out that there appear only the orders $\alpha'$, $\alpha^2$ and $\alpha'^3$. Let us stress, however, that all our kinematic packages in appendix \ref{KinPack}, are exact and \textit{not} expansions in $\alpha'$.

We next perform extensively partial fractioning and proceed to compute the integrals in (\ref{A1_1})--(\ref{A4_1}) as we explain in appendix \ref{appInt} and obtain
\beq \label{A1}
\begin{array}{ccl}
\mathbf{A}_1 &=&  \frac{1}{2}\, 4^s \,\, \bigg\{ 
(\Gamma-\Delta)\, \frac{2 \Gamma \left(\frac{s+1}{2}\right) \Gamma 
\left(\frac{s+3}{2}\right)}{\Gamma (s+3)}
+\frac{1}{2}(\Theta+\Lambda) \, \frac{\pi ^{3/2} 2^{-s-2} (s-3) \sec 
\left(\frac{\pi  s}{2}\right)}{\Gamma 
\left(\frac{3}{2}-\frac{s}{2}\right) \Gamma 
\left(\frac{s}{2}+1\right)}
\crbig
&& \qquad \quad -\frac{1}{4}(\Xi+\Sigma) \, \frac{\Gamma 
\left(\frac{s-1}{2}\right) \Gamma \left(\frac{s+3}{2}\right)}{\Gamma (s+1)} +\frac{1}{16}(\Omega_+-\Omega_-) \, \frac{\pi  \sec \left(\frac{\pi  
s}{2}\right) \Gamma \left(\frac{s+3}{2}\right)}{ \Gamma 
\left(\frac{5}{2}-\frac{s}{2}\right) \Gamma (s)} \bigg\}\,,
\end{array}
\eeq
\beq \label{A2}
\begin{array}{ccl}
\mathbf{A}_2 &=&\frac{1}{4} \, 4^{s}\,\, \,   \bigg\{- (P + \widetilde{P})\frac{\sqrt{\pi } 
2^{-s-1} s
\Gamma \left(\frac{s+1}{2}\right)}{\Gamma \left(\frac{s}{2}+2\right)}
-\, (S+\widetilde{S}) \frac{ \Gamma \left(\frac{s+1}{2}\right)
\Gamma \left(\frac{s+3}{2}\right)}{ \Gamma (s+2)}
\crbig
&& \qquad -\frac{1}{4} \, (U+\widetilde{U}) \frac{\Gamma \left(\frac{s-1}{2}\right) \Gamma
\left(\frac{s+1}{2}\right)}{ \Gamma (s+1)} -\frac{1}{4} \, (W+\widetilde{W}) \frac{\sqrt{\pi } 2^{-s} \Gamma 
\left(\frac{s-1}{2}\right)}{\Gamma \left(\frac{s}{2}\right)} \bigg\}\,,
\end{array}
\eeq
\beq \label{A3}
\begin{array}{ccl}
\mathbf{A}_3&=& \frac{1}{8} \,4^s\bigg\{    -(G+H) \,  \frac{\pi ^{3/2} 2^{-s} \sec \left(\frac{\pi  s}{2}\right)}{\Gamma \left(\frac{1}{2}-\frac{s}{2}\right)
   \Gamma \left(\frac{s}{2}+1\right)} +(K-L) \, \frac{\sqrt{\pi } 2^{-s+1} \Gamma \left(\frac{s+1}{2}\right)}{\Gamma \left(\frac{s}{2}+1\right)} 
\crbig
&& \qquad +\frac{1}{4}(N+O)\, \frac{\Gamma \left(\frac{s-1}{2}\right) \Gamma \left(\frac{s+1}{2}\right)}{ \Gamma (s)}   \bigg\} \,,
\end{array}
\eeq
\beq \label{A4}
\begin{array}{ccl}
\mathbf{A}_4 &=& \frac{1}{16} \, 4^s\bigg\{ 2 A \, \frac{\sqrt{\pi } 2^{-s} s \, \Gamma \left(\frac{s-1}{2}\right)}{\Gamma \left(\frac{s}{2}+1\right)}  -   (C+\widetilde{\Delta}) \,\frac{\sqrt{\pi } 2^{-s}  \Gamma \left(\frac{s-1}{2}\right)}{\Gamma \left(\frac{s}{2}+1\right)}  +  (E-F)\, \frac{ (s-1) \,\big[\Gamma \left(\frac{s-1}{2}\right)\big]^2}{4 \Gamma (s)} \bigg\}\,,
\end{array}
\eeq
We have thus computed the \textit{full} $\mathcal{A}(2,1)$ amplitude (at tree level): it is given by (\ref{amplf}) together with (\ref{A1})--(\ref{A4}) and the kinematic packages in  \ref{KinPack}. Let us highlight that this result is valid in the superstring critical dimension and in 4D equally well, and it contains an arbitrary $D_{\mu \nu}$, which we will choose in the next section.

\subsection{Amplitude of three massive spin--two states in superstring theory}

Next we turn to $\mathcal{A}(3,0)$ involving three massive spin--two states. To reduce the length of the notation in the formulae, we define
\beq
 \psi_i \equiv \psi(x_i) \quad , \quad  X_i \equiv X(x_i)  \quad, \quad \phi_i =\phi (x_i)  \quad , \quad c(x_i)\equiv c_i\,. 
\eeq
Using (\ref{vertop}) and (\ref{vertop2}) in (\ref{ampl2_1}) we obtain
\beq \label{ampl2_2}
\begin{array}{ccl}
\mathcal{A}(3,0) &=&  \frac{g_o^3}{4\alpha'^2}\ C_{D_2} \,\Tr(T^{a_1}T^{a_2}T^{a_3}) \,  \displaystyle \int_{\mathbf{R}}   \frac{   \prod\limits_{i=1}^{3} \md x_i}{V_{\textrm{CKG}}}  \, \alpha_{\mu_1 \nu_1}^1 \alpha_{\mu_2 \nu_2}^2\alpha_{\mu_3 \nu_3}^3 \, (\mathcal{X}_1+\mathcal{X}_2+\mathcal{X}_3)
\crbig
&& \quad + (2 \leftrightarrow 3) \,,
\end{array}
\eeq
where have defined
\beq \label{usef1}
\begin{array}{ccl}
\mathcal{X}_1 & \equiv & i\,\langle :e^{-\phi_1}: \, : e^{-\phi_2}:  \rangle\, \langle    \psi^{\nu_1}_1  \psi^{\nu_2}_2 \rangle
\crbig
&& \quad \times  \langle : \partial X_1^{\mu_1 } e^{ik_1X_1}: \, :  \partial X_2^{\mu_2} e^{ik_2X_2} : \, :  \partial X_3^{\mu_3} \partial X^{\nu_3}_3  e^{ik_3X_3} :\rangle
\crbig
\mathcal{X}_2 &\equiv &- 2i \, \alpha'\,  \langle    \psi^{\nu_1}_1 \psi^{\nu_2}_2 \, :\partial \psi^{\mu_3}_3 \psi^{\nu_3}_3 : \rangle 
\crbig
&& \quad \times  \langle :e^{-\phi_1}: \, : e^{-\phi_2}:  \rangle \,  \langle : \partial X^{\mu_1 }_1 e^{ik_1X_1}: \, :  \partial X^{\mu_2}_2 e^{ik_2X_2} : \, :  e^{ik_3X_3} :\rangle 
\crbig
\mathcal{X}_3 &\equiv & 2 \alpha' \, \langle :e^{-\phi_1}: \, : e^{-\phi_2}:  \rangle  \, \langle    \psi^{\nu_1}_1 \psi^{\nu_2}_2 \, : k_3 \psi_3 \, \psi^{\nu_3}_3 : \rangle  
\crbig
&& \quad \times  \langle : \partial X^{\mu_1 }_1 e^{ik_1X_1}: \, :  \partial X^{\mu_2}_2 e^{ik_2X_2} : \, :  \partial X^{\mu_3}_3e^{ik_3X_3} :\rangle \,,
\end{array}
\eeq
where all $\mathcal{X}_i \,, i=1,2,3\,$, carry all indices $\mu_j, \nu_j\,,j=1,2,3\,$, which we however suppress so as to maintain a shorter length of notation. We proceed to compute the correlators (\ref{usef1}) in appendix \ref{contractions_susy}; the procedure is again tedious but not so as in the case of $\mathcal{A}(2,1)$.

Using (\ref{KBgen}), the relevant Koba--Nielsen factor is given by
\beq \label{KB2}
\begin{array}{ccl}
\mathcal{E}_{(3,0)}& \equiv & \langle : e^{ik_1X_1} : \, :  e^{ik_2X_2} : \, :  e^{ik_3X_3} :  \rangle  = x_{12}^{2\alpha' k_1 k_2}\, x_{13}^{2\alpha' k_1  k_3} \, x_{23}^{2\alpha' k_2   k_3}\,,
\end{array}
\eeq
where we omit the absolute values for the ordering $x_1 > x_2 >x_3$  we have chosen. Moreover, the c--ghost constribution is
\beq
\langle c_1 c_2 c_3\rangle =x_{12} x_{13} x_{23}\,,
\eeq
so that using (\ref{produs})
\beq \label{x_dep}
\widetilde{\mathcal{E}} \, \langle c_1 c_2 c_3\rangle = x_{12}^2 x_{13}^2x_{23}^2\,.
\eeq
Using appendix \ref{contractions_susy}, we thus see that the combination (\ref{x_dep}) cancels the $x$--dependence of the $\mathcal{X}_i$ as it should: three--point amplitudes have no $x$--dependance. Using (\ref{on2}) we also see that the two possible orderings of the vertex operators contribute equally to the amplitude, as is again typical for three--point amplitudes. Putting everything together, we obtain
\beq \label{ampl2_rest}
\begin{array}{ccl}
\mathcal{A}(3,0)&=&\frac{g_o}{4\alpha'^3} \Tr(T^{a_1}\{T^{a_2},T^{a_3}\}) \,   \Big\{ 3\, (2\alpha')^2 \Tr(\alpha^1 \cdot \alpha^2 \cdot \alpha^3)+(2\alpha')^3 \times
\crbig
&& \Big[ (k_1\cdot \alpha^2 \cdot k_1) (\alpha^3 \cdot \alpha^1)+(k_2 \cdot \alpha^3 \cdot k_2) (\alpha^2 \cdot \alpha^1)+(k_3 \cdot \alpha^1 \cdot k_3) (\alpha^2 \cdot \alpha^3)
\crbig
&& + 3 \, k_1 \cdot \alpha^2\cdot  \alpha^1\cdot  \alpha^3\cdot  k_2 +  3\, k_2 \cdot \alpha^3 \cdot  \alpha^2 \cdot  \alpha^1\cdot  k_3+ 3\, k_3 \cdot \alpha^1 \cdot  \alpha^3 \cdot  \alpha^2 \cdot  k_1 \Big]
\crbig
&& +(2\alpha')^4 \,\Big[(k_1 \cdot \alpha^2 \cdot k_1 ) (k_2 \cdot \alpha^3 \cdot \alpha^1 \cdot k_3)  + (k_2 \cdot \alpha^3 \cdot k_2 ) (k_3 \cdot \alpha^1 \cdot \alpha^2 \cdot k_1)
\crbig
&& + (k_3 \cdot \alpha^1 \cdot k_3 ) (k_1 \cdot \alpha^2 \cdot \alpha^3 \cdot k_2)  
 \Big]\Big\} \,.
\end{array}
\eeq
Let us note that (\ref{ampl2_rest}) is the full amplitude at tree--level and not a truncation neither an expansion in $\alpha'$. Moreover, cyclic symmetry is manifest in (\ref{ampl2_rest}):  because of the symmetry property of the polarisation tensors, the no--momenta term is cyclically symmetric on its own, as is each line involving two--momenta terms and as also is the set of four--momenta terms.

\subsection{Amplitude of three massive spin--two states in  bosonic string theory}

Finally, using  (\ref{bosv2}), the bosonic amplitude (\ref{genampl2}) takes then the form
\beq
\begin{array}{ccl}
\widetilde{\mathcal{A}}(3,0) &=& \frac{g_o^3}{8\alpha'^3}\ C_{D_2} \Tr(T^a T^b T^c) \,\alpha_{\mu_1 \nu_1}^1 \alpha_{\mu_2 \nu_2}^2 \alpha_{\mu_3 \nu_3}^3 \times
\crbig
&&  \langle :  c_1 \partial X^{\mu_1}_1 \partial X^{\nu_1}_1 e^{ik_1 X_1} : \,:  c_2 \partial X^{\mu_2}_2  \partial X^{\nu_2}_2 e^{ik_2 X_2}   : \,: c_3  \partial X^{\mu_3}_3  \partial X^{\nu_3}_3 e^{ik_3 X_3}: \rangle\,.
\end{array}
\eeq
We then perform the contractions and give the result in appendix \ref{contractions_bosonic}, thus obtaining
\beq \label{ampl2_rest_bos}
\begin{array}{ccl}
\widetilde{\mathcal{A}}(3,0) &=& \frac{g_o}{8\alpha'^4} \, \Tr(T^a T^b T^c) \, \bigg\{ 2\alpha'^3\,\Tr(\alpha^1 \cdot \alpha^2 
\cdot \alpha^3) +\alpha'^4 \Big[ (k_1\cdot \alpha^2 \cdot k_1) \Tr(\alpha^3 \cdot \alpha^1)
\crbig
&&+(k_2 \cdot \alpha^3 \cdot k_2) \Tr(\alpha^2 \cdot \alpha^1)+(k_3 \cdot \alpha^1 \cdot k_3) \Tr(\alpha^2 \cdot \alpha^3)
\crbig
&&+4  \big(  k_1 \cdot \alpha^2\cdot  \alpha^1\cdot  \alpha^3\cdot  k_2 +  k_2 \cdot \alpha^3 \cdot  \alpha^2 \cdot  \alpha^1\cdot  k_3+  k_3 \cdot \alpha^1 \cdot  \alpha^3 \cdot  \alpha^2 \cdot  k_1 \big) \Big]
\crbig
&&+4\alpha'^5 \,\Big[(k_1 \cdot \alpha^2 \cdot k_1 ) (k_2 \cdot \alpha^3 \cdot \alpha^1 \cdot k_3)  + (k_2 \cdot \alpha^3 \cdot k_2 ) (k_3 \cdot \alpha^1 \cdot \alpha^2 \cdot k_1)
\crbig
&& + (k_3 \cdot \alpha^1 \cdot k_3 ) (k_1 \cdot \alpha^2 \cdot \alpha^3 \cdot k_2)  
 \Big]
\crbig
&&+2\alpha'^6\, (k_1 \cdot \alpha^2 \cdot k_1) (k_2 \cdot\alpha^3 \cdot k_2) (k_3 \cdot  \alpha^1 \cdot  k_3) \bigg\}\,.
\end{array}
\eeq
Again, all dependence on the vertex operator positions $x_i$ has cancelled in  
the amplitude (\ref{ampl2_rest_bos}). Furthermore,  as expected the latter is  cyclically symmetric. Notice again that the expression (\ref{ampl2_rest_bos}) is exact and not a truncation.

\section{From strings to fields}

To simplify our setup, we now consider a single spacetime--filling D--brane, namely a 4D brane in 4D spacetime, by choosing
\begin{itemize}
\item for the $D$--matrix:
\beq \label{sv}
D^\mu_\nu = \delta^\mu_\nu \quad , \quad D^{\mu \nu}=D^\mu_\lambda g^{\lambda \nu} = g^{\mu \nu} \quad , \quad (Dq)^\mu = q^\mu  \quad , \quad \mu,\nu=0,\dots,3\,,
\eeq
\item $U(1)$ as the brane gauge group, so that all traces of products of generators disappear from the amplitude prefactors.
\end{itemize}
We further make the following identifications
\beq \label{id_eff}
\varepsilon_{\mu \nu}  \rightarrow  G_{\mu \nu} \quad , \quad \alpha^{1,2}_{\mu \nu} \rightarrow M_{\mu \nu}  \quad , \quad k_\mu \, , q_\mu \rightarrow i \partial_\mu
\eeq
in the expressions of the amplitudes so as to extract the corresponding effective Lagrangians, being careful in having the partial derivative act on the appropriate external state: the momenta $k^\mu$ and $q^\mu$ become partial derivatives acting on $M_{\mu \nu}$ and $G_{\mu \nu}$ respectively.

\subsection{The universal $\textrm{G}^3$ Lagrangian}

Performing the identifications (\ref{id_eff}) in (\ref{ampl_G3}), one obtains \cite{Gross:1986mw,Lust:1987xm}
\beq \label{G3_univ_lag}
\begin{array}{ccl}
\mathcal{L}^{\textrm{eff}}_{\textrm{G}^3}& =& g_c\,  G^{\mu \nu} \big[ \partial_\mu G_{\rho \sigma} \partial_\nu G^{\rho \sigma}-2 \partial_\nu G_{\rho \sigma} \partial^\sigma G_\mu^\rho\big] + ( \textrm{higher deriv. terms})\,,
\end{array}
\eeq
i.e. the cubic terms of the Einstein--Hilbert action as expanded around a Minkowski background. Comparing (\ref{G3_univ_lag}) with the bimetric prediction (\ref{G3}), we find that
\beq \label{relation_1}
\frac{1}{m_g \sqrt{1+\alpha^2}} \overset{!}{\equiv} g_c \,.
\eeq
Before examining the meaning of this identification, let us extract the $\textrm{GM}^2$ Lagrangian.

\subsection{A consistent low-energy limit and the $\textrm{GM}^2$ Lagrangian}

We now consider the amplitude $\mathcal{A}(2,1)$, namely (\ref{amplf}) together with (\ref{A1})--(\ref{A4}). Let us write the objects $\mathbf{A}_i$ schematically as follows
\beq \label{schem}
\mathbf{A}_i = \mathcal{K}(k_1, k_2, q ; \alpha') \times \mathcal{I}(s)\,,
\eeq
where $\mathcal{K}(k_1, k_2, q ; \alpha')$ stand for the kinematic packages, while in $\mathcal{I}(s)$ we have gathered everything that depends on the momenta and $\alpha'$ solely via $s$, namely the prefactor $4^s$, whose source is the Koba--Nielsen factor, multiplied by the value of the corresponding integral; the latter involves contributions from the Koba--Nielsen factor, the c--ghost contractions as well as the denominators of all other contractions. Since our goal is to derive the low--energy effective action, which implies that eventually we will expand the full amplitude around a small value of $\alpha'$, let us examine the dependance of the $\mathbf{A}_i$ on $\alpha'$. In particular, the $\mathcal{K}(k_1, k_2, q ; \alpha')$ are \textit{exact} expressions containing terms up to a specific power of $\alpha'$ as can be seen in appendix \ref{KinPack}. In contrast, it is obvious that the $\mathcal{I}(s)$ are infinite expansions in $s$; we would like to emphasize that their dependance on $\alpha'$ is \textit{obscure}, given that the scattering involves massive states in the external legs, with the respective masses depending strongly on $\alpha'$ as in (\ref{on2}). 

These considerations imply that the limit of small $\alpha'$ should be taken with care. Let us look at this more closely. Using (\ref{on1}) and (\ref{on2}), we see that
\beq \label{observe1}
 \alpha' \, k_1 \cdot k_2 \overset{\alpha' \rightarrow 0}{\longrightarrow} 1 
 \eeq
 while
 \beq\label{observe2}
 \alpha' \, k_{1,2} \cdot q \overset{\alpha' \rightarrow 0}{\longrightarrow} 0\,.
\eeq
Using  (\ref{mandel}), this yields
\beq \label{lims}
s \overset{\alpha' \rightarrow 0}{\longrightarrow} 0 \quad \textrm{or} \quad t \overset{\alpha' \rightarrow 0}{\longrightarrow} -1
\eeq
instead of the naive limiting value $s \overset{\alpha' \rightarrow 0}{\longrightarrow} -2$. We now make the following observations:
\begin{itemize}
 \item no term entering $\mathcal{K}(k_1, k_2, q ; \alpha')$ becomes of order $1$ as $\alpha' \rightarrow 0$, as the potentially dangerous objects of the kind $\alpha' \, k_1^\mu  k_2^\nu$ (the contraction of the latter with suitable polarization tensors implied) are always suppressed by at least one power of $\alpha'$, see appendix \ref{KinPack}
 \item while the limit (\ref{lims}) is well--defined, substitution of $s$ via the first of the   equations (\ref{mandel}) in $\mathcal{I}(s)$ is not suitable for our purposes. To see this, let us note that the general expansion of $\mathcal{I}(s)$ around a small $s$ takes the form
\beq \label{int_trunc_ob}
 \mathcal{I}(s) = \sum_n a_n s^n =\sum_n b_n (\alpha' \, k_1 \cdot k_2)^n\,,
\eeq
where in the second step we have used the first of equations (\ref{mandel}) and $a_n\,,\,b_n$ are constants depending on the value of $ \mathcal{I}(s)$. To determine the effective action, we intend to truncate this series at the first non--zero contribution. However, any truncation of (\ref{int_trunc_ob}) is meaningless, since $\alpha' \, k_1 \cdot k_2 \rightarrow 1$ at every order because of (\ref{observe1}). It is thus clear that the parametrization of $s$ in terms of $k_1$ and $k_2$ loses its meaning after the truncation. \textit{Instead}, using (\ref{momc}) in the special case (\ref{sv}), we may rewrite $s$ as
\beq \label{alter_s}
s=-2 \alpha' \, k_1 \cdot q \,,
\eeq
which is well--behaved at every order, thus enabling truncation at any order.
\end{itemize}

Based on these observations, we
\begin{enumerate}
\item expand the $ \mathcal{I}(s)$  in (\ref{A1})--(\ref{A4}) around a small $s$ according to (\ref{lims}) and (\ref{alter_s}) and present the results in appendix \ref{app_expansions} 
\item use the expressions of the kinematic packages, which are already organized according to integer powers of $\alpha'$ as presented appendix \ref{KinPack}, in the special case (\ref{sv})
\item use points 1. and 2. in (\ref{schem}) and, keeping terms up to order $\alpha'^2$, we obtain from (\ref{amplf})
\end{enumerate}
\beq \label{gm2_final}
\begin{array}{ccl}
\mathcal{A}(2,1) &=& g_c \,  \Big\{  -2\Tr(\alpha^1 \cdot \alpha^2) \, \varepsilon_{\mu \nu} k_1^\mu k_2^\nu +2(\varepsilon \cdot \alpha^2 \cdot \alpha^1)_{\mu \nu}  k_1^\mu k_2^\nu
\crbig
&& +2(\varepsilon \cdot \alpha^1 \cdot \alpha^2)_{\mu \nu}  k_1^\nu k_2^\mu  + 2(\varepsilon \cdot \alpha^2 \cdot \alpha^1)_{\mu \nu}\, k_2^\mu q^\nu +2(\varepsilon \cdot \alpha^1 \cdot \alpha^2)_{\mu\nu} \, k_1^\mu q^\nu  
\crbig
&&+2\Tr(\varepsilon \cdot \alpha^1 \cdot  \alpha^2) \, (k_1 \cdot q)    + \big[ \Tr(\varepsilon \cdot \alpha^2) \alpha^1_{\mu \nu}  - 2( \alpha^1 \cdot \varepsilon \cdot \alpha^2)_{\mu \nu}
\crbig
&&+\Tr(\varepsilon \cdot \alpha^1) \alpha^2_{\mu \nu} \big] \, q^\mu   q^\nu \Big\}   +\mathcal{O}\left(\alpha'^3\right)\,.
\end{array}
\eeq
Next, we make the replacements (\ref{id_eff}) in (\ref{gm2_final}) and, using partial integrations and the conditions (\ref{on1}), (\ref{on2}), we obtain up to order $\alpha'^2$
\beq \label{result}
\begin{array}{ccl}
\mathcal{L}^{\textrm{eff}}_{\textrm{GM}^2}&=& g_c  \, \Big[  G^{\mu\nu}\big(\partial_\mu M_{\rho \sigma}  \partial_\nu M^{\rho \sigma}  -4 \partial_\nu M_{\rho \sigma}  \partial^\sigma M_{\mu}^{\rho}   \big)
\crbig
&& \qquad +  M^{\mu\nu} \big( \partial_\mu G_{\rho \sigma}  \partial_\nu M^{\rho \sigma}-\partial_\rho G_{\mu \sigma}  \partial_\nu M^{\rho \sigma} \big)  \Big] \,.
\end{array}
\eeq

We now compare (\ref{result}) with the bimetric prediction (\ref{GM2}) and observe the following:
\begin{itemize}
\item The string yields the very same two--derivative terms as bimetric theory. Let us recall that these are all possible Lorentz invariant terms involving one massless and two massive TT rank--2 tensors.

\item Apart from the $1$ (string) vs $2$ (bimetric) discrepancy in the overall prefactor of the second lines of (\ref{result}) and (\ref{GM2}),  \textit{all other} coefficients match, upon imposing
\beq \label{relation_2}
\frac{1}{m_g\,\sqrt{1+\alpha^2}} \overset{!}{\equiv} g_c \,.
\eeq
Notice that (\ref{relation_1}) and (\ref{relation_2}) are the same: if it were not for the $1$ vs $2$ discrepancy, the comparison of the effective $\textrm{G}^3$ and $\textrm{GM}^2$ Lagrangians with the respective bimetric predictions would yield identical information. In any case, this relation would not provide information on the parameters $\beta_n$ of the bimetric potential, as we have anticipated. Because of this, and to investigate the origin of the $1$ vs $2$ discrepancy, we continue with the analysis of $\mathcal{A}(3,0)$ in the next subsection.
\end{itemize}

\subsection{The $\textrm{M}^3$ Lagrangian}

We now consider the amplitude $\mathcal{A}(3,0)$ in the superstring case, namely (\ref{ampl2_rest}). The latter is an exact expression and we do not need to truncate an infinite series, neither the treatment of the previous subsection. Simply by making the replacements (\ref{id_eff}) in (\ref{ampl2_rest}), we obtain the effective Lagrangian
\beq \label{leff1}
\begin{array}{ccl}
\mathcal{L}^{\textrm{eff}}_{\textrm{M}^3}& =& \frac{g_o}{\alpha'}\, \Big\{ \big[M^3\big] +2 \alpha' M^{\mu \nu} \big[ \partial_\mu M_{\rho \sigma} \partial_\nu M^{\rho \sigma}-\textcolor{blue}{3} \partial_\nu M_{\rho \sigma} \partial^\sigma M_\mu^\rho\big]
\crbig
&& \qquad \qquad +4 \alpha'^2 \, \partial^\mu \partial^\nu M_{\rho \sigma} \partial^\rho M_\nu^\kappa \partial^\sigma M_{\mu \kappa} \Big\}\,.
\end{array}
\eeq
We now compare (\ref{leff1}) with the bimetric prediction (\ref{M3}) and observe the following:
\begin{itemize}
\item The string yields the very same non--derivative and two--derivative terms as bimetric theory. The additional four--derivative term we find is to be understood as a string tree--level correction to the field theory. 

\item There appears, however, a significant discrepancy: the (absolute value of the) relative coefficient between the two two--derivative terms in (\ref{leff1}) is $3$ (in blue), while it is $2$ in (\ref{M3}); we will refer to this as the ``$2$ vs $3$'' discrepancy.  As explained in subsection (\ref{eae}), let as recall that in the latter \textit{all} self--interactions of the massive spin-2 with two derivatives at the cubic level originate in the Einstein--Hilbert action; this is also the case in dRGT massive gravity \cite{deRham:2010kj, deRham:2010gu, deRham:2010ik}. This means that kinetic part of the Lagrangian of the massive spin--2 state of the open superstring spectrum respects diffeos at the quadratic level, but not at the cubic level; the kinematics of this state is \textit{not} of the GR type and cannot be described by a Ricci scalar.

\item Because of the previous point, we refrain from identifying the couplings that appear in (\ref{leff1}) with those in (\ref{M3}).

\item We would like to highlight that the $2$ vs $3$ discrepancy is strikingly reminiscent of the $1/2$ vs $1/3$ discrepancy between the massless and massive graviton propagators, which are given respectively by (see for example \cite{Hinterbichler:2011tt})
\beq \label{grav_prop}
\begin{array}{ccl}
\mathcal{D}_{\mu \nu ,\kappa \lambda}^G (p) &=& \frac{-i}{p^2} \Big[ \frac{1}{2}( \eta_{\mu \kappa} \eta_{\nu \lambda} +\eta_{\mu \lambda} \eta_{\nu \kappa} ) -\frac{1}{2} \eta_{\mu \nu} \eta_{\kappa \lambda}\Big]
\end{array}
\eeq
\beq
\begin{array}{ccl}
\mathcal{D}_{\mu \nu ,\kappa \lambda}^M (p)& =&  \frac{-i}{p^2+m_{\textrm{FP}}^2} \Big[ \frac{1}{2}( P_{\mu \kappa} P_{\nu \lambda} +P_{\mu \lambda} P_{\nu \kappa} ) -\frac{1}{3} P_{\mu \nu} P_{\kappa \lambda}\Big]\,,
\end{array}
\eeq
where $P_{\mu \nu} \equiv \eta_{\mu \nu}+\frac{p_\mu p_\nu}{m^2_{\textrm{FP}}}$. Interestingly, 
\beq \label{prop_lim}
\mathcal{D}_{\mu \nu ,\kappa \lambda}^M (p) \overset{m_{\textrm{FP}}\rightarrow 0}{\longrightarrow}\frac{-i}{p^2} \Big[ \frac{1}{2}( \eta_{\mu \kappa} \eta_{\nu \lambda} +\eta_{\mu \lambda} \eta_{\nu \kappa} ) -\frac{1}{3} \eta_{\mu \nu} \eta_{\kappa \lambda}\Big] + \textrm{singular terms} \,.
\eeq
Upon comparing (\ref{grav_prop}) with (\ref{prop_lim}), one observes a $1/2$ vs $1/3$ discrepancy in the coefficients of the last non--singular term, with this being a manifestation of the van Dam, Veltman, Zakharov discontinuity \cite{vanDam:1970vg, Zakharov:1970cc}. Notice that our $2$ vs $3$ discrepancy is qualitatively different, as it is observed at the level of the kinetic, not the mass, term.

\end{itemize}

As a final step, it is instructive to consider the effective cubic massive spin--2 self--interactions as extracted from the bosonic string:  by making the replacements (\ref{id_eff}) in (\ref{ampl2_rest_bos}), we obtain the effective Lagrangian
\beq \label{leff1_bos}
\begin{array}{ccl}
\mathcal{L}^{\textrm{eff}, \textrm{bos}}_{\textrm{M}^3}& =&\frac{ g_o}{\alpha'} \, \Big\{2 \big[M^3\big] +3\alpha' M^{\mu \nu} \big[ \partial_\mu M_{\rho \sigma} \partial_\nu M^{\rho \sigma}-4 \partial_\nu M_{\rho \sigma} \partial^\sigma M_\mu^\rho\big]
\crbig
&&+ 12\alpha'^2 \, \partial^\mu \partial^\nu M_{\rho \sigma} \partial^\rho M_\nu^\kappa \partial^\sigma M_{\mu \kappa} -2\alpha'^3\, \partial^\mu \partial^\nu M_{\mu \nu} \partial^\rho \partial^\sigma M_{\rho \sigma} \partial^\lambda \partial^\kappa M_{\lambda \kappa} \Big\}\,.
\end{array}
\eeq
By comparing (\ref{M3}) with (\ref{leff1_bos}), we observe a $2$ vs $4$ discrepancy. To conclude, in the case of both the superstring and the bosonic string, we do not find agreement between the bimetric and the string prediction, as we show that the kinematics of the massive spin--2 state we have used does not respect the structure of the linearized Ricci scalar.

\section{Conclusions}

In this work we have begun to investigate in what way ghost-free bimetric gravity, which around Minkowski backgrounds propagates a massless graviton as well as an additional massive spin--2 field, can arise from string theory as an effective field theory.
For this purpose, we have employed the procedure of deriving the string effective action from the explicit computation of string scattering amplitudes and
the comparison of them with the relevant terms of the anticipated field theory. In this regard, our work had to deal with several new challenges in the context of string amplitude calculations and the
associated effective field theories:
first of all, extending the work of \cite{Feng:2010yx}, 
we have computed string amplitudes with more than one massive external spin--2 string state. Second, 
we had to determine the 
on-shell conditions on the scattering amplitudes 
of bimetric gravity. Furthermore, for the comparison between the string amplitudes with both massive and massless external states and the corresponding on-shell field Langrangian, a 
very careful and subtle 
$\alpha'$-limit with respect to the kinematic variables had to be applied, since at the massive level in principle all massive string excitations become relevant.

As we discussed in this work,  massive spin--2 states arise in string theory in several ways.
They can be KK excitations of the standard graviton and they arise as stringy Regge excitations both in the closed string and also in the open string sector.
Actually the closed string spin--2 fields can be constructed from a kind of double copy construction as the product of two open string  massive vector states (so--called Proca fields).
However, the massive open string vectors as well as the associated massive, closed string spin--2 Regge modes are not present in the critical dimensions $d=10$ or $d=26$; 
they arise only
after compactification, and then the closed massive spin--2 
modes generically correspond
to massive spin--2 components of a higher spin-four supermultiplet of extended supergravity \cite{Ferrara:2018iko}. 
On the other hand, the massive open string excitations are in our opinion the most generic and most model-independent
choice
for the massive spin--2 field of bimetric gravity in the context of string theory: they are present 
already in the critical dimensions. Furthermore, when restricted to four space-dimensions with open string excitations on D3-brane(s), the relevant massive spin--2 fields build the highest component of
a massive spin--2 supermultiplet, the so--called Weyl supermultiplet. This has a close connection to (Weyl)$^2$  (super)gravity \cite{Ferrara:2018wqd}, whose precise relation to bimetric gravity deserved further investigation (see also  \cite{Gording:2018not} and \cite{Aoki:2020rae}).
Because of these reasons, we decided to investigate in this work the open string option for the massive spin--2 field of bimetric gravity, its corresponding 
covariant string vertex operator and the (massive)$^2$-massless, i.e (open)$^2$-closed, and also  the (massive)$^3$, i.e (open)$^3$, string scattering amplitudes in great detail.
The closed string option for the massive spin--2 excitation and its effective action will be discussed in a future publication \cite{LMMS_next}.

As a result of our computations, we have shown that the lowest--order terms extracted from the string amplitude precisely match the bimetric action up to numerical coefficients, in the case of both the self--interactions of the massive spin--2 and that of its interactions with the graviton.
However, in both cases, we have found a discrepancy at the level of numerical coefficients between string theory and bimetric gravity. In particular, one of the 
distinct features of bimetric gravity is that both initial metric fields $g$ and $f$ are described by a diffeomorphism invariant Einstein term. Only the potential breaks the two diffeomorphism symmetries to one  and creates
a ghost-free mass term for the linear combination $M$ around Minkowski backgrounds, while the kinematics (at least to cubic order) of both $G$ and $M$ respect linearized diffeomorphisms. In the string side however, what we find is that only the graviton's kinematics respects linearized diffeomorphisms, but not that of the open string state that we have identified with $M$; notice that we make this obvservation strictly at the level of the cubic self--interactions of $M$, which do not disturb the structure of its propagator and do not, therefore, seem to be associated with the presence of a ghost. However, the absence of a second diffeomorphism invariance in our string setup should not come as a great 
surprise, since in string theory  it is absolutely not guaranteed, or even clear, that each massive spin--2 mode should also come with its own diffeomorphism symmetry.
Whether this could be the case for the massive closed string excitations, which as afore mentioned come from a double copy construction of a massive Proca field, will be clarified in \cite{LMMS_next}. As a final comment, let us note that the tree--level disk amplitude with two massless closed and one massive open string state in the external legs is an interesting future computation, given the absence of such cubic interaction terms in bimetric theory around Minkowski backgrounds \cite{LMMS_next}.

\vspace{15mm}

\leftline{\noindent{\bf Acknowledgments}}
\vskip 1mm
\noindent
We would like to thank Ignatios Antoniadis, Eugeny Babichev, Thibault Damour, Elias Kiritsis, Marvin L\"uben, Evgeny Skvortsov and Angnis Schmidt--May for useful discussions. CM thanks the Albert Einstein Institute in Potsdam for a most warm hospitality as well as the organizers of the Rencontres Th\'eoriciennes in Paris for the opportunity to present this work. The work of DL is supported  by the Origins Excellence Cluster.

\newpage

\appendix 
\section{Calculations for $\mathcal{A}(2,1)$}

\subsection{Sample contractions}\label{sample_contrGM2}

Using the disk correlators (\ref{corrs}) and Wick's theorem, we perform the contractions in (\ref{ampl_2}). The formulae are  long and here we only present one of the shorter examples:
\beq
\begin{array}{ccl}
-\frac{\alpha'^2}{4} \langle : \partial X^\kappa (x_1) e^{ik_1X(x_1)}: \,:  \partial X^\rho (x_2) e^{ik_2X(x_2)} : \, : e^{iq\widetilde{X}(\ov{z})}: \, :e^{iq X(z)}:\rangle
\crbig
 \times \,\langle :e^{-\phi(x_1)}: \, : e^{-\phi(x_2)}:  \rangle \, \ \, \langle    \psi^\lambda(x_1) \psi^\sigma(x_2) \,: \big(q\widetilde{\psi}(\ov{z})\big) \widetilde{\psi}^\mu(\ov{z}) : \, : \big(q\psi (z)\big) \psi^\nu(z): \rangle =
\crbig
\frac{ \alpha'^3}{8} \mathcal{E} \, \bigg\{ -\frac{\big[2\alpha'  k_2^\kappa k_1^\rho g^{\lambda \sigma} -g^{\kappa \rho}g^{\lambda \sigma}  \big]l^{\nu \mu}}{(x_1-x_2)^4 (z-\ov{z})^2} - \frac{\alpha'k_1^\rho q^\kappa g^{\lambda \sigma}l^{\nu \mu}}{(x_1-x_2)^3(x_1-z)(z-\ov{z})^2}  - \frac{\alpha'   (Dq)^\kappa k_1^\rho g^{\lambda \sigma}l^{\nu \mu}}{(x_1-x_2)^3(x_1-\ov{z})(z-\ov{z})^2}\crbig
 +\frac{\alpha'  k_2^\kappa q^\rho g^{\lambda \sigma}l^{\nu \mu}}{(x_1-x_2)^3(x_2-z)(z-\ov{z})^2}   +\frac{\alpha'  k_2^\kappa (Dq)^\rho g^{\lambda \sigma}l^{\nu \mu}}{(x_1-x_2)^3(x_2-\ov{z})(z-\ov{z})^2}
 \crbig
 -\frac{\big[2\alpha'   k_2^\kappa k_1^\rho -g^{\kappa \rho} \big][ q^\lambda\widetilde{f}^{\sigma \nu \mu}+g^{\lambda \nu} \widetilde{h}^{\sigma \mu} \big] }{(x_1-x_2)^3 (x_1-z)(x_2-\ov{z})(z-\ov{z})}
 -\frac{\big[2\alpha'   k_2^\kappa k_1^\rho -g^{\kappa \rho}  \big] \big[ q^\sigma{f}^{\lambda \nu \mu}+g^{\sigma \nu}  h^{\lambda  \mu}\big] }{(x_1-x_2)^3 (x_1-\ov{z})(x_2-z)(z-\ov{z})}
\crbig
 +\frac{1}{2} \frac{\alpha'   q^\kappa q^\rho  g^{\lambda \sigma}l^{\nu \mu}}{(x_1-x_2)^2(x_1-z)(x_2-z)(z-\ov{z})^2}  + \frac{1}{2} \frac{\alpha'    q^\kappa (Dq)^\rho  g^{\lambda \sigma}l^{\nu \mu}}{(x_1-x_2)^2(x_1-z)(x_2-\ov{z})(z-\ov{z})^2}
 \crbig
    + \frac{1}{2}\frac{\alpha'    (Dq)^\kappa q^\rho g^{\lambda \sigma}l^{\nu \mu}}{(x_1-x_2)^2(x_1-\ov{z})(x_2-z)(z-\ov{z})^2}
  + \frac{1}{2}\frac{\alpha'    (Dq)^\kappa (Dq)^\rho g^{\lambda \sigma}l^{\nu \mu}}{(x_1-x_2)^2(x_1-\ov{z})(x_2-\ov{z})(z-\ov{z})^2}
  \crbig
   + \frac{ \alpha'    k_2^\kappa (Dq)^\rho q^\lambda\widetilde{f}^{\sigma \nu \mu}+\alpha'    k_2^\kappa (Dq)^\rho g^{\lambda \nu}\widetilde{h}^{\sigma \mu}}{(x_1-x_2)^2(x_1-z)(x_2-\ov{z})^2(z-\ov{z})}
    -\frac{\alpha' k_1^\rho q^\kappa q^\lambda\widetilde{f}^{\sigma \nu \mu}+ \alpha'k_1^\rho q^\kappa g^{\lambda \nu}\widetilde{h}^{\sigma \mu}}{(x_1-x_2)^2(x_1-z)^2(x_2-\ov{z})(z-\ov{z})}
\crbig
+ \frac{\alpha'    k_2^\kappa q^\rho q^\sigma{f}^{\lambda \nu \mu}+ \alpha'    k_2^\kappa q^\rho g^{\sigma \nu} h^{\lambda  \mu}}{(x_1-x_2)^2(x_1-\ov{z})(x_2-z)^2(z-\ov{z})}
  -\frac{\alpha'    (Dq)^\kappa k_1^\rho q^\sigma{f}^{\lambda \nu \mu}+\alpha'    (Dq)^\kappa k_1^\rho g^{\sigma \nu} h^{\lambda  \mu}}{(x_1-x_2)^2(x_1-\ov{z})^2(x_2-z)(z-\ov{z})} 
  \crbig
    -\frac{\alpha' k_1^\rho q^\kappa q^\sigma{f}^{\lambda \nu \mu}+\alpha' k_1^\rho q^\kappa g^{\sigma \nu} h^{\lambda  \mu}}{(x_1-x_2)^2(x_1-z)(x_1-\ov{z})(x_2-z)(z-\ov{z})}
   -\frac{ \alpha'   (Dq)^\kappa k_1^\rho q^\lambda\widetilde{f}^{\sigma \nu \mu}+\alpha'    (Dq)^\kappa k_1^\rho g^{\lambda \nu}\widetilde{h}^{\sigma \mu}}{(x_1-x_2)^2(x_1-z)(x_1-\ov{z})(x_2-\ov{z})(z-\ov{z})}
 \crbig
  + \frac{\alpha'    k_2^\kappa (Dq)^\rho q^\sigma{f}^{\lambda \nu \mu}+\alpha'   k_2^\kappa (Dq)^\rho g^{\sigma \nu} h^{\lambda  \mu}}{(x_1-x_2)^2(x_1-\ov{z})(x_2-z)(x_2-\ov{z})(z-\ov{z})}
  + \frac{ \alpha'    k_2^\kappa q^\rho q^\lambda\widetilde{f}^{\sigma \nu \mu}+\alpha'    k_2^\kappa q^\rho g^{\lambda \nu}\widetilde{h}^{\sigma \mu}}{(x_1-x_2)^2(x_1-z)(x_2-z)(x_2-\ov{z})(z-\ov{z})}
\crbig
 +\frac{1}{2} \frac{ \alpha'    q^\kappa (Dq)^\rho  q^\lambda\widetilde{f}^{\sigma \nu \mu}+ \alpha'    q^\kappa (Dq)^\rho  g^{\lambda \nu}\widetilde{h}^{\sigma \mu}}{(x_1-x_2)(x_1-z)^2(x_2-\ov{z})^2(z-\ov{z})}
+\frac{1}{2} \frac{\alpha'  (Dq)^\kappa q^\rho q^\sigma{f}^{\lambda \nu \mu}+ \alpha'  (Dq)^\kappa q^\rho g^{\sigma \nu} h^{\lambda  \mu}}{(x_1-x_2)(x_1-\ov{z})^2(x_2-z)^2(z-\ov{z})}
\crbig
 +\frac{1}{2} \frac{\alpha'    q^\kappa q^\rho  q^\lambda\widetilde{f}^{\sigma \nu \mu}+\alpha'   q^\kappa q^\rho  g^{\lambda \nu}\widetilde{h}^{\sigma \mu}}{(x_1-x_2)(x_1-z)^2(x_2-z)(x_2-\ov{z})(z-\ov{z})}
 + \frac{1}{2}\frac{\alpha'    (Dq)^\kappa (Dq)^\rho q^\lambda\widetilde{f}^{\sigma \nu \mu}+ \alpha'   (Dq)^\kappa (Dq)^\rho g^{\lambda \nu}\widetilde{h}^{\sigma \mu}}{(x_1-x_2)(x_1-z)(x_1-\ov{z})(x_2-\ov{z})^2(z-\ov{z})}
 \crbig
 +\frac{1}{2} \frac{\alpha'    q^\kappa q^\rho  q^\sigma{f}^{\lambda \nu \mu}+ \alpha'  q^\kappa q^\rho  g^{\sigma \nu} h^{\lambda  \mu}}{(x_1-x_2)(x_1-z)(x_1-\ov{z})(x_2-z)^2(z-\ov{z})}
+\frac{1}{2}  \frac{\alpha'    (Dq)^\kappa (Dq)^\rho q^\sigma{f}^{\lambda \nu \mu}+ \alpha'    (Dq)^\kappa (Dq)^\rho g^{\sigma \nu} h^{\lambda  \mu}}{(x_1-x_2)(x_1-\ov{z})^2(x_2-z)(x_2-\ov{z})(z-\ov{z})}
\crbig
 +\frac{1}{2} \frac{\alpha'    (Dq)^\kappa q^\rho q^\lambda\widetilde{f}^{\sigma \nu \mu}+ \alpha'    (Dq)^\kappa q^\rho g^{\lambda \nu}\widetilde{h}^{\sigma \mu}+\alpha'    q^\kappa (Dq)^\rho  q^\sigma{f}^{\lambda \nu \mu}+\alpha'    q^\kappa (Dq)^\rho  g^{\sigma \nu}h^{\lambda  \mu}}{(x_1-x_2)(x_1-z)(x_1-\ov{z})(x_2-z)(x_2-\ov{z})(z-\ov{z})}\bigg\}\,.
\end{array}
\eeq
In the above contraction we have introduced $l^{\mu \nu}$, $h^{\mu \nu},\widetilde{h}^{\mu \nu}$ and $f^{\mu\nu\lambda},\widetilde{f}^{\mu\nu\lambda}$, that we define as follows
\beq
\begin{array}{ccl}
l^{\mu\nu} \equiv-q\cdot Dq D^{\mu\nu}+(Dq)^\nu (Dq)^\mu
\end{array}
\eeq
\beq
\begin{array}{ccl}
h^{\mu\nu}&=-\widetilde{h}^{\mu\nu} \equiv &(Dq)^{\mu}(Dq)^{\nu}-D^{\mu\nu} q \cdot Dq
\end{array}
\eeq
\beq
\begin{array}{ccl}
f^{\mu\nu\lambda}&=-\widetilde{f}^{\mu\nu\lambda} \equiv & -(Dq)^\mu D^{\nu\lambda}+D^{\mu\lambda}(Dq)^\nu \,.
\end{array}
\eeq
For convenience, we also define the objects
\beq
c^{\sigma \nu \lambda} \equiv -g^{\sigma \nu} q^\lambda + g^{\lambda \nu}q^\sigma \quad , \quad d^{\sigma \mu \lambda} \equiv -D^{\sigma \mu} (Dq)^{\lambda}  + D^{\lambda \mu} (Dq)^{\sigma} 
\eeq 
that will appear within certain kinematic packages.

\subsection{The kinematic packages} \label{KinPack}
Here we list our resulting expressions for all kinematic packages; they are exact in $\alpha'$ (i.e. no truncation), with their constituent terms ordered from the highest power ($\alpha'^3$) to the lowest ($\alpha'^1$).
\subsubsection*{Packages for $\mathbf{A}_1$}
\beq \label{kp4,1}
\begin{array}{ccl}
\Theta^{\mu \nu \kappa \lambda \rho \sigma} & =&\frac{1}{8}\alpha'^3 \Big[-
(Dk_2)^\mu q^\kappa  (Dq)^\rho g^{\sigma \nu} q^\lambda +
(Dk_2)^\mu q^\kappa  (Dq)^\rho g^{\lambda \nu}q^\sigma 
\crbig
&&+k_1^\nu q^\kappa (Dq)^\rho D^{\sigma \mu} (Dq)^{\lambda} -   k_1^\nu q^\kappa (Dq)^\rho D^{\lambda \mu} (Dq)^{\sigma}
\Big]
\crbig
&& +\frac{1}{8}\alpha'^2 \Big[D^{\rho \mu}  q^\kappa g^{\sigma \nu} q^\lambda
-D^{\rho \mu}  q^\kappa g^{\lambda \nu}q^\sigma 
\crbig
&& -g^{\kappa \nu}
(Dq)^\rho  D^{\sigma \mu} (Dq)^{\lambda}
 +g^{\kappa \nu}
(Dq)^\rho   D^{\lambda \mu} (Dq)^{\sigma} \Big]
\end{array}
\eeq
\beq \label{kp4,2}
\begin{array}{ccl}
\Lambda^{\mu \nu \kappa \lambda \rho \sigma} &=& \frac{1}{8}\alpha'^3 \Big[-
k_2^\nu (Dq)^\kappa q^\rho D^{\sigma \mu} (Dq)^{\lambda}  +
k_2^\nu (Dq)^\kappa q^\rho D^{\lambda \mu} (Dq)^{\sigma}
\crbig
&&+(Dk_1)^\mu   (Dq)^\kappa q^\rho g^{\sigma \nu} q^\lambda -(Dk_1)^\mu   (Dq)^\kappa q^\rho g^{\lambda \nu}q^\sigma \Big]
\crbig
&& + \frac{1}{8} \alpha'^2 \Big[+g^{\rho \nu}   (Dq)^\kappa D^{\sigma \mu}
(Dq)^{\lambda} -g^{\rho \nu}   (Dq)^\kappa D^{\lambda \mu}
(Dq)^{\sigma} 
\crbig
&& -D^{\kappa \mu}q^\rho g^{\sigma \nu} q^\lambda
 +D^{\kappa \mu}q^\rho g^{\lambda \nu}q^\sigma \Big]
\end{array}
\eeq
\beq \label{kp4,3}
\begin{array}{ccl}
\Xi^{\mu \nu \kappa \lambda \rho \sigma} & =&  \frac{1}{8}\alpha'^3 \Big[-
k_2^\nu q^\kappa (Dq)^\rho D^{\sigma \mu} (Dq)^{\lambda}  +
k_1^\nu q^\kappa q^\rho  D^{\sigma \mu} (Dq)^{\lambda}  -    k_1^\nu q^\kappa q^\rho  D^{\lambda \mu}
(Dq)^{\sigma}\\
&&  -(Dk_2)^\mu   (Dq)^\kappa (Dq)^\rho g^{\sigma \nu} q^\lambda
-(Dk_2)^\mu   q^\kappa q^\rho  g^{\sigma \nu} q^\lambda  +
(Dk_1)^\mu   q^\kappa (Dq)^\rho g^{\sigma \nu} q^\lambda \\
&& +(Dk_2)^\mu   (Dq)^\kappa (Dq)^\rho g^{\lambda \nu}q^\sigma
+(Dk_2)^\mu   q^\kappa q^\rho g^{\lambda \nu}q^\sigma-
(Dk_1)^\mu   q^\kappa (Dq)^\rho g^{\lambda \nu}q^\sigma 
\crbig
&& 
 +   k_2^\nu q^\kappa (Dq)^\rho  D^{\lambda \mu}
(Dq)^{\sigma} +
k_1^\nu (Dq)^\kappa (Dq)^\rho D^{\sigma \mu} (Dq)^{\lambda} \\
&& -    k_1^\nu (Dq)^\kappa (Dq)^\rho D^{\lambda \mu}
(Dq)^{\sigma}\Big] + \frac{1}{8} \alpha'^2 \Big[ -g^{\kappa \nu} q^\rho  D^{\sigma \mu}
(Dq)^{\lambda}+ g^{\kappa \nu} q^\rho  D^{\lambda \mu}
(Dq)^{\sigma}  
\crbig
&&   + D^{\rho \mu}  (Dq)^\kappa g^{\sigma \nu}
q^\lambda   - D^{\rho \mu}  (Dq)^\kappa  g^{\lambda \nu}q^\sigma \Big]
\end{array}
\eeq
\beq \label{kp4,4}
\begin{array}{ccl}
\Sigma^{\mu \nu \kappa \lambda \rho \sigma} &=& \frac{1}{8}\alpha'^3 \Big[ -    k_2^\nu (Dq)^\kappa (Dq)^\rho D^{\sigma \mu}
(Dq)^{\lambda}+   k_2^\nu q^\kappa q^\rho D^{\lambda \mu}
(Dq)^{\sigma}-k_2^\nu q^\kappa q^\rho D^{\sigma \mu} (Dq)^{\lambda}   \\
&&  +   k_2^\nu (Dq)^\kappa (Dq)^\rho  D^{\lambda \mu}
(Dq)^{\sigma}+(Dk_2)^\mu (Dq)^\kappa q^\rho g^{\lambda
\nu}q^\sigma -(Dk_1)^\mu   q^\kappa q^\rho g^{\lambda
\nu}q^\sigma  \\
&&- (Dk_1)^\mu   (Dq)^\kappa (Dq)^\rho  g^{\lambda \nu}q^\sigma-
   k_1^\nu (Dq)^\kappa q^\rho  D^{\lambda \mu} (Dq)^{\sigma} +(Dk_1)^\mu   q^\kappa q^\rho g^{\sigma \nu}
q^\lambda 
\crbig
&&
-(Dk_2)^\mu (Dq)^\kappa q^\rho g^{\sigma \nu}
q^\lambda + k_1^\nu (Dq)^\kappa q^\rho D^{\sigma \mu} (Dq)^{\lambda}\\
&&  +(Dk_1)^\mu   (Dq)^\kappa (Dq)^\rho g^{\sigma \nu} q^\lambda  \Big] + \frac{1}{8}\alpha'^2 \Big[ +g^{\rho \nu}   q^\kappa D^{\sigma \mu}
(Dq)^{\lambda}  -g^{\rho \nu}   q^\kappa D^{\lambda \mu}
(Dq)^{\sigma} 
\crbig
&&  - D^{\kappa \mu}(Dq)^\rho   g^{\sigma \nu}
q^\lambda 
  + D^{\kappa \mu}(Dq)^\rho   g^{\lambda \nu}q^\sigma \Big]
\end{array}
\eeq
\beq \label{kp4,5}
\begin{array}{ccl}
\Gamma^{\mu \nu \kappa \lambda \rho \sigma} & =& \frac{1}{8}\alpha'^3 \Big[
q^\kappa (Dq)^\rho  q^\lambda (Dq)^{\sigma} D^{\nu
\mu}-q^\kappa (Dq)^\rho  q^\lambda D^{\sigma \mu} (Dq)^\nu
\crbig
&&  -q^\kappa (Dq)^\rho  g^{\lambda \nu} (Dq)^\sigma (Dq)^\mu 
 + q^\kappa (Dq)^\rho  g^{\lambda \nu}  D^{\sigma \mu}\, q D
q \Big]
\end{array}
\eeq
\beq \label{kp4,6}
\begin{array}{ccl}
\Delta^{\mu \nu \kappa \lambda \rho \sigma}  & =& \frac{1}{8}\alpha'^3
\Big[-(Dq)^\kappa q^\rho q^\sigma (Dq)^{\lambda} D^{\nu
\mu}+(Dq)^\kappa q^\rho q^\sigma D^{\lambda \mu} (Dq)^{\nu}
\crbig
&& +(Dq)^\kappa q^\rho g^{\sigma \nu} (Dq)^{\lambda } (Dq)^\mu 
 -   (Dq)^\kappa q^\rho g^{\sigma \nu} D^{\lambda \mu} \, q D
q \Big]
\end{array}
\eeq
\beq \label{kp4,7}
\begin{array}{ccl}
\Phi^{\mu \nu \kappa \lambda \rho \sigma} & = &  \frac{1}{16}\alpha'^3 \Big[
(Dq)^\mu   q^\kappa (Dq)^\rho g^{\sigma \nu} q^\lambda -
(Dq)^\mu   q^\kappa (Dq)^\rho  g^{\lambda \nu}q^\sigma  \\
&& +  (Dq)^\nu q^\kappa (Dq)^\rho  D^{\lambda \mu}
(Dq)^{\sigma}   +q^\kappa q^\rho q^\lambda (Dq)^{\sigma}
D^{\nu \mu} -q^\kappa q^\rho q^\lambda D^{\sigma \mu} (Dq)^\nu
 \\
&&  -q^\kappa q^\rho g^{\lambda \nu} (Dq)^\sigma (Dq)^\mu
 +  q^\kappa q^\rho  g^{\lambda \nu} D^{\sigma \mu}\, q D q+
(Dq)^\kappa (Dq)^\rho   q^\lambda (Dq)^{\sigma} D^{\nu \mu}
\\
 && - (Dq)^\kappa (Dq)^\rho   q^\lambda D^{\sigma \mu} (Dq)^\nu
-  (Dq)^\kappa (Dq)^\rho  g^{\lambda \nu} (Dq)^\sigma (Dq)^\mu
\crbig
&& +   (Dq)^\kappa (Dq)^\rho g^{\lambda \nu} D^{\sigma \mu}\, q D
q      -
(Dq)^\nu q^\kappa (Dq)^\rho D^{\sigma \mu} (Dq)^{\lambda}  \Big]
\crbig
\end{array}
\eeq
\beq \label{kp4,8}
\begin{array}{ccl}
\Psi^{\mu \nu \kappa \lambda \rho \sigma} & = &  \frac{1}{16}\alpha'^3 \Big[
(Dq)^\mu    (Dq)^\kappa q^\rho g^{\sigma \nu} q^\lambda
-(Dq)^\mu    (Dq)^\kappa q^\rho  g^{\lambda \nu}q^\sigma  \\
&&  +   (Dq)^\nu (Dq)^\kappa q^\rho D^{\lambda \mu}
(Dq)^{\sigma} -q^\kappa q^\rho q^\sigma (Dq)^{\lambda} D^{\nu
\mu}+q^\kappa q^\rho q^\sigma D^{\lambda \mu} (Dq)^{\nu} 
\\
&& +q^\kappa q^\rho g^{\sigma \nu} (Dq)^{\lambda } (Dq)^\mu  -q^\kappa q^\rho 
g^{\sigma \nu} D^{\lambda \mu} \, q D q
-(Dq)^\kappa (Dq)^\rho q^\sigma (Dq)^{\lambda} D^{\nu \mu}
\\
&& +(Dq)^\kappa (Dq)^\rho q_2^\sigma D^{\lambda \mu} (Dq)^{\nu} 
 +(Dq)^\kappa (Dq)^\rho  g^{\sigma \nu} (Dq)^{\lambda }
(Dq)^\mu 
\crbig
&& -(Dq)^\kappa (Dq)^\rho  g^{\sigma \nu} D^{\lambda \mu} \,
q D q -
(Dq)^\nu (Dq)^\kappa q^\rho D^{\sigma \mu} (Dq)^{\lambda}\Big]
\end{array}
\eeq
\beq \label{kp4,9}
\begin{array}{ccl}
\Omega^{\mu \nu \kappa \lambda \rho \sigma} & =& \frac{1}{16}\alpha'^3 \Big[ -
(Dq)^\nu  q^\kappa q^\rho D^{\sigma \mu} (Dq)^{\lambda}  -  
(Dq)^\nu  (Dq)^\kappa (Dq)^\rho D^{\sigma \mu} (Dq)^{\lambda}
 \\
&& -(Dq)^\mu  q^\kappa q^\rho g^{\lambda \nu}q^\sigma +
 (Dq)^\nu  q^\kappa q^\rho   D^{\lambda \mu} (Dq)^{\sigma} +
(Dq)^\mu  (Dq)^\kappa (Dq)^\rho g^{\sigma \nu} q^\lambda  
\\
&& - (Dq)^\mu  (Dq)^\kappa (Dq)^\rho g^{\lambda \nu}q^\sigma +(Dq)^\kappa q^\rho
q^\lambda(Dq)^{\sigma} D^{\nu \mu}+(Dq)^\mu  q^\kappa q^\rho g^{\sigma \nu} q^\lambda \\
&&  - (Dq)^\kappa q^\rho g^{\lambda \nu}(Dq)^\sigma (Dq)^\mu  +
(Dq)^\kappa q^\rho g^{\lambda \nu}  D^{\sigma \mu}\, q D q
\\
&& +q^\kappa (Dq)^\rho  q^\sigma D^{\lambda \mu} (Dq)^{\nu}+
q^\kappa (Dq)^\rho  g^{\sigma \nu} (Dq)^{\lambda } (Dq)^\mu   
\crbig
&& +    (Dq)^\nu  (Dq)^\kappa (Dq)^\rho D^{\lambda \mu}
(Dq)^{\sigma}-q^\kappa (Dq)^\rho  g^{\sigma \nu}  D^{\lambda \mu} \, q D q 
\crbig
&& -q^\kappa (Dq)^\rho q^\sigma(Dq)^{\lambda} D^{\nu \mu} -(Dq)^\kappa q^\rho q^\lambda D^{\sigma \mu}
(Dq)^\nu\Big]
\end{array}
\eeq
and
\beq \label{kpp4}
\begin{array}{ccl}
\Gamma\equiv  \alpha^1_{\kappa \lambda} \alpha^2_{\rho 
\sigma} \varepsilon_{\mu \nu} \, \Big[\Gamma^{\mu \nu \kappa \lambda \rho \sigma}+\Theta^{\mu \nu 
\kappa \lambda
\rho \sigma} \Big], \,\,\,\,\,\,\,\,\, \Delta\equiv  \alpha^1_{\kappa \lambda} \alpha^2_{\rho 
\sigma} \varepsilon_{\mu \nu} \,  \Big[ \Delta^{\mu \nu \kappa \lambda \rho
\sigma} - \Lambda^{\mu \nu \kappa \lambda \rho
\sigma} \Big]
\end{array}
\eeq
\beq \label{kpp4,2}
\begin{array}{ccl}
\Theta&\equiv & \alpha^1_{\kappa \lambda} \alpha^2_{\rho 
\sigma}\varepsilon_{\mu \nu} \,  \Big[\Theta^{\mu \nu \kappa \lambda
\rho \sigma} -\Phi^{\mu \nu \kappa \lambda \rho
\sigma}-\Xi^{\mu \nu \kappa \lambda \rho
\sigma} \Big]
\end{array}
\eeq
\beq \label{kpp4,3}
\begin{array}{ccl}
\Lambda&\equiv & \alpha^1_{\kappa \lambda} \alpha^2_{\rho 
\sigma}\varepsilon_{\mu \nu} \,  \Big[\Lambda^{\mu \nu \kappa \lambda \rho
\sigma}+\Psi^{\mu \nu \kappa \lambda
\rho \sigma} -\Sigma^{\mu \nu \kappa \lambda \rho
\sigma} \Big]
\end{array}
\eeq
\beq \label{kpp4,4}
\begin{array}{ccl}
\Xi&\equiv & \alpha^1_{\kappa \lambda} \alpha^2_{\rho 
\sigma}\varepsilon_{\mu \nu} \,   \Big[\Lambda^{\mu \nu \kappa \lambda \rho
\sigma}+\Psi^{\mu \nu \kappa \lambda
\rho \sigma}-\Omega^{\mu \nu \kappa
\lambda \rho \sigma}+\Xi^{\mu \nu \kappa \lambda \rho
\sigma} - 2\Sigma^{\mu \nu \kappa \lambda \rho
\sigma} \Big]
\end{array}
\eeq
\beq \label{kpp4.5}
\begin{array}{ccl}
\Sigma&\equiv & \alpha^1_{\kappa \lambda} \alpha^2_{\rho 
\sigma} \varepsilon_{\mu \nu} \,  \Big[\Theta^{\mu \nu \kappa \lambda
\rho \sigma} -\Phi^{\mu \nu \kappa \lambda \rho
\sigma}+\Omega^{\mu \nu \kappa
\lambda \rho \sigma}-2\Xi^{\mu \nu \kappa \lambda \rho
\sigma}+\Sigma^{\mu \nu \kappa \lambda \rho
\sigma} \Big]
\end{array}
\eeq
\beq \label{kpp4,6}
\begin{array}{ccl}
\Omega_+&\equiv & \alpha^1_{\kappa \lambda} 
\alpha^2_{\rho \sigma} \varepsilon_{\mu \nu} \,  \Big[- \Lambda^{\mu \nu \kappa \lambda \rho
\sigma}+\Theta^{\mu \nu \kappa \lambda
\rho \sigma}-\Phi^{\mu \nu \kappa \lambda \rho
\sigma}-\Psi^{\mu \nu \kappa \lambda
\rho \sigma}+2\Omega^{\mu \nu \kappa
\lambda \rho \sigma}
\crbig
&& \hspace{2.1cm} -3\Xi^{\mu \nu \kappa \lambda \rho
\sigma}+3\Sigma^{\mu \nu \kappa \lambda \rho
\sigma}\Big]
\end{array}
\eeq
\beq \label{kpp4,7}
\begin{array}{ccl}
\Omega_-&\equiv & \alpha^1_{\kappa \lambda} 
\alpha^2_{\rho \sigma} \varepsilon_{\mu \nu} \,  \Big[ -\Lambda^{\mu \nu \kappa \lambda \rho
\sigma}+\Theta^{\mu \nu \kappa \lambda
\rho \sigma}-\Phi^{\mu \nu \kappa \lambda \rho
\sigma}-\Psi^{\mu \nu \kappa \lambda
\rho \sigma}+2\Omega^{\mu \nu \kappa
\lambda \rho \sigma}
\crbig
&& \hspace{2.1cm} -3\Xi^{\mu \nu \kappa \lambda \rho
\sigma}+3\Sigma^{\mu \nu \kappa \lambda \rho
\sigma}\Big]
\end{array}
\eeq

\subsubsection*{Packages for $\mathbf{A}_2$}
\beq\label{kp3,1}
\begin{array}{ccl}
P^{\mu \nu \kappa \rho}  & =&   \frac{1}{4} \,\alpha'^3\big[ k_1^\nu (Dk_2)^\mu 
q^\kappa (Dq)^\rho]-\frac{1}{4}\alpha'^2 [D^{\rho \mu} k_1^\nu q^\kappa +g^{\kappa 
\nu} (Dq)^\rho (Dk_2)^\mu]+\frac{1}{4}\alpha' g^{\kappa \nu} D^{\rho \mu}
\end{array}
\eeq
\beq\label{kp3,2}
\begin{array}{ccl}
\widetilde{P}^{\mu \nu \kappa \rho}  & =& \frac{1}{4} \,\alpha'^3\big[ (Dk_1)^\mu  k_2^\nu 
(Dq)^\kappa q^\rho]- \frac{1}{4}\alpha'^2[(Dk_1)^\mu g^{\rho \nu} (Dq)^\kappa+ 
D^{\kappa \mu} q^\rho k_2^\nu] + \frac{1}{4}\alpha'   D^{\kappa \mu} 
g^{\rho \nu}
\end{array}
\eeq
\beq\label{kp3,3}
\begin{array}{ccl}
Q^{\mu \nu \kappa \rho \lambda \sigma}  & =& \frac{1}{4} \alpha'^3 \Big[ [ -  k_1^\nu q^\kappa 
(Dk_1)^\mu (Dq)^\rho +  k_1^\nu q^\kappa(Dk_2)^\mu q^\rho+(Dk_2)^\mu  
(Dq)^\rho k_1^\nu (Dq)^\kappa
\crbig
&&  - (Dk_2)^\mu (Dq)^\rho  k_2^\nu q^\kappa] g^{\lambda \sigma}-    k_1^\nu k_2^\kappa (Dq)^\rho D^{\lambda 
\mu} (Dq)^{\sigma} +  D^{\sigma \mu} (Dq)^{\lambda} k_1^\nu k_2^\kappa 
(Dq)^\rho
\crbig
&& +(Dk_2)^\mu   k_2^\kappa (Dq)^\rho g^{\lambda \nu}q^\sigma  + 
(Dk_2)^\mu   q^\kappa k_1^\rho g^{\lambda \nu}q^\sigma +  k_1^\nu q^\kappa k_1^\rho D^{\sigma \mu} (Dq)^{\lambda}
 \crbig
&& - (Dk_2)^\mu   k_2^\kappa (Dq)^\rho g^{\sigma \nu} q^\lambda     - k_1^\nu 
q^\kappa k_1^\rho D^{\lambda \mu} (Dq)^{\sigma}- g^{\sigma \nu} 
q^\lambda  (Dk_2)^\mu   q^\kappa k_1^\rho \Big]
\crbig
&& +\frac{1}{4}\alpha'^2 \Big[ [ g^{\kappa \nu} (Dq)^\rho (Dk_1)^\mu - g^{\kappa \nu} 
q^\rho (Dk_2)^\mu+ D^{\rho \mu}  k_2^\nu q^\kappa - D^{\rho \mu} k_1^\nu 
(Dq)^\kappa ] g^{\lambda \sigma}
\crbig
&&+g^{\kappa \nu}k_1^\rho D^{\lambda \mu} 
(Dq)^{\sigma}- D^{\sigma \mu} (Dq)^{\lambda} g^{\kappa \nu}k_1^\rho- 
D^{\rho \mu}   k_2^\kappa g^{\lambda \nu}q^\sigma  + g^{\sigma \nu} 
q^\lambda  D^{\rho \mu} k_2^\kappa \Big]
\end{array}
\eeq
\beq\label{kp3,4}
\begin{array}{ccl}
\widetilde{Q}^{\mu \nu \kappa \rho \lambda \sigma}  &=& \frac{1}{4}\alpha'^3 \Big[ [ - (Dk_1)^\mu(Dq)^\kappa   
k_1^\nu  q^\rho -(Dk_1)^\mu  (Dq)^\kappa k_2^\nu q^\rho + (Dk_1)^\mu  
(Dq)^\kappa   k_2^\nu  (Dq)^\rho
\crbig
&&   + (Dk_1)^\mu k_2^\nu q^\kappa q^\rho ] g^{\lambda \sigma} +D^{\lambda \mu} (Dq)^{\sigma} k_2^\nu 
k_2^\kappa q^\rho+ D^{\lambda \mu} (Dq)^{\sigma}  k_2^\nu (Dq)^\kappa 
k_1^\rho 
\crbig
&& -  D^{\sigma \mu} (Dq)^{\lambda} k_2^\nu k_2^\kappa q^\rho  - D^{\sigma \mu} (Dq)^{\lambda} k_2^\nu (Dq)^\kappa k_1^\rho - g^{\lambda \nu}q^\sigma (Dk_1)^\mu   (Dq)^\kappa k_1^\rho 
\crbig
&& -  g^{\lambda \nu}q^\sigma (Dk_1)^\mu   k_2^\kappa q^\rho + g^{\sigma \nu} 
q^\lambda (Dk_1)^\mu   (Dq)^\kappa k_1^\rho+ g^{\sigma \nu} q^\lambda 
(Dk_1)^\mu   k_2^\kappa q^\rho  \Big]
\crbig
&& +\frac{1}{4}\alpha'^2 \Big[ [ D^{\kappa \mu}   q^\rho k_1^\nu - D^{\kappa \mu}   
(Dq)^\rho k_2^\nu + g^{\rho \nu}  (Dk_2)^\mu (Dq)^\kappa - g^{\rho \nu}  
(Dk_1)^\mu q^\kappa ] g^{\lambda \sigma}
\crbig
&&  - D^{\lambda \mu} (Dq)^{\sigma} g^{\rho 
\nu}    k_2^\kappa+ D^{\sigma \mu} (Dq)^{\lambda} g^{\rho \nu}    
k_2^\kappa+ g^{\lambda \nu}q^\sigma D^{\kappa \mu}k_1^\rho  - g^{\sigma 
\nu} q^\lambda D^{\kappa \mu}k_1^\rho \Big]
\end{array}
\eeq
\beq\label{kp3,5}
\begin{array}{ccl}
R^{\mu \nu \kappa \rho \lambda \sigma} & =&\frac{1}{4}\alpha'^3 \Big[ [- (Dk_1)^\mu k_1^\nu  
q^\kappa q^\rho -(Dk_1)^\mu  k_1^\nu   (Dq)^\kappa (Dq)^\rho  + 
(Dk_1)^\mu  k_2^\nu q^\kappa (Dq)^\rho 
\crbig
&& -(Dk_2)^\mu k_2^\nu q^\kappa q^\rho -(Dk_2)^\mu  k_2^\nu  (Dq)^\kappa (Dq)^\rho  + (Dk_2)^\mu k_1^\nu 
(Dq)^\kappa q^\rho ] g^{\lambda \sigma}
\crbig
&& -  D^{\lambda \mu} (Dq)^{\sigma} k_1^\nu 
(Dq)^\kappa k_1^\rho  - D^{\lambda \mu} (Dq)^{\sigma}  k_1^\nu 
k_2^\kappa q^\rho+    D^{\lambda \mu} (Dq)^{\sigma} k_2^\nu q^\kappa 
k_1^\rho 
\crbig
&&  +   D^{\lambda \mu} (Dq)^{\sigma}   k_2^\nu k_2^\kappa (Dq)^\rho  + D^{\sigma \mu} (Dq)^{\lambda} k_1^\nu (Dq)^\kappa k_1^\rho  + 
D^{\sigma \mu} (Dq)^{\lambda}  k_1^\nu k_2^\kappa q^\rho 
\crbig 
&& - D^{\sigma \mu} 
(Dq)^{\lambda}  k_2^\nu q^\kappa k_1^\rho -   D^{\sigma \mu} 
(Dq)^{\lambda}   k_2^\nu k_2^\kappa (Dq)^\rho - g^{\lambda \nu}q^\sigma (Dk_1)^\mu  q^\kappa k_1^\rho 
\crbig 
 &&-g^{\lambda \nu}q^\sigma (Dk_1)^\mu   k_2^\kappa (Dq)^\rho+ g^{\lambda 
\nu}q^\sigma (Dk_2)^\mu   (Dq)^\kappa k_1^\rho+ g^{\lambda \nu}q^\sigma 
(Dk_2)^\mu   k_2^\kappa q^\rho \,
\crbig
&&  +  g^{\sigma \nu} q^\lambda  (Dk_1)^\mu  q^\kappa k_1^\rho +  
g^{\sigma \nu} q^\lambda  (Dk_1)^\mu   k_2^\kappa (Dq)^\rho- g^{\sigma 
\nu} q^\lambda  (Dk_2)^\mu   (Dq)^\kappa k_1^\rho
\crbig
&& -  g^{\sigma \nu} 
q^\lambda  (Dk_2)^\mu   k_2^\kappa q^\rho \Big]
\crbig
&&+ \frac{1}{4}\alpha'^2[+ g^{\kappa \nu} q^\rho (Dk_1)^\mu+D^{\kappa \mu}(Dq)^\rho 
k_1^\nu +g^{\rho \nu}(Dk_2)^\mu q^\kappa + D^{\rho \mu}k_2^\nu (Dq)^\kappa] g^{\lambda \sigma}
\end{array}
\eeq
\beq\label{kp3,6}
\begin{array}{ccl}
S^{\mu \nu \kappa \rho \lambda \sigma} & =&-\frac{1}{8} \alpha'^3 \Big[-  [(Dq)^\nu q^\kappa  (Dk_2)^\mu 
(Dq)^\rho+ (Dq)^\mu (Dq)^\rho k_1^\nu q^\kappa]g^{\lambda \sigma}
\crbig
&&+  k_1^\rho q^\kappa  q^\lambda 
(Dq)^{\sigma} D^{\nu \mu} -  k_2^\kappa (Dq)^\rho q^\lambda D^{\sigma 
\mu} (Dq)^\nu -  k_2^\kappa (Dq)^\rho g^{\lambda \nu} (Dq)^\sigma 
(Dq)^\mu
\crbig
&&
 - k_1^\rho q^\kappa  q^\lambda D^{\sigma \mu} (Dq)^\nu  +k_2^\kappa (Dq)^\rho q^\lambda (Dq)^{\sigma} D^{\nu \mu}  - k_1^\rho 
q^\kappa   g^{\lambda \nu} (Dq)^\sigma (Dq)^\mu 
\crbig
&& +  k_2^\kappa (Dq)^\rho 
g^{\lambda \nu}  D^{\sigma \mu}\, q D q + k_1^\rho q^\kappa  g^{\lambda 
\nu}  D^{\sigma \mu}\, q D q \Big]
\crbig
&&- \frac{1}{8}\alpha'^2 [(Dq)^\nu 
q^\kappa  D^{\rho \mu} + (Dq)^\mu (Dq)^\rho g^{\kappa \nu} ]g^{\lambda \sigma}
\end{array}
\eeq
\beq\label{kp3,7}
\begin{array}{ccl}
\widetilde{S}^{\mu \nu \kappa \rho \lambda \sigma} & =&\frac{1}{8}\alpha'^3 \Big[ (Dq)^\mu (Dq)^\kappa k_2^\nu 
q^\rho+ (Dq)^\nu  q^\rho (Dk_1)^\mu (Dq)^\kappa]g^{\lambda \sigma}
\crbig
&& +[  k_2^\kappa q^\rho+   (Dq)^\kappa k_1^\rho 
\big]   \big[-  q^\sigma (Dq)^{\lambda} D^{\nu \mu}+  q^\sigma 
D^{\lambda \mu} (Dq)^{\nu}+ g^{\sigma \nu} (Dq)^{\lambda } (Dq)^\mu
\crbig
&&  - g^{\sigma \nu} D^{\lambda \mu} \, q D q \big]\Big]-\frac{1}{8}\alpha'^2[ 
(Dq)^\mu (Dq)^\kappa g^{\rho \nu} + (Dq)^\nu  q^\rho D^{\kappa \mu}]g^{\lambda \sigma}
\end{array}
\eeq
\beq\label{kp3,8}
\begin{array}{ccl}
T^{\mu \nu \kappa \rho \lambda \sigma}& =& \,\frac{1}{8} \alpha'^3\bigg[ [(Dq)^\nu (Dq)^\kappa 
(Dk_2)^\mu  (Dq)^\rho + (Dq)^\mu  (Dq)^\rho  k_1^\nu (Dq)^\kappa 
\crbig
&& - 
(Dq)^\mu  (Dq)^\rho k_2^\nu q^\kappa \, -(Dq)^\nu q^\kappa (Dk_1)^\mu  (Dq)^\rho  +(Dq)^\nu 
q^\kappa(Dk_2)^\mu  q^\rho  
\crbig
&& +  (Dq)^\mu  q^\rho k_1^\nu q^\kappa ] g^{\lambda \sigma} -\big[  (Dq)^\kappa k_1^\rho + k_2^\kappa q^\rho 
\big] \, \Big[ q^\lambda \big[ (Dq)^{\sigma} D^{\nu \mu}-D^{\sigma \mu} 
(Dq)^\nu \big] 
\crbig
&& +  g^{\lambda \nu} \big[-(Dq)^\sigma (Dq)^\mu  + 
D^{\sigma \mu}\, q D q  \big] \Big]
\crbig
&& - \big[ q^\kappa k_1^\rho+ k_2^\kappa (Dq)^\rho \big] \, 
\Big[  (Dq)^\nu  \, \big[-D^{\sigma \mu} (Dq)^{\lambda}  + D^{\lambda 
\mu} (Dq)^{\sigma} \big] 
\crbig
&& +(Dq)^\mu \,  \big[ -g^{\sigma \nu} q^\lambda + 
g^{\lambda \nu}q^\sigma\big]\Big] \bigg]- \frac{1}{8}\alpha'^2 [ (Dq)^\nu (Dq)^\kappa D^{\rho \mu}+(Dq)^\mu q^\rho 
g^{\kappa \nu}]g^{\lambda \sigma}
\end{array}
\eeq
\beq\label{kp3,9}
\begin{array}{ccl}
\widetilde{T}^{\mu \nu \kappa \rho \lambda \sigma} & =&\frac{1}{8}\alpha'^3 \bigg[ [(Dq)^\mu q^\kappa k_2^\nu  q^\rho 
+ (Dq)^\nu   (Dq)^\rho (Dk_1)^\mu (Dq)^\kappa - (Dq)^\mu  (Dq)^\kappa 
k_1^\nu q^\rho
\crbig
&& + (Dq)^\mu  (Dq)^\kappa k_2^\nu  (Dq)^\rho - (Dq)^\nu q^\rho (Dk_1)^\mu q^\kappa +  (Dq)^\nu q^\rho (Dk_2)^\mu  (Dq)^\kappa]g^{\lambda \sigma}
\crbig
&&+ \big[
k_1^\rho q^\kappa  + k_2^\kappa (Dq)^\rho  
\big]\, \Big[ q^\sigma \big[-(Dq)^{\lambda} D^{\nu \mu}+D^{\lambda \mu} 
(Dq)^{\nu} \big]+ g^{\sigma \nu} \big[ (Dq)^{\lambda } (Dq)^\mu 
\crbig
&& - D^{\lambda \mu} \, q D q \big] \Big]  + \big[ -(Dq)^\mu q^\kappa k_1^\rho 
-(Dq)^\mu  k_2^\kappa (Dq)^\rho \big] \,\big[ -g^{\sigma \nu} q^\lambda 
+ g^{\lambda \nu}q^\sigma\big]
\crbig
&& +  \big[  (Dq)^\nu  k_2^\kappa q^\rho +    (Dq)^\nu q^\kappa 
k_1^\rho \big] \,\big[-D^{\sigma \mu} (Dq)^{\lambda} + D^{\lambda \mu} 
(Dq)^{\sigma} \big] \Bigg]
\crbig
&&
-\frac{1}{8}\alpha'^2[(Dq)^\mu q^\kappa g^{\rho \nu}+(Dq)^\nu   (Dq)^\rho D^{\kappa 
\mu}  ] g^{\lambda \sigma}
\end{array}
\eeq
\beq\label{kp3,10}
\begin{array}{ccl}
U^{\mu \nu \kappa \rho} & =&   \,\frac{1}{16}\alpha'^3 [q_2^\kappa (Dq)^\rho 
(Dq)^\mu (Dq)^\nu  +q^\kappa (Dq)^\rho  q D 
q \, D^{\nu \mu} - q^\kappa (Dq)^\rho    (Dq)^\nu (Dq)^\mu  ] 
\crbig
&&- \frac{1}{8}\alpha'^2 \,\, q^\kappa (Dq)^\rho D^{\nu \mu} 
\end{array}
\eeq
\beq\label{kp3,11}
\begin{array}{ccl}
\widetilde{U}^{\mu \nu \kappa \rho} & =&  \, \frac{1}{16}\alpha'^3 [ (Dq)^\kappa q^\rho 
(Dq)^\mu (Dq)^\nu  +(Dq)^\kappa q^\rho  q D q \, 
D^{\nu \mu} -(Dq)^\kappa q^\rho  (Dq)^\nu (q D)^\mu ] 
\crbig
&&-\frac{1}{8}\alpha'^2 \,\,(Dq)^\kappa q^\rho D^{\nu \mu} 
\end{array}
\eeq
\beq\label{kp3,12}
\begin{array}{ccl}
 W^{\mu \nu \kappa \rho} & =& \frac{1}{16}\alpha'^3  \, \Big[ (Dq)^\mu (Dq)^\nu 
(Dq)^\kappa (Dq)^\rho + (Dq)^\mu (Dq)^\nu q^\kappa q^\rho 
\crbig
&& - \big[  q^\kappa q^\rho +  (Dq)^\kappa (Dq)^\rho 
\big]  \big[ -q Dq  \, D^{\nu \mu} +  (Dq)^\nu (Dq)^\mu \big] \Big]
\crbig
&&- \frac{1}{8} \, 
\alpha'^2\big[ D^{\nu \mu} (Dq)^\rho (Dq)^\kappa - D^{\nu \mu} q^\rho 
q^\kappa \big]
\end{array}
\eeq

and
\beq \label{kpp3}
\begin{array}{ccl}
P \equiv \alpha_{\kappa \lambda}^1 \alpha_{\rho \sigma}^2 \varepsilon_{\mu \nu}g^{\lambda \sigma}  \, P^{\mu \nu \kappa
\rho}, \,\,\,\,\,\,\, \widetilde{P} \equiv \alpha_{\kappa \lambda}^1 \alpha_{\rho \sigma}^2 \varepsilon_{\mu \nu}g^{\lambda \sigma}  \, \widetilde{P}^{\mu \nu \kappa\rho} 
\end{array}
\eeq
\beq \label{kpp3,2}
\begin{array}{ccl}
S&\equiv& \alpha_{\kappa \lambda}^1 \alpha_{\rho \sigma}^2 \varepsilon_{\mu \nu} \, \big[S^{\mu \nu \kappa \rho \lambda \sigma}- Q^{\mu \nu \kappa \rho \lambda \sigma} \big]
\end{array}
\eeq
\beq \label{kpp3,3}
\begin{array}{ccl}
\widetilde{S}&\equiv& \alpha_{\kappa \lambda}^1 \alpha_{\rho \sigma}^2 \varepsilon_{\mu \nu}  \,\big[\widetilde{S}^{\mu \nu \kappa \rho \lambda \sigma}- \widetilde{Q}^{\mu \nu \kappa \rho \lambda \sigma} \big]
\end{array}
\eeq
\beq \label{kpp3,4}
\begin{array}{ccl}
U&\equiv& \alpha_{\kappa \lambda}^1 \alpha_{\rho \sigma}^2 \varepsilon_{\mu \nu}  \, \big[ U^{\mu \nu \kappa \rho} g^{ \lambda \sigma}- Q^{\mu \nu \kappa \rho \lambda \sigma}+ R^{\mu \nu \kappa \rho \lambda \sigma}-  T^{\mu \nu \kappa \rho \lambda \sigma}\big]
\end{array}
\eeq
\beq \label{kpp3,5}
\begin{array}{ccl}
\widetilde{U}&\equiv& \alpha_{\kappa \lambda}^1 \alpha_{\rho \sigma}^2 \varepsilon_{\mu \nu} \,\big[ \widetilde{U}^{\mu \nu \kappa \rho} g^{ \lambda \sigma}- \widetilde{Q}^{\mu \nu \kappa \rho \lambda \sigma}+ R^{\mu \nu \kappa \rho \lambda \sigma}- \widetilde{T}^{\mu \nu \kappa \rho \lambda \sigma} \big]
\end{array}
\eeq
\beq \label{kpp3,6}
\begin{array}{ccl}
W&\equiv& \alpha_{\kappa \lambda}^1 \alpha_{\rho \sigma}^2 \varepsilon_{\mu \nu}  \, \big[ W^{\mu \nu \kappa \rho} g^{ \lambda \sigma} +Q^{\mu \nu \kappa \rho \lambda \sigma}+ \widetilde{Q}^{\mu \nu \kappa \rho \lambda \sigma}-2  R^{\mu \nu \kappa \rho \lambda \sigma}
\crbig
&& \hspace{1.8cm}+  T^{\mu \nu \kappa \rho \lambda \sigma}+ \widetilde{ T}^{\mu \nu \kappa \rho \lambda \sigma} \big]
\end{array}
\eeq
\beq \label{kpp3,7}
\begin{array}{ccl}
\widetilde{W}&\equiv& \alpha_{\kappa \lambda}^1 \alpha_{\rho \sigma}^2 \varepsilon_{\mu \nu} \,\big[ W^{\mu \nu \kappa \rho} g^{ \lambda \sigma} +Q^{\mu \nu \kappa \rho \lambda \sigma}+ \widetilde{Q}^{\mu \nu \kappa \rho \lambda \sigma}-2  R^{\mu \nu \kappa \rho \lambda \sigma}
\crbig
&&\hspace{1.8cm} +  T^{\mu \nu \kappa \rho \lambda \sigma}+  \widetilde{ T}^{\mu \nu \kappa \rho \lambda \sigma} \big]
\end{array}
\eeq

\subsubsection*{Packages for $\mathbf{A}_3$}
\beq \label{kp2,1}
\begin{array}{ccl}
G^{\mu \nu \kappa \rho} &=& \frac{\alpha'^3}{2}  \, \big[  (Dk_2)^\mu k_1^\nu k_2^\kappa (Dq)^\rho \ + (Dk_2)^\mu k_1^\nu q^\kappa k_1^\rho  \big]  - \frac{\alpha'^2}{2}  \, \big[D^{\rho \mu} k_1^\nu k_2^\kappa  + g^{\kappa \nu}k_1^\rho (Dk_2)^\mu    \big] 
\end{array}
\eeq
\beq \label{kp2,2}
\begin{array}{ccl}
H^{\mu \nu \kappa \rho }&=& \frac{\alpha'^3}{2} \, \big[ (Dk_1)^\mu k_2^\nu k_2^\kappa q^\rho  +  (Dk_1)^\mu k_2^\nu (Dq)^\kappa k_1^\rho   \big]  - \frac{\alpha'^2}{2}  \, \big[g^{\rho \nu} (Dk_1)^\mu k_2^\kappa + D^{\kappa \mu}k_1^\rho k_2^\nu  \big] 
\end{array}
\eeq
\beq \label{kp2,3}
\begin{array}{ccl}
I^{\mu \nu \kappa \rho \lambda \sigma}&=& \frac{1}{2} \alpha'^3 \,\Big\{ \big[ -  (Dk_2)^\mu k_2^\nu k_2^\kappa (Dq)^\rho - (Dk_1)^\mu k_1^\nu q^\kappa k_1^\rho -  (Dk_1)^\mu k_1^\nu k_2^\kappa (Dq)^\rho 
\crbig
&&   +(Dk_2)^\mu k_1^\nu (Dq)^\kappa k_1^\rho   +  (Dk_2)^\mu k_1^\nu k_2^\kappa q^\rho  -(Dk_2)^\mu k_2^\nu q^\kappa k_1^\rho  \big] \, g^{\lambda \sigma} -k_1^\nu k_2^\kappa k_1^\rho \, d^{\sigma  \mu \lambda }
\crbig
&&  + (Dk_2)^\mu   k_2^\kappa k_1^\rho \,c^{\sigma \nu \lambda} \Big\}  +\frac{1}{2} \alpha'^2 \, \big[D^{\rho \mu}k_2^\nu k_2^\kappa \, g^{\lambda \sigma} 
\crbig
&& + g^{\kappa \nu}(Dk_1)^\mu k_1^\rho  \, g^{\lambda \sigma}
  + \frac{1}{2}g^{\kappa \rho}   k_1^\nu \, d^{\sigma  \mu \lambda } 
 - \frac{1}{2} g^{\kappa \rho} (Dk_2)^\mu \, c^{\sigma \nu \lambda}  \big] 
\end{array}
\eeq
\beq \label{kp2,4}
\begin{array}{ccl}
J^{\mu \nu \kappa \rho \lambda \sigma}&=& \frac{1}{2} \alpha'^3 \, \Big\{ \big[-(Dk_1)^\mu k_1^\nu (Dq)^\kappa k_1^\rho   - (Dk_1)^\mu k_1^\nu k_2^\kappa q^\rho  + (Dk_1)^\mu k_2^\nu q^\kappa k_1^\rho
 \crbig
&& +  (Dk_1)^\mu k_2^\nu k_2^\kappa (Dq)^\rho  - (Dk_2)^\mu k_2^\nu (Dq)^\kappa k_1^\rho -  (Dk_2)^\mu k_2^\nu k_2^\kappa q^\rho \big]  \, g^{\lambda \sigma} \crbig
&&- (Dk_1)^\mu   k_2^\kappa k_1^\rho \, c^{\sigma \nu \lambda}  +  k_2^\nu k_2^\kappa k_1^\rho \, d^{\sigma \mu \lambda }\Big\}  +\frac{1}{2} \alpha'^2 \, \big[ D^{\kappa \mu}k_1^\rho k_1^\nu  \, g^{\lambda \sigma} 
 \crbig
&&+  g^{\rho \nu} (Dk_2)^\mu k_2^\kappa g^{\lambda \sigma}
  + \frac{1}{2}g^{\kappa \rho}(Dk_1)^\mu \, c^{\sigma \nu \lambda} 
 -\frac{1}{2} g^{\kappa \rho}   k_2^\nu \, d^{\sigma \mu \lambda } \big] 
\end{array}
\eeq
\beq \label{kp2,5}
\begin{array}{ccl}
K^{\mu \nu \kappa \rho \lambda \sigma}&=& \frac{1}{8} \alpha'^3 \,\Big\{ \big[-  (Dk_2)^\mu (Dq)^\nu k_2^\kappa (Dq)^\rho -  (Dq)^\mu k_1^\nu q^\kappa k_1^\rho  - (Dk_2)^\mu (Dq)^\nu q^\kappa k_1^\rho
 \crbig
&& - (Dq)^\mu k_1^\nu k_2^\kappa (Dq)^\rho  \big] g^{\lambda \sigma}  +   k_2^\kappa k_1^\rho q^\lambda \,\widetilde{f}^{\sigma \nu \mu}+  k_2^\kappa k_1^\rho g^{\lambda \nu}\, \widetilde{h}^{\sigma \mu} 
\Big\} 
\crbig
&& + \frac{1}{8} \alpha'^2 \, \big[ g^{\kappa \nu}k_1^\rho (Dq)^\mu+ D^{\rho \mu}(Dq)^\nu k_2^\kappa  g^{\lambda \sigma} - \frac{1}{2}g^{\kappa \rho} q^\lambda \, \widetilde{f}^{\sigma \nu \mu} - \frac{1}{2} g^{\kappa \rho}g^{\lambda \nu}  \, \widetilde{h}^{\sigma \mu}  \big] 
\end{array}
\eeq
\beq \label{kp2,6}
\begin{array}{ccl}
L^{\mu \nu \kappa \rho \lambda \sigma}&=& \frac{1}{8} \alpha'^3 \, \Big\{ \big[ (Dq)^\mu k_2^\nu k_2^\kappa q^\rho  + (Dk_1)^\mu (Dq)^\nu (Dq)^\kappa k_1^\rho  + (Dk_1)^\mu (Dq)^\nu k_2^\kappa q^\rho  
\crbig
&& +  (Dq)^\mu k_2^\nu (Dq)^\kappa k_1^\rho \big]  g^{\lambda \sigma}  +  k_2^\kappa k_1^\rho q^\sigma \, {f}^{\lambda \nu \mu}+ k_2^\kappa k_1^\rho g^{\sigma \nu} \, h^{\lambda  \mu}  \Big\}  
\crbig
&& + \frac{1}{8}\alpha'^2 \, \big[- D^{\kappa \mu} k_1^\rho (Dq)^\nu-g^{\rho \nu}(Dq)^\mu k_2^\kappa g^{\lambda \sigma} - \frac{1}{2} g^{\kappa \rho} q^\sigma \,{f}^{\lambda \nu \mu}- \frac{1}{2} g^{\kappa \rho}g^{\sigma \nu} \, h^{\lambda  \mu}  \big] 
\end{array}
\eeq
\beq \label{kp2,7}
\begin{array}{ccl}
M^{\mu \nu \kappa \rho \lambda \sigma}&=& \frac{1}{8}  \alpha'^3 \, \Big\{ \big[ - (Dk_2)^\mu (Dq)^\nu (Dq)^\kappa k_1^\rho  + (Dk_1)^\mu (Dq)^\nu k_2^\kappa (Dq)^\rho 
 \crbig
&& + (Dq)^\mu k_2^\nu q^\kappa k_1^\rho + (Dk_1)^\mu (Dq)^\nu q^\kappa k_1^\rho   -(Dq)^\mu k_1^\nu (Dq)^\kappa k_1^\rho - (Dq)^\mu k_1^\nu k_2^\kappa q^\rho 
\crbig
&&  + (Dq)^\mu k_2^\nu k_2^\kappa (Dq)^\rho   -(Dk_2)^\mu (Dq)^\nu k_2^\kappa q^\rho  \big] g^{\lambda \sigma}  +   (Dq)^\nu k_2^\kappa k_1^\rho d^{\sigma \mu \lambda } 
\crbig
&&   -  (Dq)^\mu   k_2^\kappa k_1^\rho c^{\sigma \nu \lambda} \Big\} + \frac{1}{16} \, \alpha'^2 \, \big[ - g^{\kappa \rho}   (Dq)^\nu d^{\sigma \mu \lambda }  + g^{\kappa \rho}(Dq)^\mu c^{\sigma \nu \lambda} \big] 
\end{array}
\eeq
\beq \label{kp2,8}
\begin{array}{ccl}
N^{\mu \nu \kappa \rho }&=& \frac{1}{32} \alpha'^3 \, \big[ - (Dq)^\mu (Dq)^\nu k_2^\kappa q^\rho  - (Dq)^\mu (Dq)^\nu (Dq)^\kappa k_1^\rho +  (Dq)^\kappa k_1^\rho l^{\nu \mu}   
 \crbig 
&& +  k_2^\kappa q^\rho l^{\nu \mu} \big]  + \frac{1}{16} \alpha'^2 \, \big[ D^{\nu \mu}q^\rho k_2^\kappa  + D^{\nu \mu}k_1^\rho (Dq)^\kappa  \big] 
\end{array}
\eeq
\beq \label{kp2,9}
\begin{array}{ccl}
O^{\mu \nu \kappa \rho }&=& \frac{1}{32} \alpha'^3 \, \big[- (Dq)^\mu (Dq)^\nu k_2^\kappa (Dq)^\rho   - (Dq)^\mu (Dq)^\nu q^\kappa k_1^\rho
\crbig
&&  + k_1^\rho q^\kappa l^{\nu \mu}  +   k_2^\kappa (Dq)^\rho l^{\nu \mu} \big] +\frac{1}{16} \alpha'^2 \, \big[  D^{\nu \mu}(Dq)^\rho k_2^\kappa  + D^{\nu \mu}k_1^\rho q^\kappa \big] 
\end{array}
\eeq

and

\beq \label{kpp2}
\begin{array}{ccl}
G  &=&  \alpha_{\kappa \lambda}^1 \alpha_{\rho \sigma}^2 \varepsilon_{\mu \nu}g^{ \lambda \sigma}  \, G^{\mu \nu \kappa \rho} \quad , \quad H =\alpha_{\kappa \lambda}^1 \alpha_{\rho \sigma}^2 \varepsilon_{\mu \nu}g^{ \lambda \sigma}  \, H^{\mu \nu \kappa \rho} 
\crbig
K & \equiv & \, \alpha_{\kappa \lambda}^1 \alpha_{\rho \sigma}^2 \varepsilon_{\mu \nu}\,  \big[ K^{\mu \nu \kappa \rho \lambda \sigma } + \frac{1}{2} I^{\mu \nu \kappa \rho \lambda \sigma }   \big]
\quad,\quad
L = \, \alpha_{\kappa \lambda}^1 \alpha_{\rho \sigma}^2 \varepsilon_{\mu \nu}\, \big[ L^{\mu \nu \kappa \rho \lambda \sigma } - \frac{1}{2} J^{\mu \nu \kappa \rho \lambda \sigma }  \big]
\crbig
N & \equiv & \, \alpha_{\kappa \lambda}^1 \alpha_{\rho \sigma}^2 \varepsilon_{\mu \nu}\,  \big[ 4N^{\mu \nu \kappa \rho} g^{\lambda \sigma} + 2 M^{\mu \nu \kappa \rho \lambda \sigma} - I^{\mu \nu \kappa \rho\lambda \sigma} + J^{\mu \nu \kappa \rho \lambda \sigma} \big]
\crbig
O & \equiv & \, \alpha_{\kappa \lambda}^1 \alpha_{\rho \sigma}^2 \varepsilon_{\mu \nu}\,  \big[ 4O^{\mu \nu \kappa \rho} g^{\lambda \sigma} -2 M^{\mu \nu \kappa \rho \lambda \sigma} + I^{\mu \nu \kappa \rho\lambda \sigma} - J^{\mu \nu \kappa \rho \lambda \sigma} \big]
\end{array}
\eeq

\subsubsection*{Packages for $\mathbf{A}_4$}
\beq \label{kp1}
\begin{array}{ccl}
A^{\mu \nu  \kappa \rho  } &=& \frac{1}{16} \alpha'^3 \, \big[  l^{\nu \mu}  - (Dq)^\mu (Dq)^\nu   \big]  k_2^\kappa k_1^\rho 
\crbig
&&+\frac{1}{8}\alpha'^2 \, \Big[ D^{\nu \mu} k_1^\rho k_2^\kappa  -\frac{1}{4}  \big( l^{\nu \mu}  -  (Dq)^\mu (Dq)^\nu  \big) g^{\kappa \rho}  \Big]  - \frac{1}{16} \alpha' D^{\nu \mu}g^{\kappa \rho}
\end{array}
\eeq
\beq \label{kp1,1}
\begin{array}{ccl}
B^{\mu \nu  \kappa \rho } & =& - \alpha'^3 \, \big[(Dk_1)^\mu k_1^\nu + (Dk_2)^\mu k_2^\nu \big] k_2^\kappa k_1^\rho    + \frac{1}{2} \alpha'^2 \,  \big[(Dk_1)^\mu k_1^\nu + (Dk_2)^\mu k_2^\nu \big] g^{\kappa \rho}
\end{array}
\eeq
\beq \label{kp1,2}
\begin{array}{ccl}
C^{\mu \nu  \kappa \rho } & =&  \alpha'^3 \, (Dk_1)^\mu k_2^\nu k_2^\kappa k_1^\rho- \frac{1}{2} \alpha'^2 \,(Dk_1)^\mu k_2^\nu g^{\kappa \rho}
\end{array}
\eeq
\beq \label{kp1,3}
\begin{array}{ccl}
\widetilde{\Delta}^{\mu \nu  \kappa \rho } & =& \alpha'^3 \,(Dk_2)^\mu k_1^\nu k_2^\kappa k_1^\rho- \frac{1}{2} \alpha'^2 \,  (Dk_2)^\mu k_1^\nu g^{\kappa \rho}
\end{array}
\eeq
\beq \label{kp1,4}
\begin{array}{ccl}
E^{\mu \nu  \kappa \rho } &=&-\frac{\alpha'^3}{2} \, \big[ (Dk_1)^\mu (Dq)^\nu +(Dq)^\mu k_2^\nu  \big] k_2^\kappa k_1^\rho + \frac{\alpha'^2}{4} \, \big[(Dk_1)^\mu (Dq)^\nu + (Dq)^\mu k_2^\nu \big] g^{\kappa \rho}
\end{array}
\eeq
\beq \label{kp1,5}
\begin{array}{ccl}
F^{\mu \nu  \kappa \rho } &=& \frac{\alpha'^3}{2}  \, \big[ (Dk_2)^\mu (Dq)^\nu +(Dq)^\mu k_1^\nu  \big] k_2^\kappa k_1^\rho - \frac{\alpha'^2}{4} \, \big[(Dk_2)^\mu (Dq)^\nu +(Dq)^\mu k_1^\nu \big] g^{\kappa \rho}
\end{array}
\eeq
and
\beq
\begin{array}{ccl} \label{kpp1}
A & \equiv& \alpha_{\kappa \lambda}^1 \alpha_{\rho \sigma}^2 \varepsilon_{\mu \nu}g^{\lambda \sigma} \, A^{\mu \nu  \kappa \rho} \,,\,
C = \alpha_{\kappa \lambda}^1 \alpha_{\rho \sigma}^2 \varepsilon_{\mu \nu}g^{\lambda \sigma} \, C^{\mu \nu  \kappa \rho} 
\,,\,
\widetilde{\Delta}  =\alpha_{\kappa \lambda}^1 \alpha_{\rho \sigma}^2 \varepsilon_{\mu \nu}g^{\lambda \sigma} \, \widetilde{\Delta}^{\mu \nu  \kappa \rho}
\crbig
E &\equiv &  \alpha_{\kappa \lambda}^1 \alpha_{\rho \sigma}^2 \varepsilon_{\mu \nu}\, g^{\lambda \sigma} \big[2E^{\mu \nu  \kappa \rho }+B^{\mu \nu  \kappa \rho } \big]
\quad , \quad
F =  \alpha_{\kappa \lambda}^1 \alpha_{\rho \sigma}^2 \varepsilon_{\mu \nu}\, g^{\lambda \sigma} \big[2F^{\mu \nu  \kappa \rho }-B^{\mu \nu  \kappa \rho } \big]
\end{array}
\eeq

\subsection{Computation of the relevant integrals} \label{appInt}

To calculate the integrals of $x$ that appear in our calculation, we first observe that
\beq
\begin{array}{ccl}
\int_{-\infty}^{+\infty} dx |x|^{s+2}(x^2+1)^{-s}\frac{(x+i)(x-i)}{x^4} &=& 2  \int_{0}^{+\infty} dx \, x^{s-2}(x^2+1)^{-s}(x+i)(x-i)
\crbig
&=& 2 \frac{2^{-s}\sqrt{\pi} \, s \, \Gamma\big(\frac{s-1}{2}\big)}{\Gamma\big(\frac{s}{2} +1\big)} \quad \textrm{if} \quad \Re(s)>1\,.
\end{array}
\eeq 
Similarly,
\beq
\begin{array}{ccl}
\int_{-\infty}^{+\infty} dx |x|^{s+2}(x^2+1)^{-s}\frac{(x+i)(x-i)}{x^4(x+i)^2} = \int_{-\infty}^{+\infty} dx |x|^{s+2}(x^2+1)^{-s}\frac{(x+i)(x-i)}{x^4(x-i)^2} = I_C + \ov{I}_C
\end{array}
\eeq
\beq
\begin{array}{ccl}
\int_{-\infty}^{+\infty} dx |x|^{s+2}(x^2+1)^{-s}\frac{(x+i)(x-i)}{x^4(x+i)} =- \int_{-\infty}^{+\infty} dx |x|^{s+2}(x^2+1)^{-s}\frac{(x+i)(x-i)}{x^4(x-i)} =   I_E- \ov{I}_E  
\end{array}
\eeq
\beq
\begin{array}{ccl}
 \int\limits_{-\infty}^{\infty} dx \,  |x|^{s+2}  (x^2+1)^{-s} \frac{(x+i)(x-i)}{x^3(x-i)^3}  = \int\limits_{-\infty}^{\infty} dx \,  |x|^{s+2}  (x^2+1)^{-s} \frac{(x+i)(x-i)}{x^3(x+i)^3}   = I_G +\ov{I}_G 
\end{array}
\eeq
\beq
\begin{array}{ccl}
 \int\limits_{-\infty}^{\infty} dx \,  |x|^{s+2}  (x^2+1)^{-s} \frac{(x+i)(x-i)}{x^3(x-i)^2}  =- \int\limits_{-\infty}^{\infty} dx \,  |x|^{s+2}  (x^2+1)^{-s} \frac{(x+i)(x-i)}{x^3(x+i)^2}   = I_K -\ov{I}_K 
\end{array}
\eeq
\beq
\begin{array}{ccl}
 \int\limits_{-\infty}^{\infty} dx \,  |x|^{s+2}  (x^2+1)^{-s} \frac{(x+i)(x-i)}{x^3(x-i)}  = \int\limits_{-\infty}^{\infty} dx \,  |x|^{s+2}  (x^2+1)^{-s} \frac{(x+i)(x-i)}{x^3(x+i)}= I_O +\ov{I}_O
\end{array}
\eeq
\beq
\begin{array}{ccl}
 \int\limits_{-\infty}^{\infty} dx \,  |x|^{s+2}  (x^2+1)^{-s} \frac{(x+i)(x-i)}{x^2(x-i)^4}  = \int\limits_{-\infty}^{\infty} dx \,  |x|^{s+2}  (x^2+1)^{-s} \frac{(x+i)(x-i)}{x^2(x+i)^4} =I_P +\ov{I}_P
\end{array}
\eeq
\beq
\begin{array}{ccl}
 \int\limits_{-\infty}^{\infty} dx \,  |x|^{s+2}  (x^2+1)^{-s} \frac{(x+i)(x-i)}{x^2(x-i)^3}  =- \int\limits_{-\infty}^{\infty} dx \,  |x|^{s+2}  (x^2+1)^{-s} \frac{(x+i)(x-i)}{x^2(x+i)^3} =I_S -\ov{I}_S
\end{array}
\eeq
\beq
\begin{array}{ccl}
 \int\limits_{-\infty}^{\infty} dx \,  |x|^{s+2}  (x^2+1)^{-s} \frac{(x+i)(x-i)}{x^2(x-i)^2}  = \int\limits_{-\infty}^{\infty} dx \,  |x|^{s+2}  (x^2+1)^{-s} \frac{(x+i)(x-i)}{x^2(x+i)^2} =I_U + \ov{I}_U
\end{array}
\eeq
\beq
\begin{array}{ccl}
 \int\limits_{-\infty}^{\infty} dx \,  |x|^{s+2}  (x^2+1)^{-s} \frac{(x+i)(x-i)}{x^2(x-i)}  =- \int\limits_{-\infty}^{\infty} dx \,  |x|^{s+2}  (x^2+1)^{-s} \frac{(x+i)(x-i)}{x^2(x+i)} =I_W -\ov{I}_W
\end{array}
\eeq
\beq
\begin{array}{ccl}
 \int\limits_{-\infty}^{\infty} dx \,  |x|^{s+2}  (x^2+1)^{-s} \frac{(x+i)(x-i)}{x(x-i)^4}  = -\int\limits_{-\infty}^{\infty} dx \,  |x|^{s+2}  (x^2+1)^{-s} \frac{(x+i)(x-i)}{x(x+i)^4} =I_\Gamma -\ov{I}_\Gamma
\end{array}
\eeq
\beq
\begin{array}{ccl}
 \int\limits_{-\infty}^{\infty} dx \,  |x|^{s+2}  (x^2+1)^{-s} \frac{(x+i)(x-i)}{x(x-i)^3}  = \int\limits_{-\infty}^{\infty} dx \,  |x|^{s+2}  (x^2+1)^{-s} \frac{(x+i)(x-i)}{x(x+i)^3} =I_\Theta +\ov{I}_\Theta
\end{array}
\eeq
\beq
\begin{array}{ccl}
 \int\limits_{-\infty}^{\infty} dx \,  |x|^{s+2}  (x^2+1)^{-s} \frac{(x+i)(x-i)}{x(x-i)^2}  = -\int\limits_{-\infty}^{\infty} dx \,  |x|^{s+2}  (x^2+1)^{-s} \frac{(x+i)(x-i)}{x(x+i)^2} =I_\Sigma - \ov{I}_\Sigma
\end{array}
\eeq
\beq
\begin{array}{ccl}
 \int\limits_{-\infty}^{\infty} dx \,  |x|^{s+2}  (x^2+1)^{-s} \frac{(x+i)(x-i)}{x(x-i)}  =\int\limits_{-\infty}^{\infty} dx \,  |x|^{s+2}  (x^2+1)^{-s} \frac{(x+i)(x-i)}{x(x+i)} =I_{\Omega_-} +\ov{I}_{\Omega_-}
\end{array}
\eeq

where
\beq
\begin{array}{ccl}
I_C &\equiv &  -\frac{1}{2} \bigg[\frac{2^{-s}\sqrt{\pi} \Gamma\big(\frac{s-1}{2} \big)}{\Gamma \big( \frac{s}{2}+1 \big)} + i\frac{\Gamma\big( \frac{s}{2}\big)^2}{\Gamma(s)} \bigg] \quad \textrm{if} \quad \Re(s)>1
\end{array}
\eeq
\beq
\begin{array}{ccl}
I_E & \equiv  & \frac{\Gamma \big(\frac{s}{2} \big)^2}{2\Gamma(s)} -i \frac{(s-1)\Gamma\big(\frac{s-1}{2} \big)^2}{4\Gamma(s)} \quad \textrm{if} \quad \Re(s)>1
\end{array}
\eeq
\beq
\begin{array}{ccl}
I_G & \equiv  & -\frac{\pi ^{3/2} 2^{-s-1} \sec \left(\frac{\pi  s}{2}\right)}{\Gamma \left(\frac{1}{2}-\frac{s}{2}\right)
   \Gamma \left(\frac{s}{2}+1\right)}+\frac{i \pi  (s-1) \csc \left(\frac{\pi  s}{2}\right) \Gamma
   \left(\frac{s}{2}\right)}{\Gamma \left(-\frac{s}{2}-1\right) \Gamma (s+3)} \quad \textrm{if} \quad \Re(s)>0
\end{array}
\eeq
\beq
\begin{array}{ccl}
I_K & \equiv  &\frac{i \sqrt{\pi } 2^{-s} \Gamma \left(\frac{s+1}{2}\right)}{\Gamma \left(\frac{s}{2}+1\right)} \quad \textrm{if} \quad \Re(s)>0
\end{array}
\eeq
\beq
\begin{array}{ccl}
I_O & \equiv  &\frac{\Gamma \left(\frac{s-1}{2}\right) \Gamma \left(\frac{s+1}{2}\right)+i \Gamma
   \left(\frac{s}{2}\right)^2}{2 \Gamma (s)}\quad \textrm{if}\quad \Re(s)>1
\end{array}
\eeq
\beq
\begin{array}{ccl}
I_P & \equiv  &-\frac{2^{-s-2} \sqrt{\pi } s \Gamma \left(\frac{s+1}{2}\right)}{\Gamma
   \left(\frac{s}{2}+2\right)} \quad \textrm{if}\quad \Re(s)>-1
\end{array}
\eeq
\beq
\begin{array}{ccl}
I_S & \equiv  & \frac{(1-s) \Gamma \left(\frac{s}{2}+1\right) \Gamma \left(\frac{s}{2}\right)+2 i \Gamma
   \left(\frac{s+1}{2}\right) \Gamma \left(\frac{s+3}{2}\right)}{2 \Gamma (s+2)}\quad \textrm{if}\quad \Re(s)>0
\end{array}
\eeq
\beq
\begin{array}{ccl}
I_U & \equiv  & \frac{\Gamma \left(\frac{s-1}{2}\right) \Gamma \left(\frac{s+1}{2}\right)+2 i \Gamma \left(\frac{s}{2}+1\right)
   \Gamma \left(\frac{s}{2}\right)}{2 \Gamma (s+1)}\quad \textrm{if}\quad \Re(s)>1
\end{array}
\eeq
\beq
\begin{array}{ccl}
I_W & \equiv  &-\frac{\pi  \csc \left(\frac{\pi  s}{2}\right) \Gamma \left(\frac{s}{2}+1\right)}{2 \Gamma
   \left(2-\frac{s}{2}\right) \Gamma (s)}-\frac{i \pi ^{3/2} 2^{-s} \sec \left(\frac{\pi  s}{2}\right)}{\Gamma
   \left(\frac{3}{2}-\frac{s}{2}\right) \Gamma \left(\frac{s}{2}\right)}\quad \textrm{if}\quad \Re(s)>2
\end{array}
\eeq
\beq
\begin{array}{ccl}
I_\Gamma &\equiv & \frac{(1-s) \Gamma \left(\frac{s}{2}+2\right) \Gamma \left(\frac{s}{2}\right)+2 i \Gamma \left(\frac{s+1}{2}\right) \Gamma
   \left(\frac{s+3}{2}\right)}{\Gamma (s+3)}\quad \textrm{if}\quad \Re(s)>0
\end{array}
\eeq
\beq
\begin{array}{ccl}
I_\Theta &\equiv & \sqrt{\pi } 2^{-s-2} \left(\frac{\pi  (s-3) \sec \left(\frac{\pi  s}{2}\right)}{\Gamma \left(\frac{3}{2}-\frac{s}{2}\right)
   \Gamma \left(\frac{s}{2}+1\right)}+\frac{i (s+3) \Gamma \left(\frac{s}{2}\right)}{\Gamma \left(\frac{s+3}{2}\right)}\right)\quad \textrm{if}\quad \Re(s)>1
\end{array}
\eeq
\beq
\begin{array}{ccl}
I_\Sigma &\equiv & \frac{\sqrt{\pi } 2^{-s} \Gamma \left(\frac{s}{2}-1\right)}{\Gamma \left(\frac{s+1}{2}\right)}+\frac{i \Gamma
   \left(\frac{s-1}{2}\right) \Gamma \left(\frac{s+3}{2}\right)}{\Gamma (s+1)}\quad \textrm{if}\quad \Re(s)>2
\end{array}
\eeq
\beq
\begin{array}{ccl}
I_{\Omega_-} &\equiv & \frac{\pi  \left(\frac{\sec \left(\frac{\pi  s}{2}\right) \Gamma \left(\frac{s+3}{2}\right)}{\Gamma
   \left(\frac{5}{2}-\frac{s}{2}\right)}-\frac{i \csc \left(\frac{\pi  s}{2}\right) \Gamma \left(\frac{s}{2}+1\right)}{\Gamma
   \left(2-\frac{s}{2}\right)}\right)}{2 \Gamma (s)}\quad \textrm{if}\quad \Re(s)>3
\end{array}
\eeq

\subsection{Expansions} \label{app_expansions}

\beq
\begin{array}{ccl}
\mathbf{A}_1 &=& \,\, (\Gamma-\Delta) \, \big[\frac{\pi 
}{4}+\frac{1}{2}  \pi  \, \alpha' \, k_1 \cdot k_3+\frac{1}{6} 
 \pi  \left(6+\pi ^2\right) \, \alpha'^2 \, (k_1 \cdot k_3)^2+\mathcal{O}\left(\alpha'^3\right)  \big]
  \crbig
&&  -(\Theta+\Lambda) \, \big[\frac{3\pi}{8}- \pi \,  \alpha' \, k_1 \cdot 
k_3+\frac{1}{4} \pi (16+\pi^2) \,\alpha'^2 \, ( k_1 \cdot k_3)^2 +\mathcal{O}\left(\alpha'^3\right)  \big]
\crbig
&&+(\Xi+\Sigma)\, \big[ \frac{\pi}{8} - \pi \, \alpha' \,  k_1 
\cdot  k_3 +\frac{1}{12} \pi (48+\pi^2) \, \alpha'^2 \, (k_1 \cdot k_3)^2  +\mathcal{O}\left(\alpha'^3\right)  \big]
\crbig
&&+(\Omega_+-\Omega_-) \, \big[-\frac{1}{12}\pi \, \alpha'  \, k_1 
\cdot k_3+\frac{7}{9} \pi \, \alpha'^2  \,( k_1 \cdot k_3)^2+\mathcal{O}\left(\alpha'^3\right)  \big]
\crbig
&=&  \frac{\pi \, \alpha'^2}{16}  \big[\Tr(\varepsilon \cdot \alpha^1) \alpha^2_{\mu \nu} q^\mu q^\nu + \Tr(\varepsilon \cdot \alpha^2) \alpha^1_{\mu \nu} q^\mu q^\nu -2(\alpha^1 \cdot \varepsilon \cdot \alpha^2 )_{\mu \nu} q^\mu q^\nu \big]+ \mathcal{O}\left(\alpha'^3\right)
\end{array}
\eeq
\beq
\begin{array}{ccl}
\mathbf{A}_2 &=&  (P + \widetilde{P})\,\,\big[\frac{1}{2} \pi  \,  \alpha' \, k_1 \cdot k_3 + \pi \, \alpha'^2  \,( k_1 \cdot k_3)^2+\mathcal{O}\left(\alpha'^3\right) \big]
\crbig
&&+\, (S+\widetilde{S}) \,\, \big[-\frac{\pi}{8}  -\frac{1}{12} \pi^3\,  \alpha'^2  \, ( k_1 \cdot k_3)^2   +\mathcal{O}\left(\alpha'^3\right) \big]
\crbig
&&+ (U+\widetilde{U}) \,\, \big[\frac{\pi}{8}- \frac{1}{2}  \pi\, \alpha' \, k_1 \cdot k_3 +\frac{1}{12} \pi (24+\pi^2)\,\alpha'^2 \, ( k_1 \cdot k_3)^2 +\mathcal{O}\left(\alpha'^3\right)  \big]
\crbig
&& + (W+\widetilde{W}) \,\,\big[-\frac{\pi}{4} \, \alpha' \,  k_1 \cdot 
k_3  + \pi \, \alpha'^2 \,( k_1 \cdot k_3)^2 +\mathcal{O}\left(\alpha'^3\right) \big]
\crbig
&=&  \frac{\pi}{8}\alpha'^2 \, \big[\Tr(\varepsilon \cdot \alpha^1 \cdot \alpha^2) \, k_1 \cdot q +(\varepsilon \cdot \alpha^1 \cdot \alpha^2)_{\mu \nu}  k_1^\mu q^\nu +(\varepsilon \cdot \alpha^2 \cdot \alpha^1)_{\mu \nu}  k_2^\mu q^\nu \big]  
\crbig
&& + \mathcal{O}(\alpha'^3)
\end{array}
\eeq
\beq
\begin{array}{ccl}
\mathbf{A}_3 &=&    (G+H) \,  \big[ -\frac{\pi}{8} - \frac{1}{12}\pi^3 \, \alpha'^2 \, (k_1 \cdot k_3)^2  +\mathcal{O}\left(\alpha'^3\right) \big]
\crbig
&& +(K-L)\, \big[\frac{\pi}{4}+\frac{1}{6}\pi^3 \,\alpha'^2 \, (k_1 \cdot k_3)^2  +\mathcal{O}\left(\alpha'^3\right) \big]
\crbig
&& +(N+O)\,  \big[\frac{1}{4}\pi  \, \alpha' \,  k_1 \cdot k_3 -\pi \alpha'^2 \,(k_1 \cdot k_3)^2   +\mathcal{O}\left(\alpha'^3\right)  \big] 
\crbig
&=& \frac{\pi}{8} \alpha'^2 \, \big[(\varepsilon \cdot \alpha^2 \cdot \alpha^1)_{\mu \nu}  k_1^\mu k_2^\nu +(\varepsilon \cdot \alpha^1 \cdot \alpha^2)_{\mu \nu}  k_1^\nu k_2^\mu \big] + \mathcal{O}(\alpha'^3)
\end{array}
\eeq
\beq \label{expampl}
\begin{array}{ccl}
\mathbf{A}_4 &=& A \, \big[ \pi \, \alpha' \, k_1 \cdot k_3   -4 \pi \, \alpha'^2  (k_1 \cdot k_3)^2 \, +\mathcal{O}\left(\alpha'^3\right) \big]
\crbig
&&  +  (C+\widetilde{\Delta}) \, \big[\frac{\pi}{8} -\frac{1}{2} \pi \, \alpha' \, k_1 \cdot k_3   +\frac{1}{12} \pi (24+\pi^2) \, \alpha'^2 \, (k_1 \cdot k_3)^2 +\mathcal{O}\left(\alpha'^3\right) \big]
\crbig
&& +  (E-F)\, \big[ \frac{1}{4}\pi \,\alpha' \,  k_1 \cdot k_3   - \pi \, \alpha'^2 \, (k_1 \cdot k_3)^2+\mathcal{O}\left(\alpha'^3\right) \big] 
\crbig
&=&  -\frac{\pi}{8} \alpha'^2 \,\Tr(\alpha^1 \cdot \alpha^2) \, \varepsilon_{\mu \nu} k_1^\mu k_2^\nu  + \mathcal{O}(\alpha'^3)\,,
\end{array}
\eeq
where $q$ is related to $k_3$ and $k_4$ via the definitions (\ref{def_momenta}) in the special case (\ref{sv}). 

\section{Calculations for the $\mathcal{A}(3,0)$ amplitude} \label{app1}

\subsection{Contractions for the supersymmetric case} \label{contractions_susy}

Performing the contractions in (\ref{usef1}) using the correlators (\ref{corrs}), we obtain
\beq \label{Xi2}
\begin{array}{ccl}
\mathcal{X}_1&=& i\widetilde{\mathcal{E}}\frac{1}{x_{12}^2 x_{13}^2x_{23}^2}\Big[  (2\alpha')^2  \big[ g^{\nu_1 \nu_2} g^{\mu_1 \mu_3} g^{\mu_2 \nu_3} +  g^{\nu_1 \nu_2} g^{\mu_1 \nu_3} g^{\mu_2 \mu_3}\big] 
\crbig
&& +(2\alpha')^3 \big[g^{\nu_1 \nu_2} g^{\mu_1 \mu_2} k_1^{\mu_3} k_1^{\nu_3}- g^{\nu_1 \nu_2} g^{\mu_1 \mu_3} k_1^{\nu_3} k_1^{\mu_2} - g^{\nu_1 \nu_2} g^{\mu_1 \nu_3} k_1^{\mu_3} k_1^{\mu_2} 
\crbig
&& + g^{\nu_1 \nu_2} g^{\mu_2 \mu_3} k_2^{\mu_1} k_1^{\nu_3}+ g^{\nu_1 \nu_2} g^{\mu_2 \nu_3} k_2^{\mu_1} k_1^{\mu_3}\big] -(2\alpha')^4 \, g^{\nu_1 \nu_2}  k_1^{\mu_3} k_1^{\nu_3}k_1^{\mu_2} k_2^{\mu_1}\Big]
\end{array}
\eeq
\beq
\begin{array}{ccl}
\mathcal{X}_2 &=&  i\widetilde{\mathcal{E}}\frac{1}{x_{12}^2 x_{13}^2x_{23}^2}\Big[  (2\alpha')^2 \, g^{\nu_1 \mu_3} g^{\mu_1 \mu_2} g^{\nu_2 \nu_3} -(2\alpha')^3 g^{\nu_1 \mu_3} g^{\nu_2 \nu_3} k_1^{\mu_2} k_2^{\mu_1} \Big]
\end{array}
\eeq
\beq 
\begin{array}{ccl}
\mathcal{X}_3 & =& i\widetilde{\mathcal{E}}\frac{1}{x_{12}^2 x_{13}^2x_{23}^2}\Big\{  (2\alpha')^3 \big[- g^{\mu_1 \mu_2} g^{\nu_2 \nu_3} k_1^{\mu_3} k_3^{\nu_1}+ g^{\mu_1 \mu_2} g^{\nu_1 \nu_3} k_1^{\mu_3} k_3^{\nu_2}+ g^{\mu_1 \mu_3} g^{\nu_2 \nu_3} k_1^{\mu_2} k_3^{\nu_1}
\crbig
&&-  g^{\mu_1 \mu_3} g^{\nu_1 \nu_3} k_1^{\mu_2} k_3^{\nu_2}- g^{\mu_2 \mu_3} g^{\nu_2 \nu_3} k_2^{\mu_1} k_3^{\nu_1}+ g^{\mu_2 \mu_3} g^{\nu_1 \nu_3} k_2^{\mu_1} k_3^{\nu_2}\big] 
\crbig
&&+(2\alpha')^4 \big[ g^{\nu_2 \nu_3} k_3^{\nu_1} k_1^{\mu_3} k_1^{\mu_2} k_2^{\mu_1} -g^{\nu_1 \nu_3} k_3^{\nu_2} k_1^{\mu_3} k_1^{\mu_2} k_2^{\mu_1}  \big] \Big\} \,,
\end{array}
\eeq
where we have used the momentum conservation (\ref{consM3}) and the on--shell conditions. For  $\mathcal{X}_2$, we have also used the fact that the polarization tensors are symmetric, in particular for the case of $\alpha^3_{\mu_3 \nu_3}$. This was necessary, for both the term with no momenta and the one with two momenta, in order to extract the overall $x$--dependence of $\mathcal{X}_2$. 

\subsection{Contractions for the bosonic case} \label{contractions_bosonic}

\beq \label{Xi2_2}
\begin{array}{ccl}
- \langle :  \partial X^{\mu_1}_1 \partial X^{\nu_1}_1 e^{ik_1 X_1}: \,:  \partial X^{\mu_2}_2  \partial X^{\nu_2}_2 e^{ik_2 X_2}  :  \, :  \partial X^{\mu_3}_3  \partial X^{\nu_3}_3 e^{ik_3 X_3}: \rangle
\crbig
=\frac{(2\alpha')^3\,\widetilde{\mathcal{E}}}{x_{12}^2 x_{13}^2 x_{23}^2} \bigg\{  \big[ g^{\mu_1 \mu_2} g^{\nu_1 \mu_3} g^{\nu_2 \nu_3}+ g^{\mu_1 \mu_2} g^{\nu_1 \nu_3} g^{\nu_2 \mu_3} + g^{\mu_1 \nu_2} g^{\nu_1 \mu_3} g^{\mu_2 \nu_3}+ g^{\mu_1 \nu_2} g^{\nu_1 \nu_3} g^{\mu_2 \mu_3}
\crbig
+g^{\mu_1 \mu_3} g^{\nu_1 \mu_2} g^{\nu_2 \nu_3}+ g^{\mu_1 \mu_3} g^{\nu_1 \nu_2} g^{\mu_2 \nu_3} + g^{\mu_1 \nu_3} g^{\nu_1 \mu_2} g^{\nu_2 \mu_3}+ g^{\mu_1 \nu_3} g^{\nu_1 \nu_2} g^{\mu_2 \mu_3} \big]
\crbig
+4\alpha' \Big[g^{\mu_1 \mu_3} g^{\nu_1 \nu_3} k_1^{\mu_2} k_1^{\nu_2}+ g^{\mu_1 \mu_2} g^{\nu_1 \nu_2} k_2^{\mu_3} k_2^{\nu_3}+ g^{\mu_2 \mu_3} g^{\nu_2 \nu_3} k_3^{\mu_1} k_3^{\nu_1}
\crbig
+ g^{\mu_1 \mu_2} g^{\nu_2 \mu_3} k_1^{\nu_3} k_2^{\nu_1}+ g^{\mu_1 \mu_2} g^{\nu_2 \nu_3} k_1^{\mu_3} k_2^{\nu_1}- g^{\mu_1 \mu_2} g^{\nu_1 \mu_3} k_1^{\nu_3} k_1^{\nu_2}- g^{\mu_1 \mu_2} g^{\nu_1 \nu_3} k_1^{\mu_3} k_1^{\nu_2}
\crbig
- g^{\mu_1 \nu_2} g^{\nu_1 \mu_3} k_1^{\nu_3} k_1^{\mu_2}- g^{\mu_1 \nu_2} g^{\nu_1 \nu_3} k_1^{\mu_3} k_1^{\mu_2}- g^{\mu_1 \mu_3} g^{\mu_2 \nu_3} k_1^{\nu_2} k_2^{\nu_1}- g^{\mu_1 \mu_3} g^{\nu_2 \nu_3} k_1^{\mu_2} k_2^{\nu_1}
\crbig
+ g^{\nu_1 \mu_2} g^{\nu_2 \mu_3} k_1^{\nu_3} k_2^{\mu_1}+ g^{\nu_1 \mu_2} g^{\nu_2 \nu_3} k_1^{\mu_3} k_2^{\mu_1}- g^{\nu_1 \mu_3} g^{\mu_2 \nu_3} k_1^{\nu_2} k_2^{\mu_1}- g^{\nu_1 \mu_3} g^{\nu_2 \nu_3} k_1^{\mu_2} k_2^{\mu_1} \Big]
\crbig
-16\alpha'^2 \Big[ g^{\mu_1 \mu_2} k_1^{\mu_3} k_1^{\nu_3} k_1^{\nu_2} k_2^{\nu_1} - g^{\mu_1 \mu_3} k_1^{\mu_2} k_1^{\nu_2} k_1^{\nu_3} k_2^{\nu_1}+g^{\mu_2 \mu_3} k_2^{\mu_1} k_2^{\nu_1} k_1^{\nu_3} k_1^{\nu_2} \Big]
\crbig
+ 8\alpha'^3 \, k_1^{\mu_2} k_1^{\nu_2} k_2^{\mu_3} k_2^{\nu_3} k_3^{\mu_1} k_3^{\nu_1} \bigg\}\,.
\end{array}
\eeq
Notice that (\ref{Xi2})--(\ref{Xi2_2}) are valid strictly when contracted with the polarization tensors within the amplitude and that we are able to factorize their $x$--dependence and write it as the overall prefactor $\frac{1}{x_{12}^2 x_{13}^2x_{23}^2}$.

\end{document}